\documentclass[10pt, oneside]{article}
\pdfoutput=1
\usepackage[percent]{overpic} 
\usepackage{amsmath}
\usepackage{multirow} 
\usepackage{fancyvrb} 
\usepackage{hyperref}
\usepackage[hyperpageref]{backref}
\usepackage{enumitem} 
\usepackage{easylist}
\usepackage{graphicx}							
\usepackage{amssymb}
\usepackage[usenames]{color}
\usepackage{epsf}
\usepackage[usenames,dvipsnames]{pstricks}
\usepackage{epsfig}
\usepackage{pst-grad} 
\usepackage{url}
\usepackage{wrapfig}
\usepackage{float} 
\title{Isotropic Cellular Automata: \\ \large \it the DDLab iso-rule paradigm}
\author{Andrew Wuensche%
\thanks{andy@ddlab.org,  http://www.ddlab.org}%
\hspace{2ex}{\it \small Discrete Dynamics Lab.}\\
Jos\'e Manuel G\'omez Soto%
\thanks{jmgomezuam@gmail.com, http://matematicas.reduaz.mx/$\sim$jmgomez}%
\hspace{2ex}{\it \small Universidad Aut\'onoma de Zacatecas.}\\
\hspace{2ex}{\it \small  Unidad Acad\'emica de Matem\'aticas. Zacatecas, Zac. M\'exico.}}

\date{\small Feb 2021}	

\begin{document}

\maketitle

\vspace{-3ex}
\begin{abstract}
\noindent

\noindent To respect physics and nature, cellular automata (CA) models
of self-organisation, emergence, computation and logical universality
should be isotropic, having equivalent dynamics in all directions.  We
present a novel paradigm, the iso-rule, a concise expression for
isotropic CA by the output table for each isotropic neighborhood
group, allowing an efficient method of navigating and exploring
iso-rule-space.  We describe new functions and tools in DDLab to
generate iso-groups and iso-rules, for multi-value as
well as binary, in one, two and three dimensions. These methods
include filing, filtering, mutating, analysing dynamics by
input-frequency and entropy, identifying the critical iso-groups for
glider-gun/eater dynamics, and automatically classifying
iso-rule-space.  We illustrate these ideas and methods for two
dimensional CA on square and hexagonal lattices.

\end{abstract}

\begin{center}
{\it keywords: DDLab, cellular automata, isotropy, iso-groups, iso-rules, glider-guns, 
logical universality, input-frequency, filtering, mutation.}
\end{center}

\section{Introduction} 
\label{Introduction}

\begin{figure}[htb]
\textsf{\footnotesize
\begin{center}
\begin{minipage}[c]{.8\linewidth}
\begin{minipage}[c]{.47\linewidth}
\includegraphics[width=1\linewidth,bb=9 15 247 255, clip=]{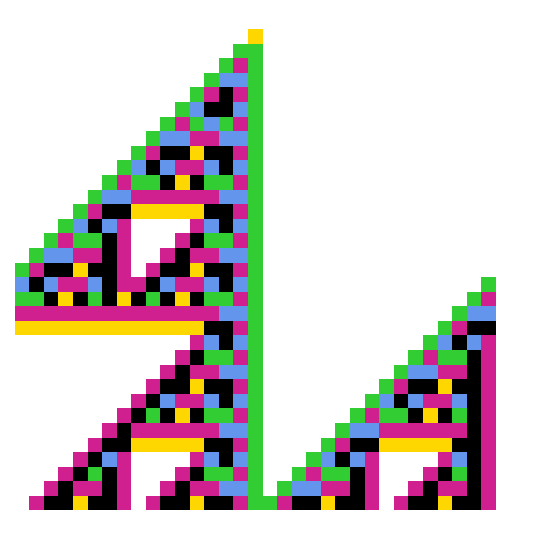}\\[-5ex] 
\begin{center}(a) the unslanted initial view\end{center}
\end{minipage}
\hfill
\begin{minipage}[c]{.47\linewidth} 
\includegraphics[width=1\linewidth,bb=10 15 247 244, clip=]{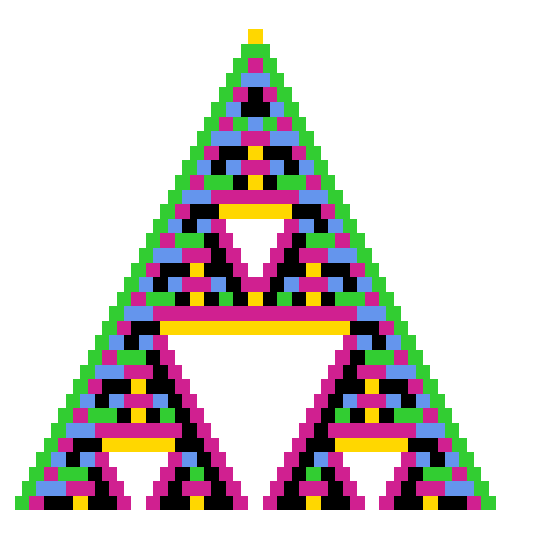}\\[-5ex]
\begin{center}(b) time-steps shifted 1/2 cell right\end{center}
\end{minipage}
\end{minipage}
\end{center}
}
\vspace{-3ex}
\caption[1d CA]
{\textsf{1d $v8k2$ space-time pattern of an isotropic CA, size 33, 
The rcode size=64,  The iso-rule (size=36): 140761540026777563655513706072505220,
as a graphic: \includegraphics[width=.5\linewidth,bb=16 22 452 40, clip=]{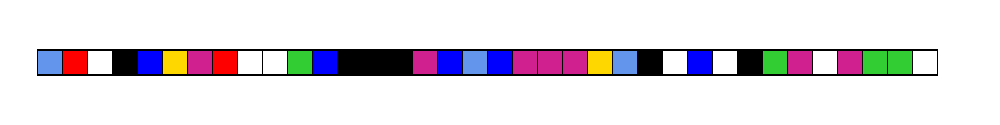}
and in hexadecimal: 07 07 c6 c0 16 ff dc f5 b4 be 30 ea 8a 90.
(a) is the initial view because even n-templates are skewed right 
(figure~\ref{1d n-templates}).
(b) successive time-steps shifted 1/2 cell right to restore symmetry\cite[EDD:32.9.1]{EDD}.}}
\label{1d v8k2 space-time pattern}
\end{figure}

\noindent Isotropy, rotational invariance, equivalence, lack of
directional bias or preference arguably underlie physics and nature at
the most basic level.  By contrast, an isotropic ``physics'' of a
cellular autamata universe belongs in a special and very limited
category in the entirety of CA rule-spaces. As in the
game-of-Life\cite{Gardner1970}, Precursor\cite{Gomez2017},
Spiral\cite{Adamatzky&Wuensche2006} and other comparable rules
discussed in this paper, equivalent dynamics in all directions and
orientations seems a proper constraint for CA models of
self-organisation, emergence, computation and logical universality,
where a glider-gun and its rotations/reflections work equivalently.

A counter example is the anisotropic X-rule\cite{Gomez2015,UC2DCA-webpage} which has
an operational glider-gun only when orientated East-West --- the same
structure rotated $90^{\circ}$ becomes a simple reflector.
Despite the interesting glider-guns and logical-gates created in the
X-rule, its anisotropic behaviour seems unnatural.

\begin{figure}[htb]
   \begin{center}
   \begin{minipage}[t]{.92\linewidth}
   \begin{minipage}[t]{.45\linewidth}
      \includegraphics[width=1\linewidth,bb=8 5 240 211, clip=]{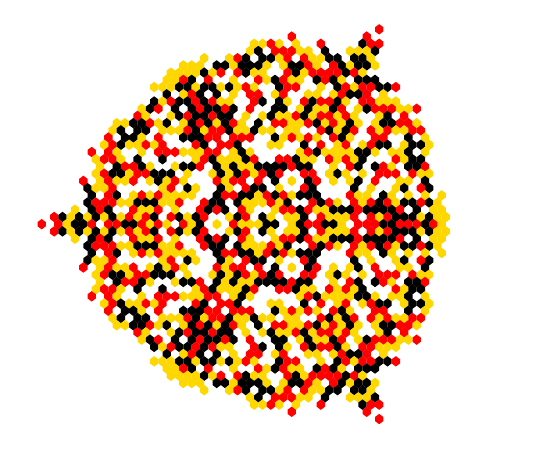}\\[-1.5ex] 
      \parbox{1\linewidth}{ \begin{center} \textsf{$v4k4$ hex 2d}  \end{center}}
   \end{minipage}
   \hfill
   \begin{minipage}[t]{.45\linewidth}
         \includegraphics[width=1\linewidth,bb=18 19 364 321, clip=]{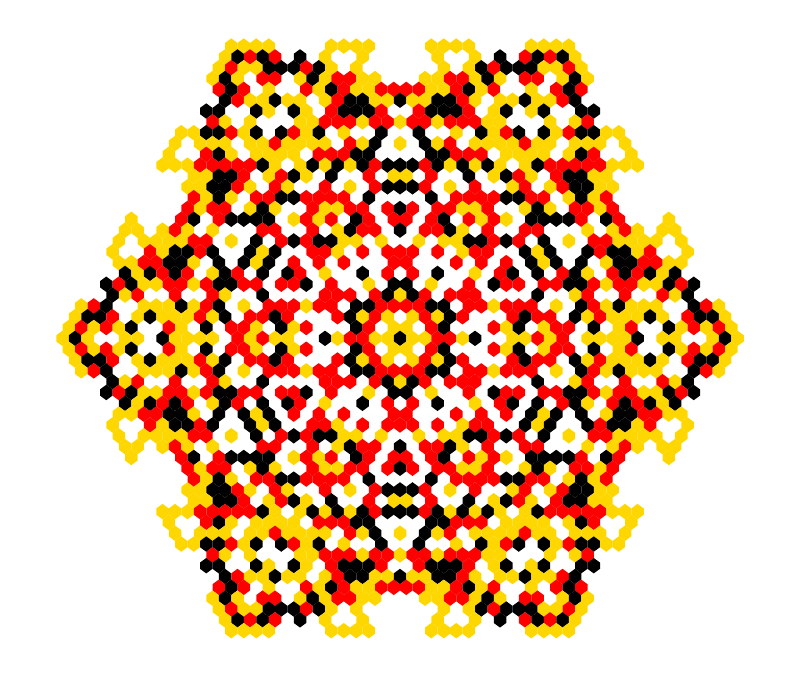}\\[-1.5ex] 
         \parbox{1\linewidth}{ \begin{center} \textsf{$v4k6$ hex 2d}  \end{center}}
   \end{minipage}\\
   \begin{minipage}[t]{.45\linewidth}
         \includegraphics[width=1\linewidth,bb=19 29 290 298, clip=]{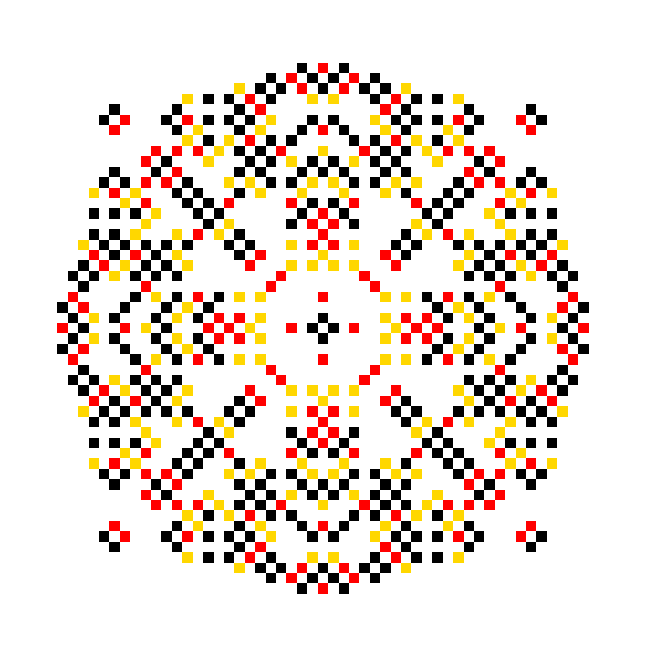}\\[-1.7ex] 
         \parbox{1\linewidth}{ \begin{center} \textsf{$v4k4s$ square 2d} \end{center}}      
   \end{minipage}
   \hfill
   \begin{minipage}[t]{.45\linewidth}
         \includegraphics[width=1\linewidth,bb=4 14 305 312, clip=]{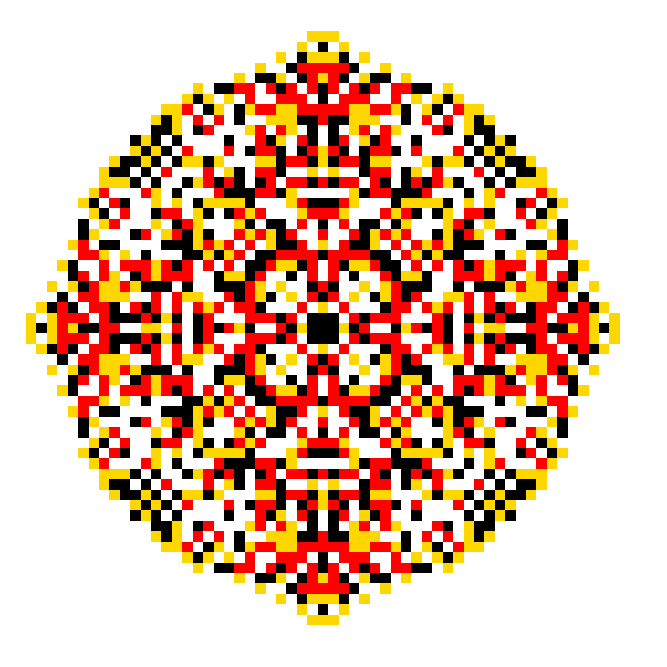}\\[-1.7ex] 
         \parbox{1\linewidth}{ \begin{center} \textsf{$v4k5s$ square 2d} \end{center}}      
   \end{minipage}\\
    \begin{minipage}[t]{.48\linewidth}
        \includegraphics[width=1\linewidth,bb=124 113 383 373, clip=]{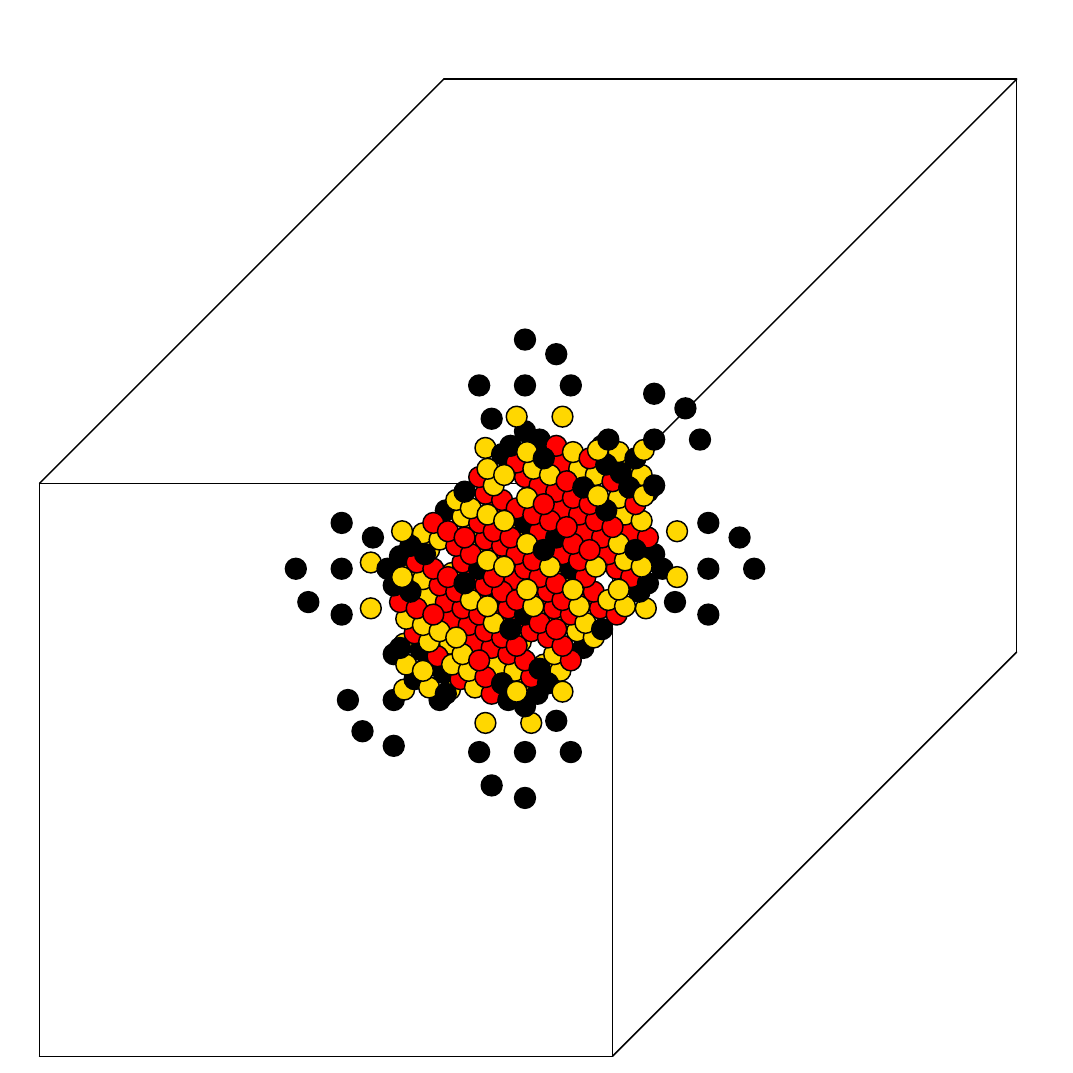}\\[-1.5ex]
        \parbox{1\linewidth}{ \begin{center} \textsf{$v5k6$ cubic 3d}  \end{center}}
   \end{minipage}
   \hfill
   \begin{minipage}[t]{.48\linewidth}
      \includegraphics[width=1\linewidth,bb=133 131 363 352, clip=]{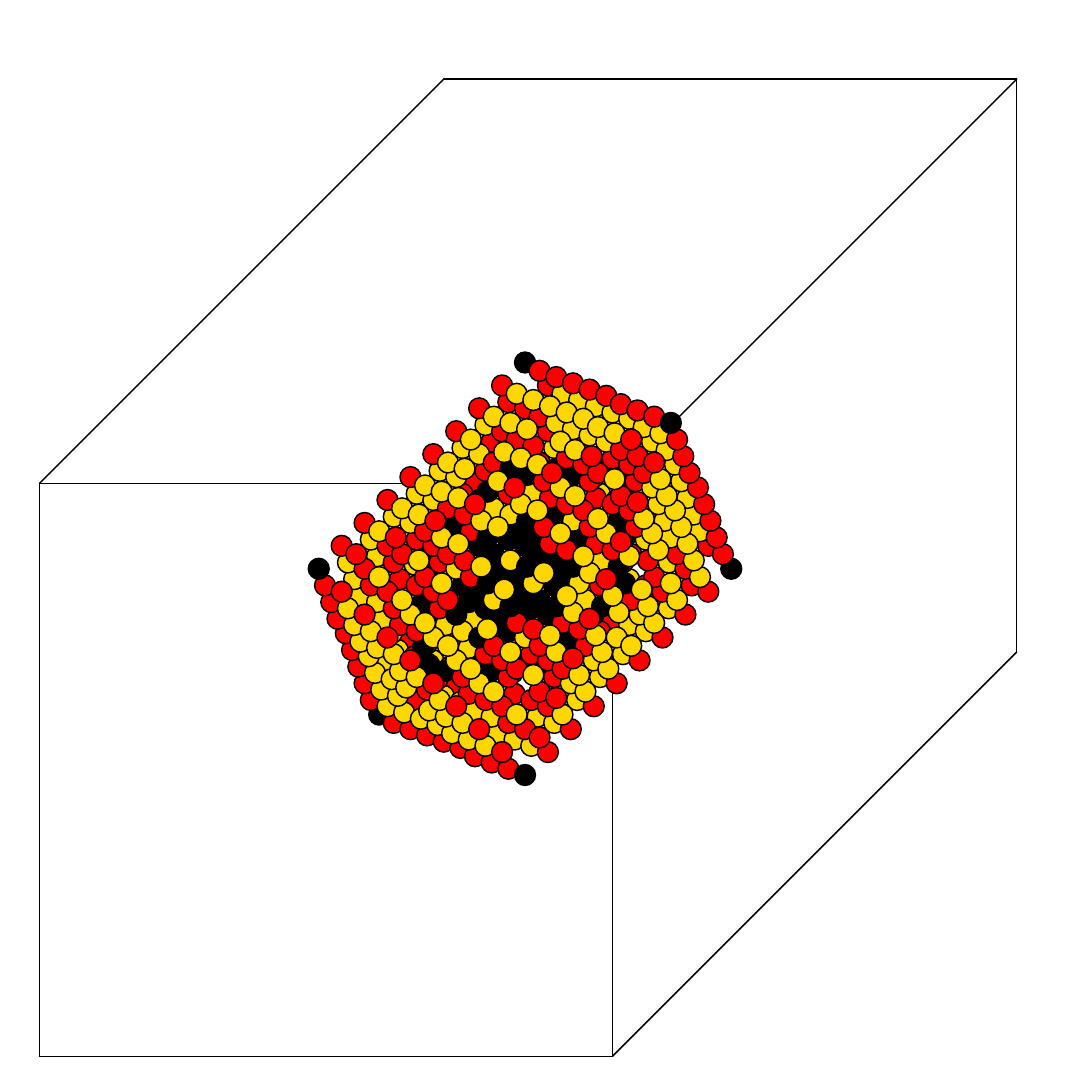}\\[-1.5ex]
      \parbox{1\linewidth}{ \begin{center} \textsf{$v5k7$ cubic 3d} \end{center}}      
   \end{minipage}   
   \end{minipage}
   \end{center} 
    \vspace{-3ex}   
    \caption [Isotropic space-time patterns]
             {\textsf{Examples of space-time pattern snapshots for
                 $v$=4 2d and 3d isotropic CA from
                 a singleton seed, a $v$$>$0 single cell against a zero background.
                 The initial symmetry must be preserved. Isotropic CA, where rotated and 
                 reflected neighborhoods have the same output, are arguably closer
                 to natural physics.}}
     \label{iso-space-time-patterns}
\end{figure}

CA rule-spaces in general are not isotropic, where rules are
defined by rule-tables of length $v^k$ where $v$=value-range and
$k$=neighborhood size, with a rule-space size of $v^{v^k}$.  Isotropic
rules, also recognisable in that symmetric patterns must conserve symmetry as time
evolves --- as from singleton seeds in 
figures~\ref{1d v8k2 space-time pattern} and \ref{iso-space-time-patterns} 
--- make up a tiny proportion of a general rule-space.
 
\begin{figure}[htb]
\textsf{\footnotesize
\begin{center}
\begin{minipage}[c]{1\linewidth}
\begin{minipage}[c]{.32\linewidth} 
  \includegraphics[width=1\linewidth,bb=6 259 135 292, clip=]{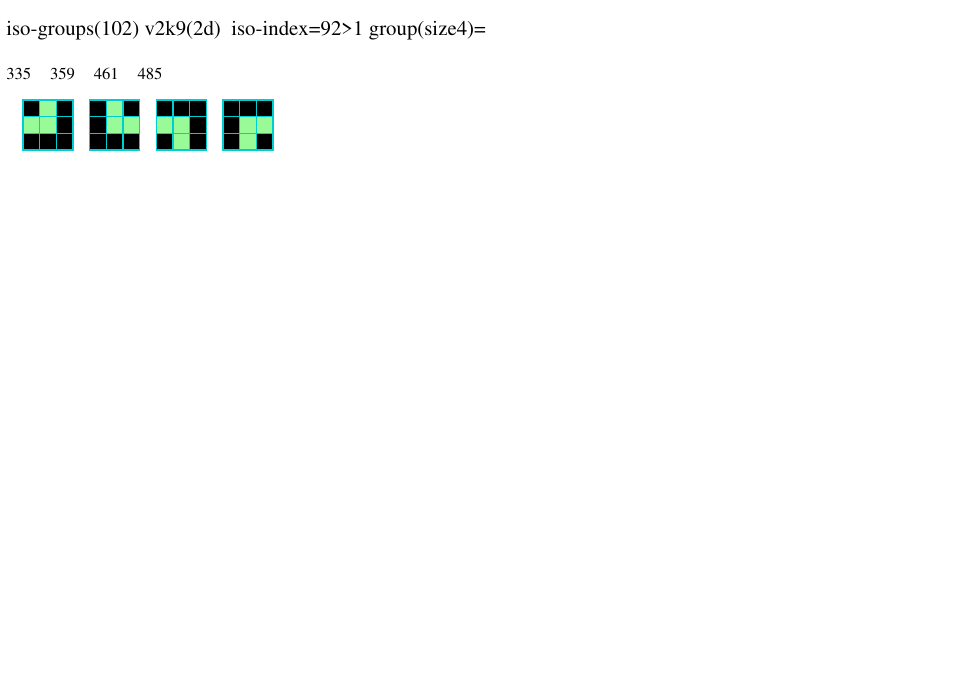}\\[-5ex]
\begin{center}(a) 2d $v2k8$ (92/101)\end{center}
\end{minipage}
\hfill
\begin{minipage}[c]{.32\linewidth}  
  \includegraphics[width=1\linewidth,bb=3 261 99 291, clip=]{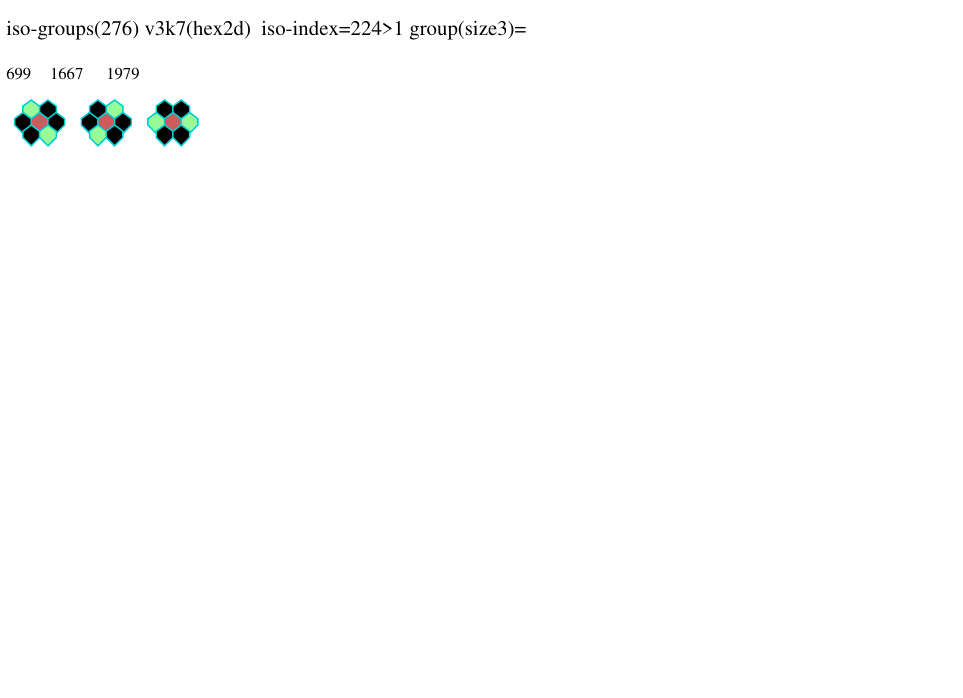}\\[-5ex]
\begin{center}(b) 2d hex (224/275)\end{center}
\end{minipage}
\hfill
\begin{minipage}[c]{.32\linewidth} 
  \includegraphics[width=1\linewidth,bb=10 222 192 284, clip=]{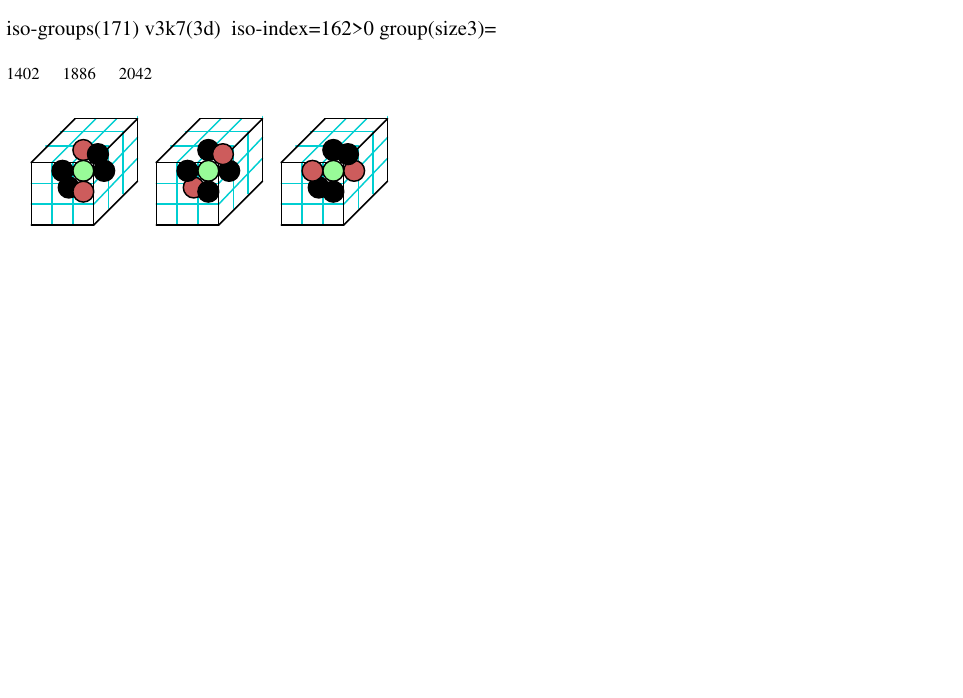}\\[-5ex]
\begin{center}(b) 3d (162/170)\end{center}
\end{minipage}
\end{minipage}
\end{center}
}
\vspace{-3ex}
\caption[iso-group examples]
{\textsf{Examples of iso-groups showing (group-index/max-index).
}}
\label{iso-group examples}
\vspace{-1ex}
\end{figure}

Although some rule categories, survival/birth, totalistic,
reaction-diffusion, are isotropic by default, these iso-subsets can be
transformed into a general expression of isotropic CA, where 
the ``iso-groups'' of equivalent neighborhoods by all possible spins and flips
share the same output --- figure~\ref{iso-group examples} 
gives examples.

\begin{figure}[htb]
  \small{\textsf{ 
\begin{minipage}[c]{.95\linewidth}
\begin{minipage}[c]{.33\linewidth} 
\fbox{\includegraphics[width=1\linewidth,bb=88 94 320 271, clip=]{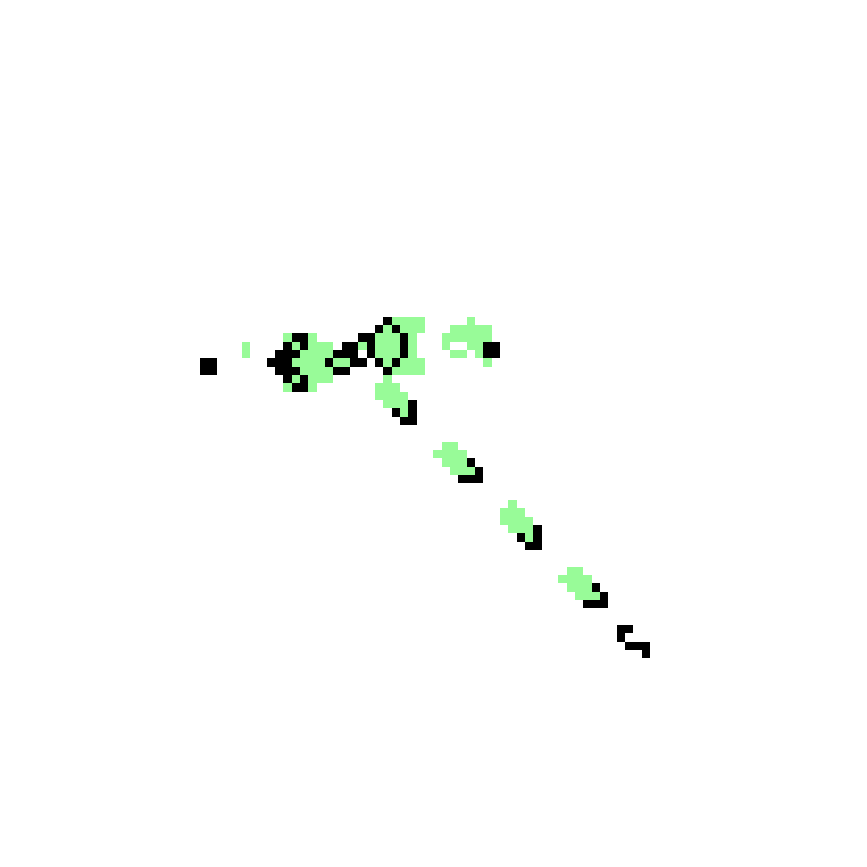}}\\ 
\textsf{Life\cite{Berlekamp1982,Gardner1970}: $p$=30}
\end{minipage}
\hfill
\begin{minipage}[c]{.39\linewidth} 
\fbox{\includegraphics[width=1\linewidth,bb=16 163 315 258, clip=]{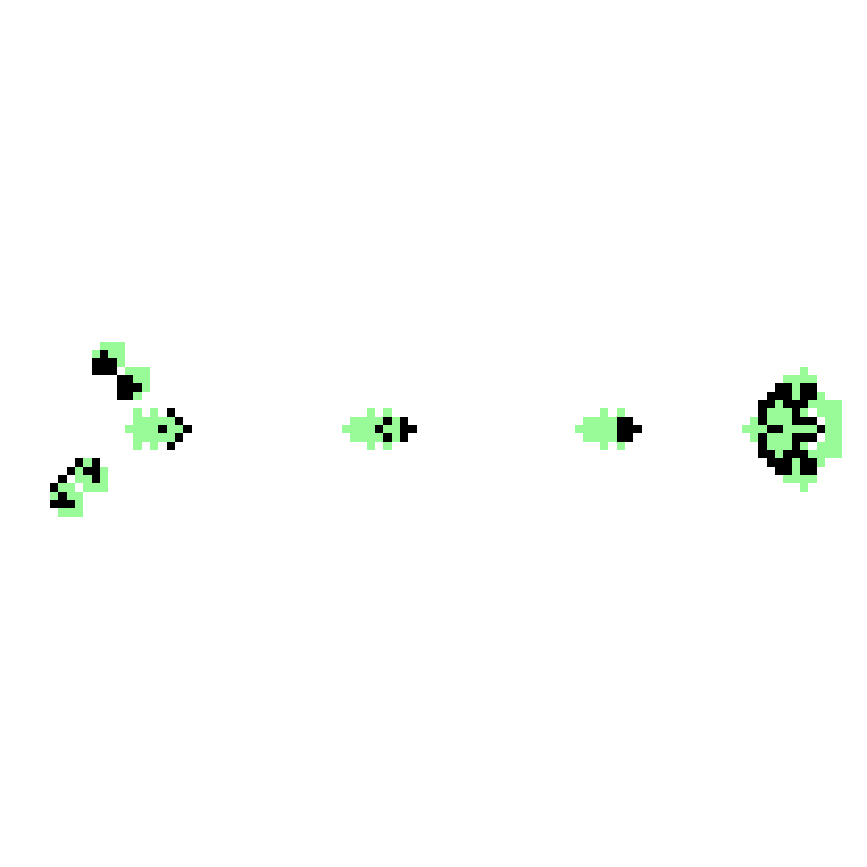}}\\[-1pt]
\fbox{\includegraphics[width=1\linewidth,bb=48 169 351 255, clip=]{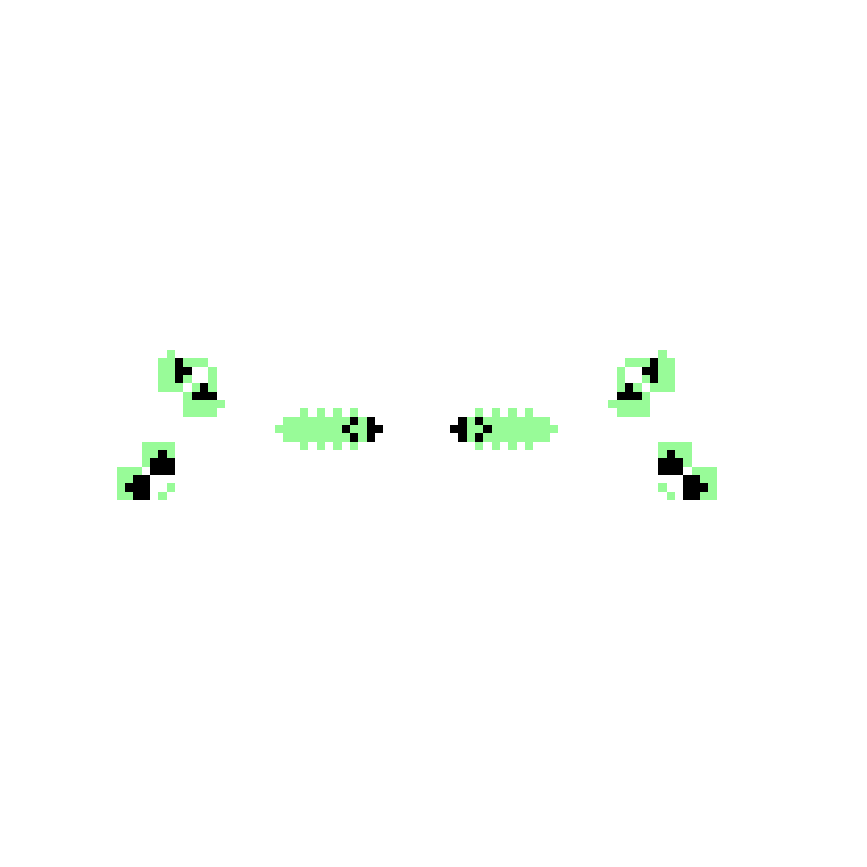}}\\
\textsf{Eppstein\cite{Eppstein2010}: $p$=68}
\end{minipage}
\hfill
\begin{minipage}[c]{.215\linewidth} 
\fbox{\includegraphics[width=1\linewidth,bb=112 125 293 339, clip=]{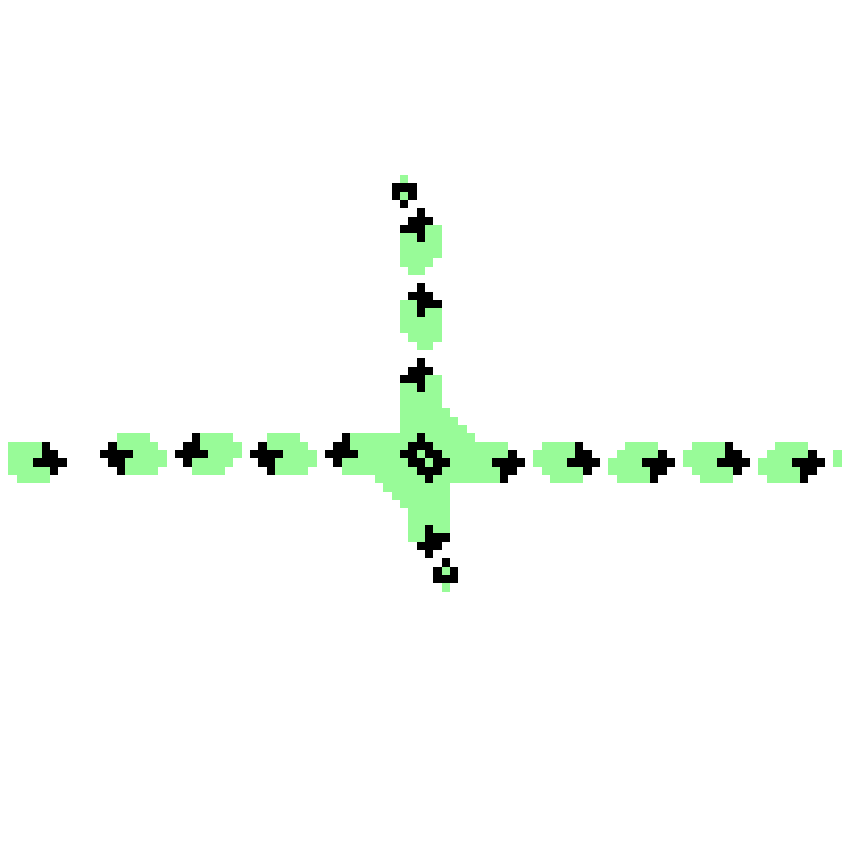}}\\
\textsf{Sapin\cite{Sapin2004}: $p$=18}
\end{minipage}
\end{minipage}\\[1ex]
\textsf{(a) Conway's survival/birth (s23/b3) game-of-Life\cite{Berlekamp1982,Gardner1970}
and Eppstein's s236/b3 rule\cite{Eppstein2010}.
Sapin's R-rule\cite{Sapin2004} evolved by genetic algorithm from iso-groups.
Life and SapinR are logically universal with glider streams stopped by eaters,
Eppstein's by head-on collisions only --- lower panel.}\\[1ex]
00 00 00 00 00 60 03 1c 61 c6 7f 86 a0 --- Life\\
04 89 86 1a 00 6d 23 1e 61 e6 7f 86 a0 --- Eppstein\\
11 34 1c 2c 52 36 7d 3b e0 f8 7e 0a a0 --- SapinR\\[1ex]
\noindent \includegraphics[width=.95\linewidth,bb=2 6 522 20, clip=]{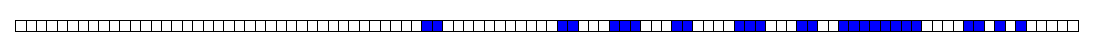}L
\includegraphics[width=.95\linewidth,bb=2 6 522 20, clip=]{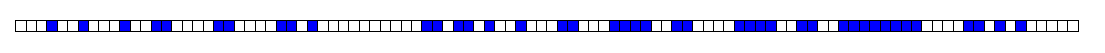}E
\includegraphics[width=.95\linewidth,bb=2 6 522 20, clip=]{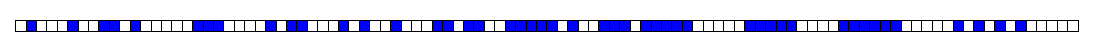}R\\[4ex]
\begin{minipage}[c]{.95\linewidth}
\begin{minipage}[c]{.28\linewidth} 
\fbox{\includegraphics[width=1\linewidth,bb=167 189 402 482, clip=]{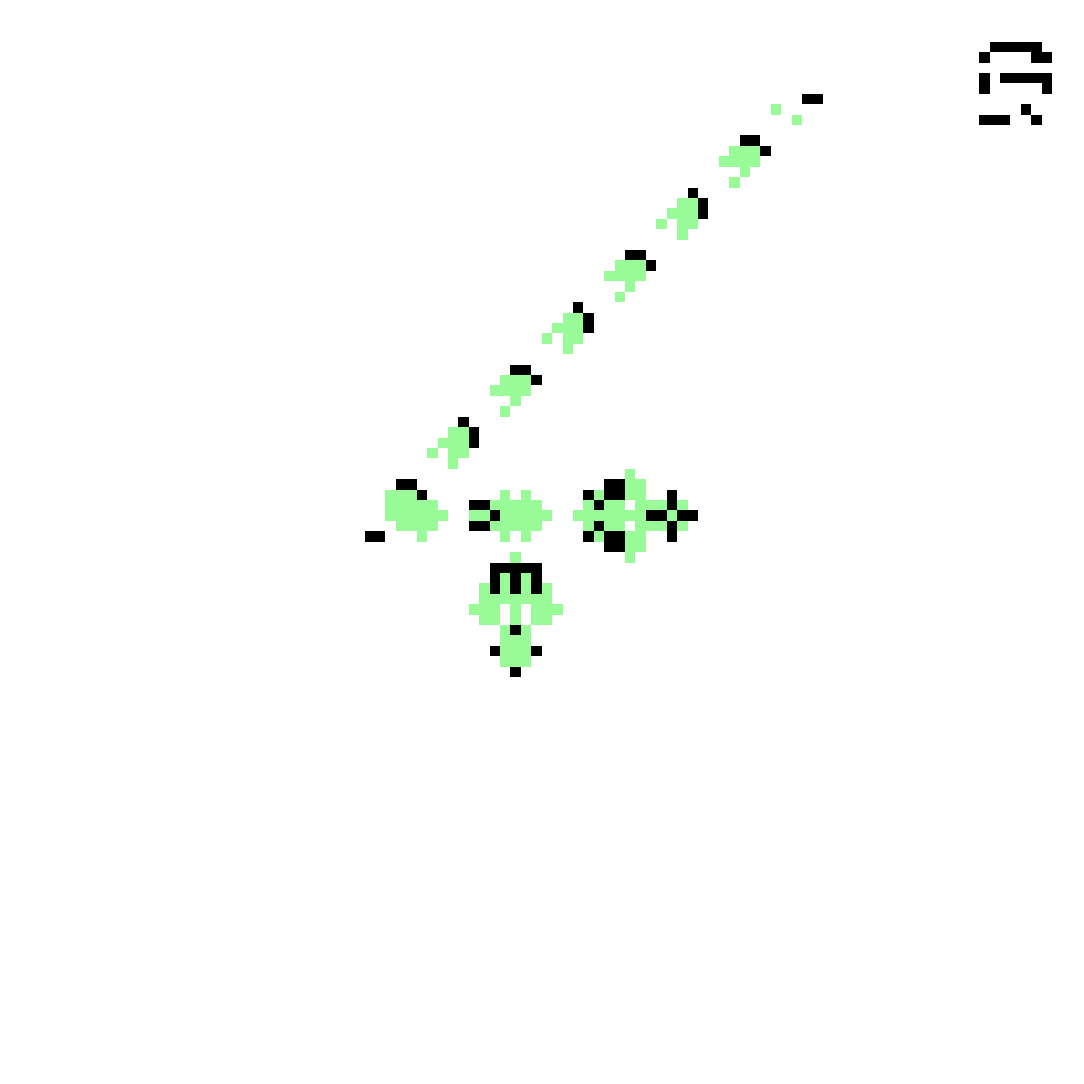}}\\ 
Variant\cite{Gomez2020}: $p$=22
\end{minipage}
\hfill
\begin{minipage}[c]{.36\linewidth} 
\fbox{\includegraphics[width=1\linewidth,bb=102 108 413 413, clip=]{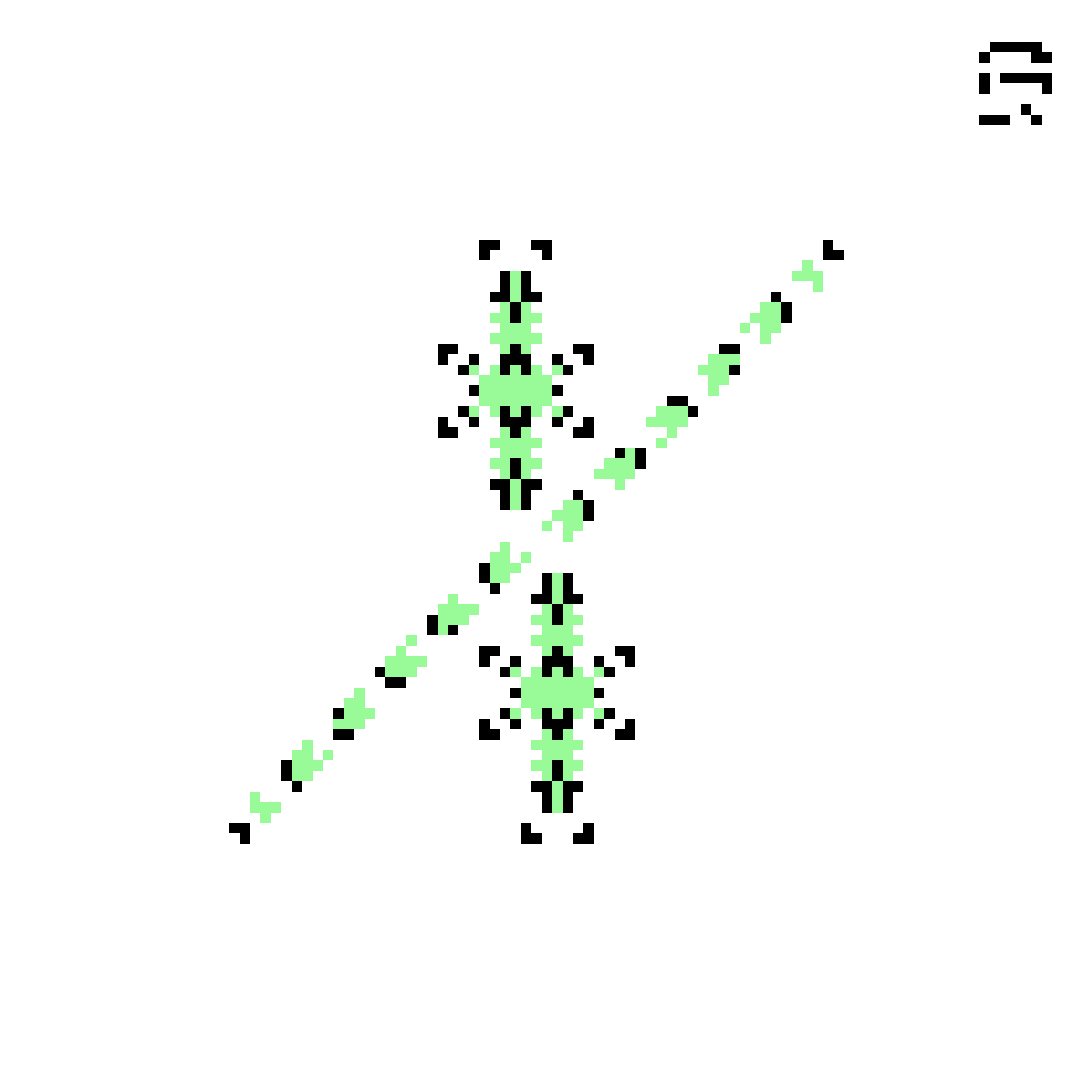}}\\ 
Precursor\cite{Gomez2017}: $p$=19
\end{minipage}
\hfill
\begin{minipage}[c]{.24\linewidth} 
\fbox{\includegraphics[width=1\linewidth,bb=40 124 247 422, clip=]{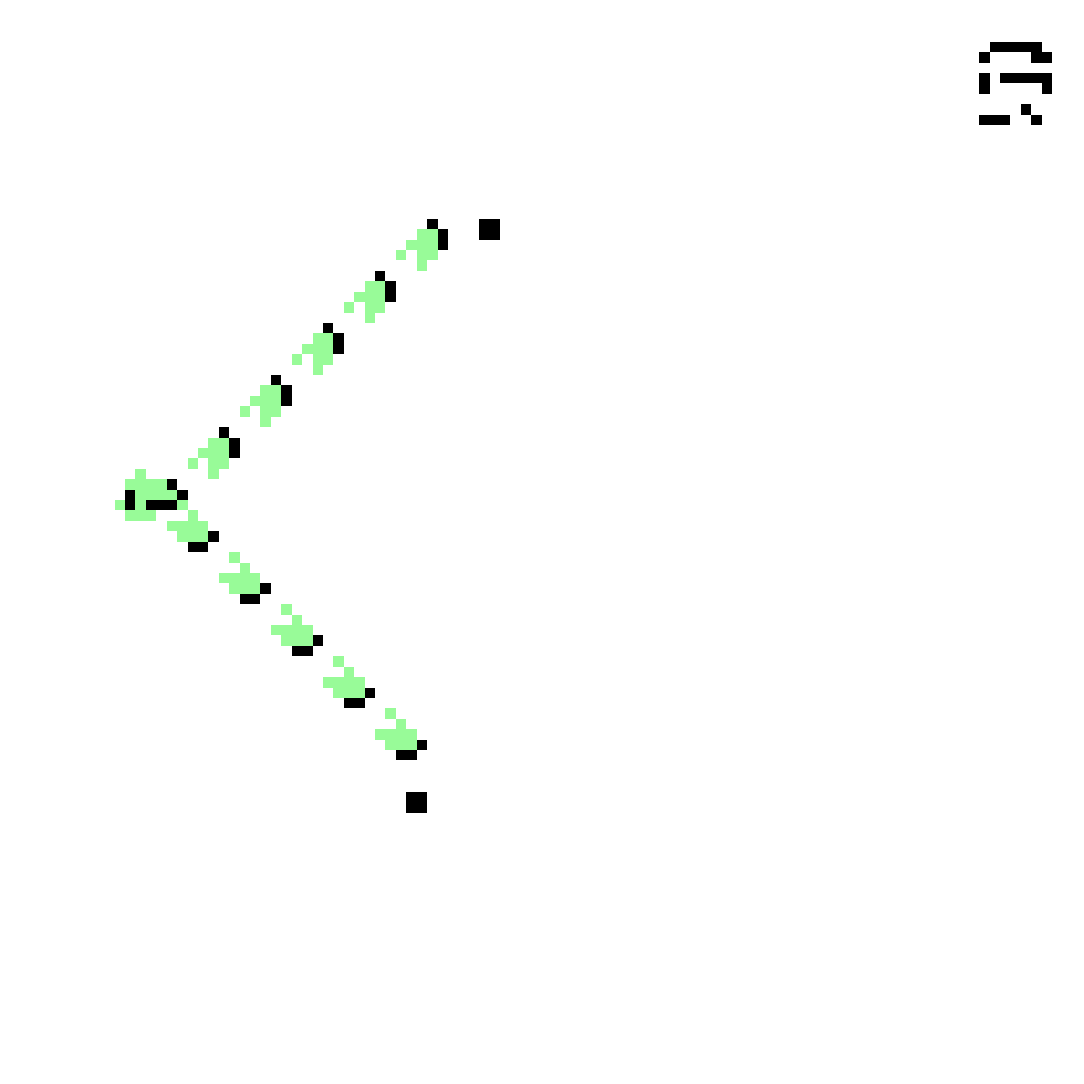}}\\ 
Sayab\cite{Gomez2018}: $p$=20
\end{minipage}
\end{minipage}\\[1ex]
\textsf{(b) Three binary logically universal iso-rules\cite{UC2DCA-webpage} belong to a family
with different glider-guns. Gliders streams are stopped by eaters. 
The Variant\cite{Gomez2020} and Precursor\cite{Gomez2017} rules are closely related differing by
two outputs. The Sayab rule\cite{Gomez2018} is a distant cousin differing from the Precursor
by 33 outputs.}\\[1ex]
24 c0 04 42 83 01 80 2c a4 29 04 e0 70 --- Variant\\
24 c0 04 42 83 01 80 24 a4 69 04 e0 70 --- Precursor\\
24 01 13 1a 14 20 50 2c 45 05 48 e0 50 --- Sayab\\[1ex]
\includegraphics[width=.95\linewidth,bb=2 6 522 20, clip=]{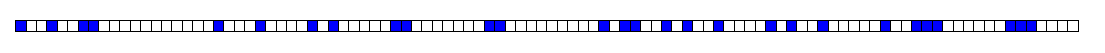}V
\includegraphics[width=.95\linewidth,bb=2 6 522 20, clip=]{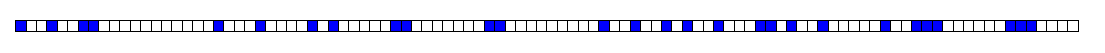}P
\includegraphics[width=.95\linewidth,bb=2 6 522 20, clip=]{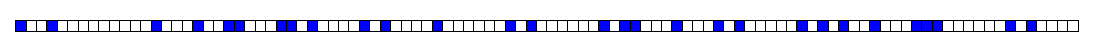}S
}}
\caption{
\textsf{Significant binary ($v2k9$) glider-gun iso-rules on 3$\times$3
  Moore n-templates with 102 iso-groups. The iso-rule's
  hexadecimal and bit expressions allow comparison of bit
  differences/similarities. The glider-guns in SapinR and the
  Sayab rule emerge spontaneously, the others are elaborately
  constructed artifacts.  $p$=glider-gun period and firing
  frequency. Green trails denote motion.}}
\label{significant binary iso-rules}
\end{figure}

Especially significant are iso-rules analogous to Conway's famous
survival/birth game-of-Life\cite{Berlekamp1982,Gardner1970}, the first
rule with logical gates constructed from
glider/eater dynamics made with the first glider-gun discovered by Gosper.
Figure~\ref{significant binary iso-rules} illustrates Life and  other significant
rules, including glider-gun iso-rules not based on survival/birth
where logical universality has been demonstrated.
\clearpage

\clearpage

\begin{figure}[htb]
\textsf{\footnotesize
\begin{center}
\begin{minipage}[c]{.85\linewidth}
\begin{minipage}[c]{.3\linewidth}
\includegraphics[width=1\linewidth,bb=280 144 414 295,  clip=]{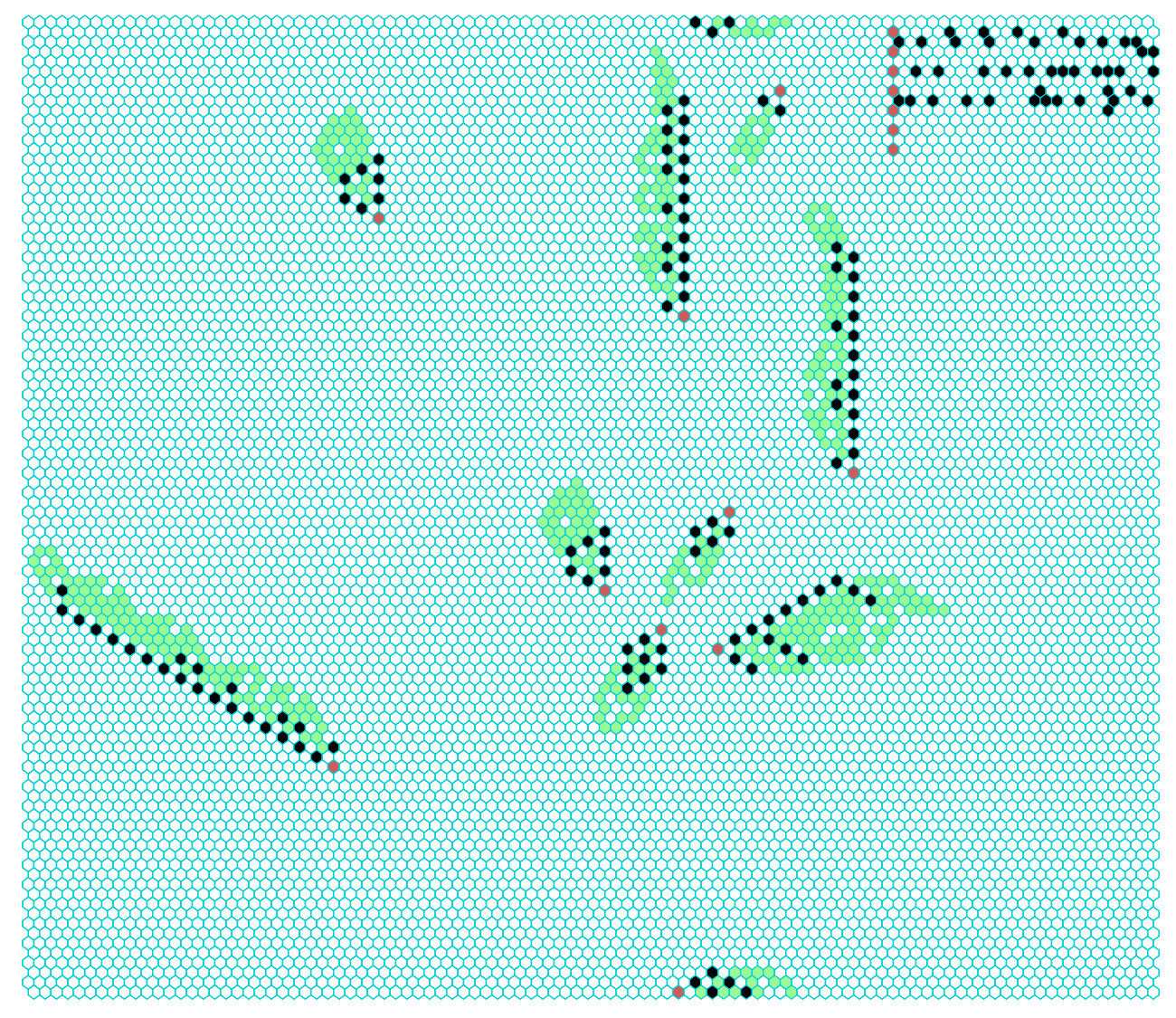}\\[-5ex]
\begin{center}(a) v3k3x1.vco, g1\\(hex)00a864\end{center}
\end{minipage}
\hfill
\begin{minipage}[c]{.3\linewidth}
\includegraphics[width=1\linewidth,bb=242 20 352 145, clip=]{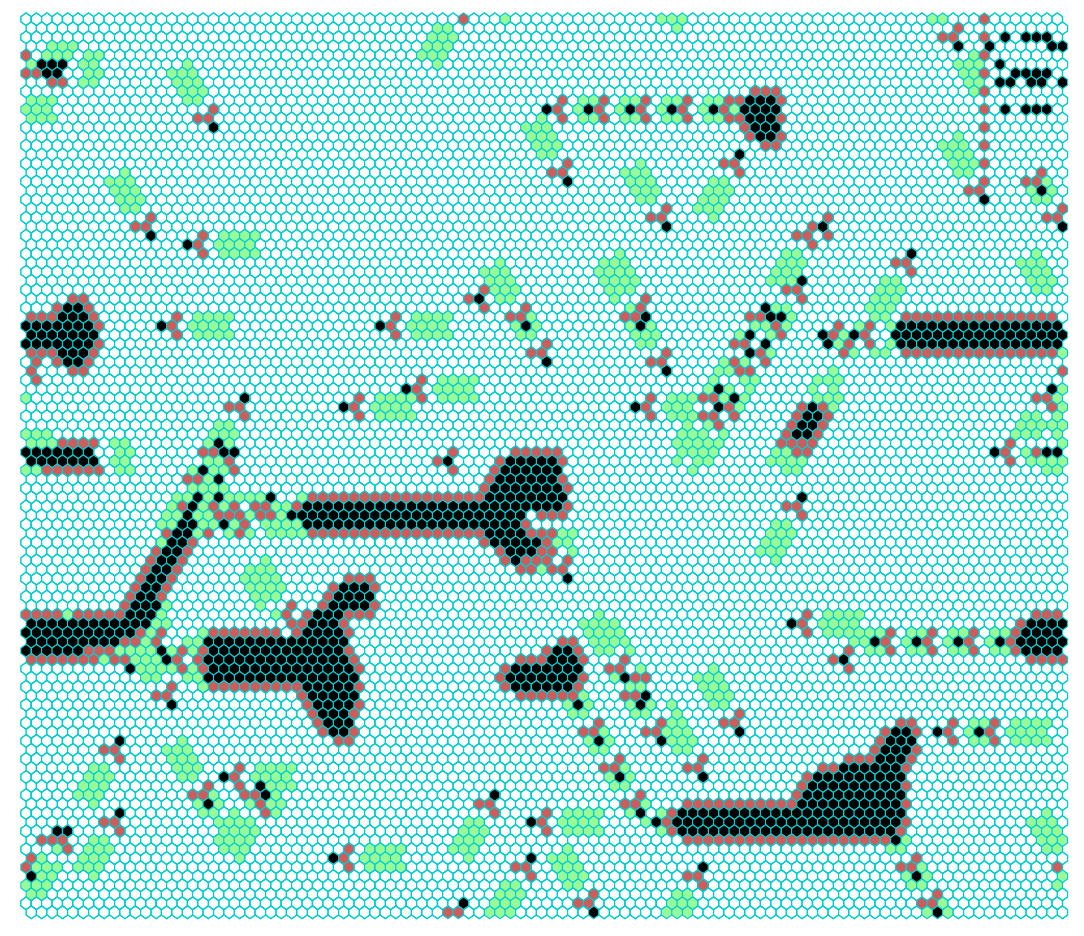}\\[-5ex]
\begin{center}(b) v3k4t1.vco, g1\\(hex)2a945900\end{center}
\end{minipage}
\hfill
\begin{minipage}[c]{.3\linewidth}
\includegraphics[width=1\linewidth,bb=315 109 471 284, clip=]{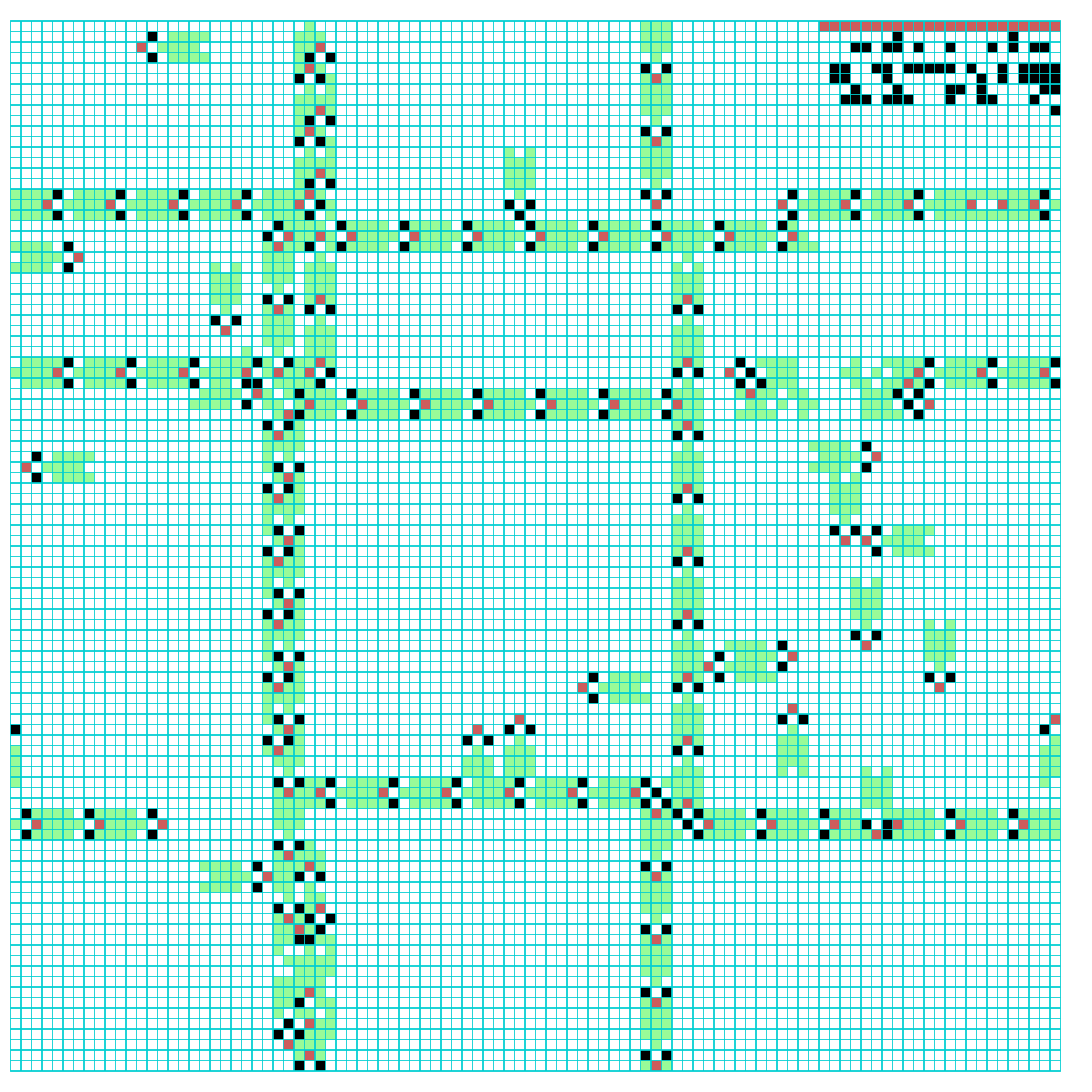}\\[-5ex]
\begin{center}(c) v3k4x1.vco\\(hex)2282a1a4\end{center}
\end{minipage}\\[1ex]
\begin{minipage}[c]{.3\linewidth}
\includegraphics[width=1\linewidth,,bb=142 230 276 386, clip=]{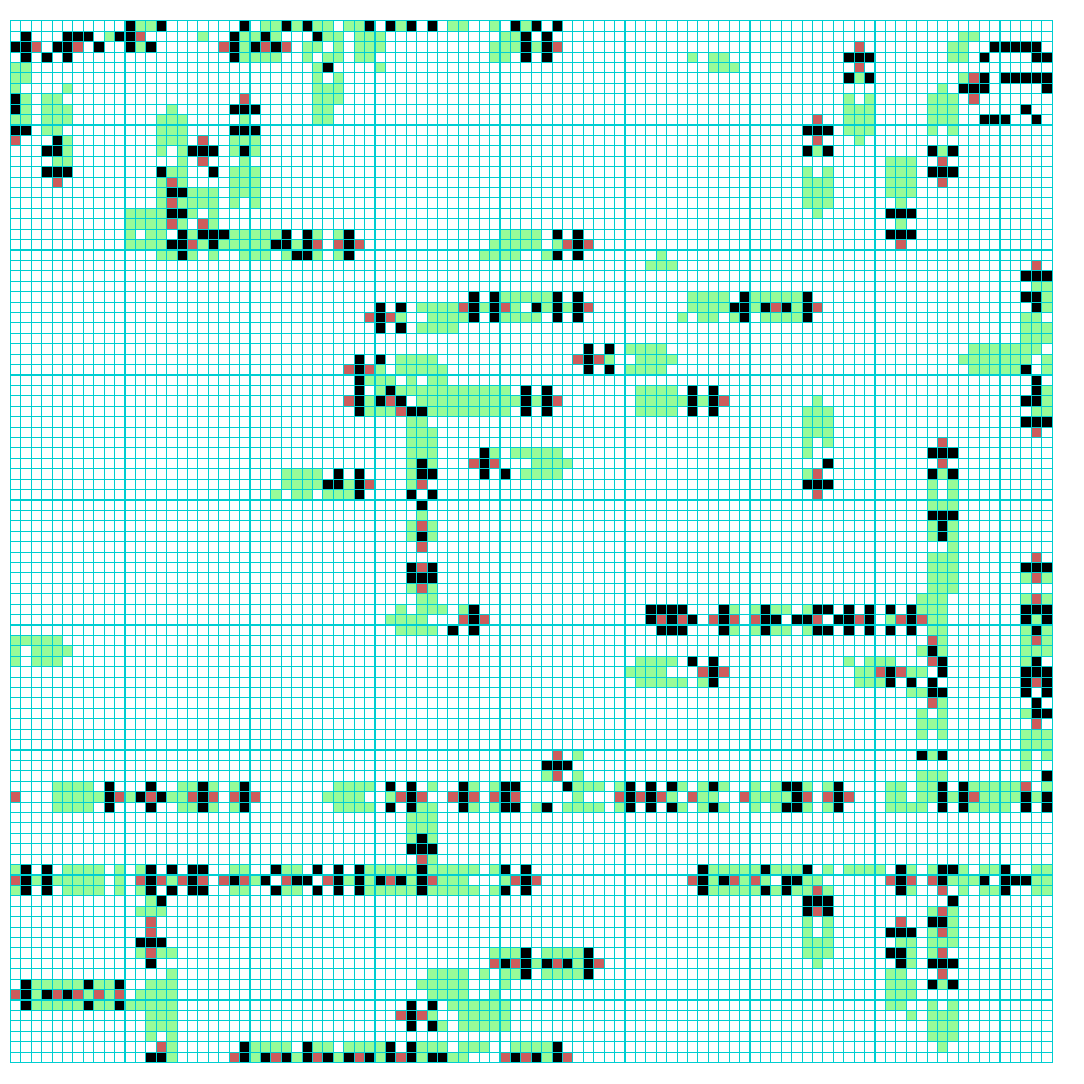}\\[-5ex]
\begin{center}(d) v3k5x1.vco, g1\\(hex)004a8a2a8254\\ \textcolor{white}{xx} \end{center}
\end{minipage}
\hfill
\begin{minipage}[c]{.3\linewidth}
\includegraphics[width=1\linewidth,bb=268 32 401 186, clip=]{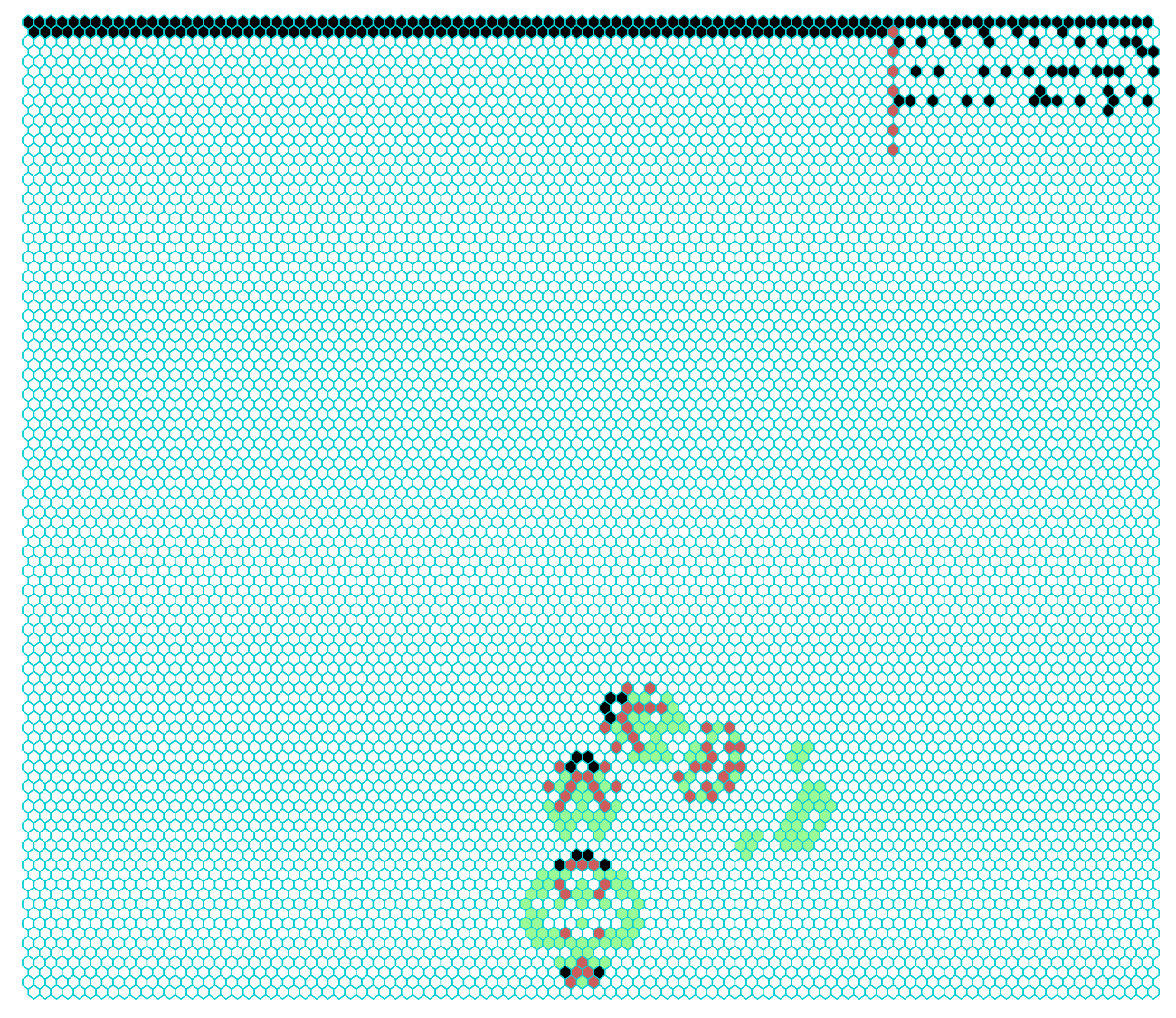}\\[-5ex]
\begin{center}(e) v3k6n6.vco, g16\\(hex)01059059560040\\ \textcolor{white}{xx} \end{center}
\end{minipage}
\hfill
\begin{minipage}[c]{.3\linewidth}
\includegraphics[width=1\linewidth,bb=263 235 386 380, clip=]{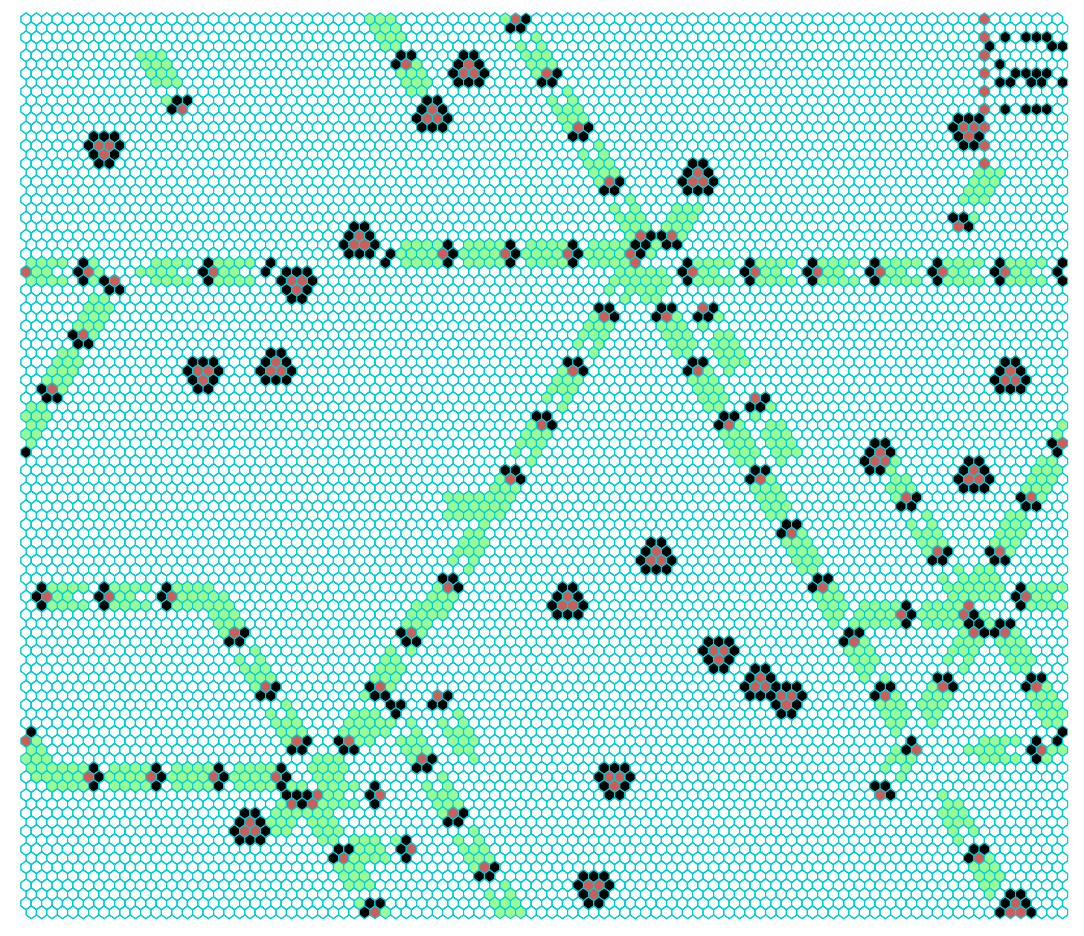}\\[-5ex]
\begin{center}(f) v3k7w1.vco, g1\\(hex)020609a2982a68aa64\\
The Spiral rule\cite{Wuensche&Adamatzky2006}\end{center}
\end{minipage}
\end{minipage}
\end{center}
}
\vspace{-2ex}
\caption[2d v3 Glider examples]
{\textsf{Glider dynamics discovered in 3-value 2d  k-totalistic rules,
on \mbox{n-templates} $k$=3 to 7, in figure~\ref{2d n-templates}.
Cell values: 0=white, 1=red, 2=black.
Green trails indicate motion.
Examples b, c, e, and f include glider-guns. The rules can be loaded in DDLab
by their file-names, in hexadecimal, or from the rule 
collections index g(x)\cite[EDD:3.5.1]{EDD}.
The dynamics emerge spontaneously, including the glider-gun in the logically
universal Spiral rule. Figures from \cite{CApulsingDDLab}.}}
\label{2d v3 Glider examples}
\vspace{-1ex}
\end{figure}

\normalsize The CA iso-rule notation provides a practical balance
between a full lookup-table on which isotropy may be imposed and an
abbreviated notation that must be isotropic --- survival/birth or
totalistic.  The iso-rule notation is concise, but not too
concise. Insights can be gained into glider-gun mechanics by observing
iso-group activity, frequency and entropy. Iso-rules permit navigating
and exploring iso-group mutants to establish their related families,
and to discover new significant iso-rules in iso-rule space.

We describe new methods\cite{DDLab-iso-rule-update}
for defining and automatically generating
iso-rules on the basis of iso-groups with predefined n-templates
in 1, 2 and 3 dimensions, and with value-ranges (colors) from 2 (binary)
up to 8 values as in figure~\ref{1d v8k2 space-time pattern}.  The
methods include editing, filing, filtering, mutating, analysing
dynamics by input-frequency and entropy, identifying the critical and
neutral iso-groups for glider-gun/eater dynamics, and automatically
classifying \mbox{iso-rule-space}. This is seen in the context of the
superset of the general rule-table, and in iso-subsets in a narrower
sense, k-totalistic, t-totalistic, outer-totalistic, survival/birth
and reaction-diffusion.  General rule-tables and iso-subsets can
be transformed into iso-rules. Binary Moore neighborhood rules,
and initial states, are compatible with ``Golly''\cite{Golly,ConwayLife-forum}.

We present the ideas and methods mainly for 2d square and
hexagonal examples as in figures~\ref{significant binary iso-rules}
and \ref{2d v3 Glider examples}, but also include 1d and 3d.
Glider-rules that feature gliders emerging spontaneously are readily found by
classifying rule-space by input-entropy variability\cite[EDD:33]{EDD}, 
with examples in \cite{Wuensche99,Wuensche05,Gomez2015}, but 
spontaneously emergent glider-guns\cite{Wuensche05,Wuensche&Adamatzky2006,Gomez2018} 
are very rare. There are just a few examples of constructed 
glider-guns made from sub-components\cite{Gardner1970,Gomez2015,Gomez2017,Gomez2020}.  
A rule (and its family of mutants) that features both emergent gliders
and eaters\footnote{Eaters
are localised configurations that can stop a glider stream.  Other
important localised configuration include reflectors, deflectors and
oscillators. We have use ``eaters'' as a shorthand for all these.},
makes a starting point for the very hard task of building a glider-gun ---
then building the logical gates for logically universal dynamics follows more readily.

Mutant iso-rule-space can be navigated and explored with the 
program ``Discrete Dynamics Lab'' (DDLab)\cite{Wuensche-DDLab} --- its many methods for
studying space-time patterns\cite[EDD:23-30]{EDD} and attractor
basins\cite[EDD:31-32]{EDD} now apply to the new iso-rule paradigm.
DDLab is documented in the book ``Exploring Discrete Dynamics''(EDD)\cite{EDD},
and we have usually indicated the relevant section when citing EDD.
Both DDLab and EDD are updated and maintained online.

\section{n-templates, 1d, 2d and 3d}
\label{n-templates, 1d, 2d and 3d}
\begin{figure}[htb]
\begin{minipage}[t]{1\linewidth}
\begin{minipage}[t]{.15\linewidth}
\vspace{-8ex}
\begin{flushright}
\textsf{odd-$k$\\[-.5ex]
3 to 11\\[-.5ex]
symmetric}
\end{flushright}
\end{minipage}
\hfill
\begin{minipage}[t]{.9\linewidth}
\includegraphics[height=.105\linewidth, bb=80 99 109 143, clip=]{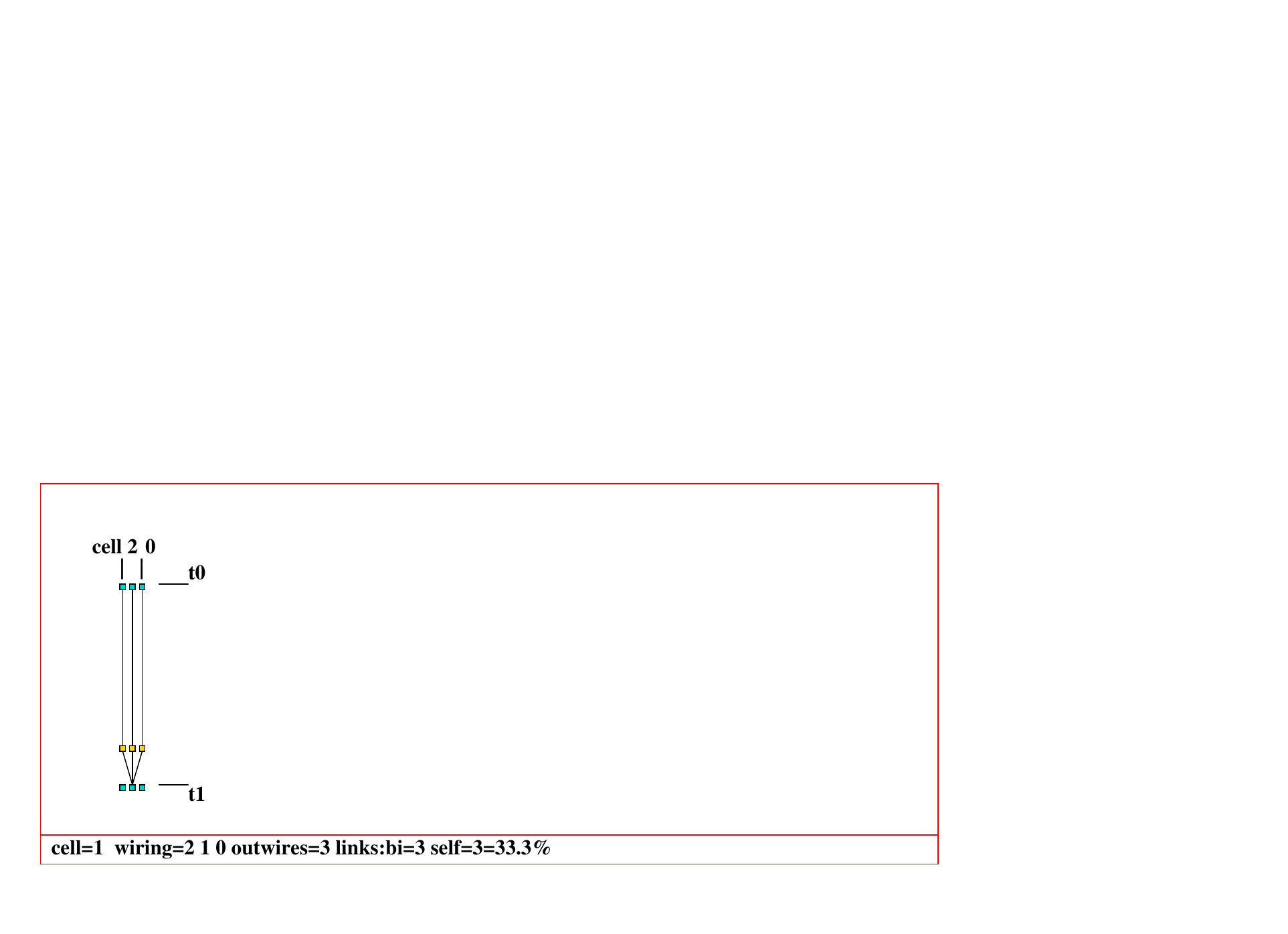}
\includegraphics[height=.105\linewidth, bb=80 99 123 143, clip=]{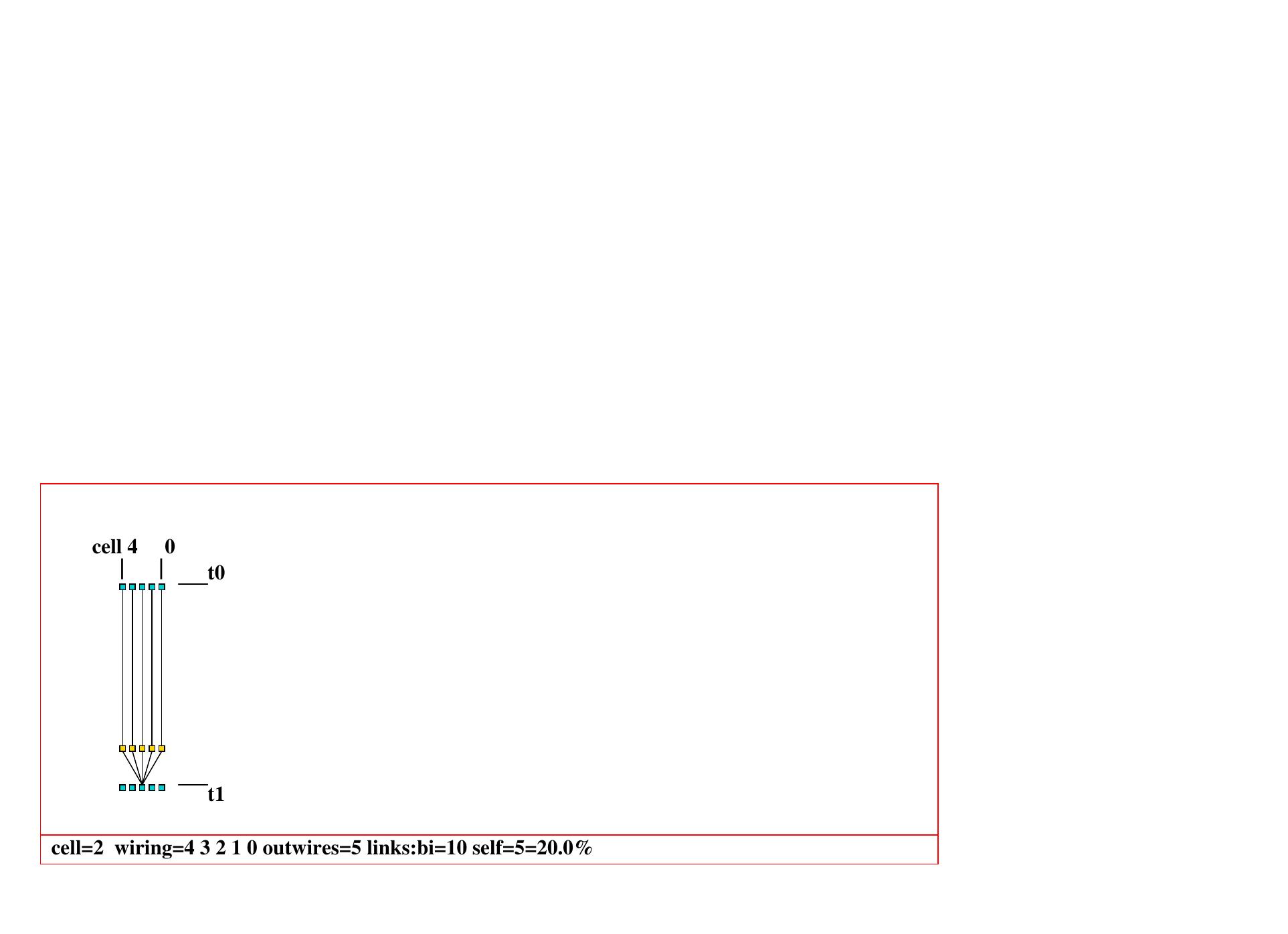}
\includegraphics[height=.105\linewidth, bb=80 99 137 143, clip=]{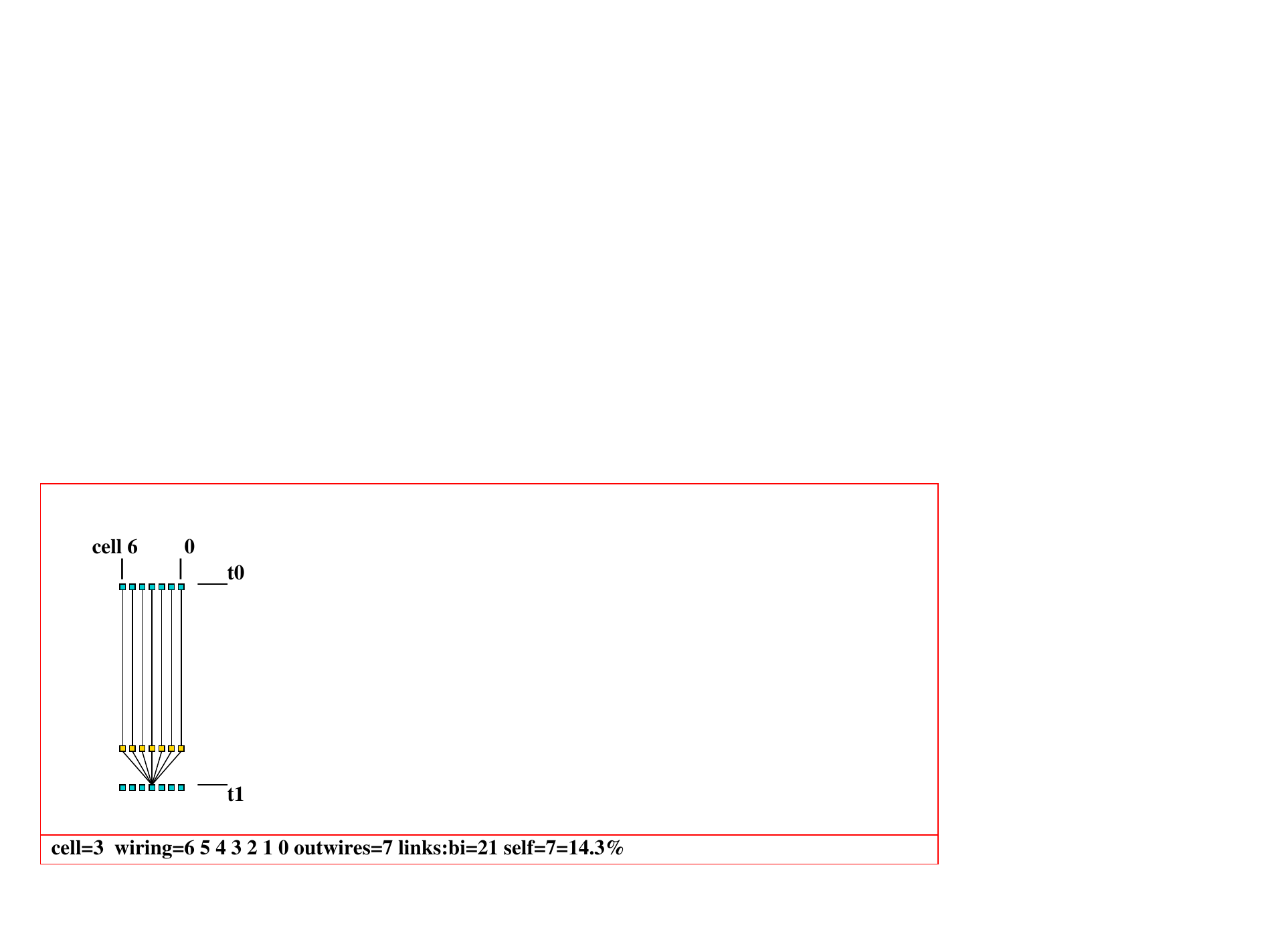}
\includegraphics[height=.105\linewidth, bb=80 99 151 143, clip=]{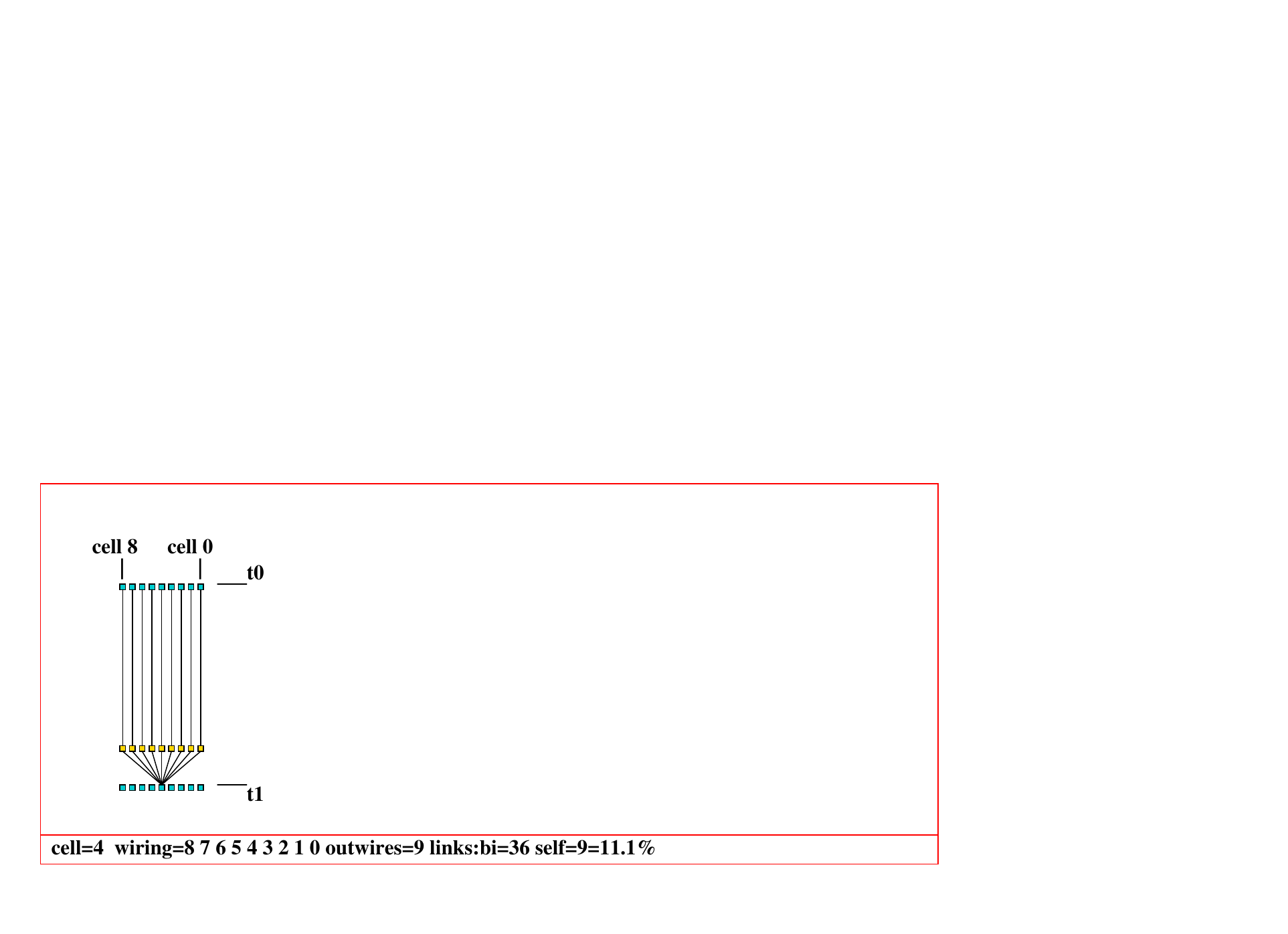}
\includegraphics[height=.105\linewidth, bb=80 99 165 143, clip=]{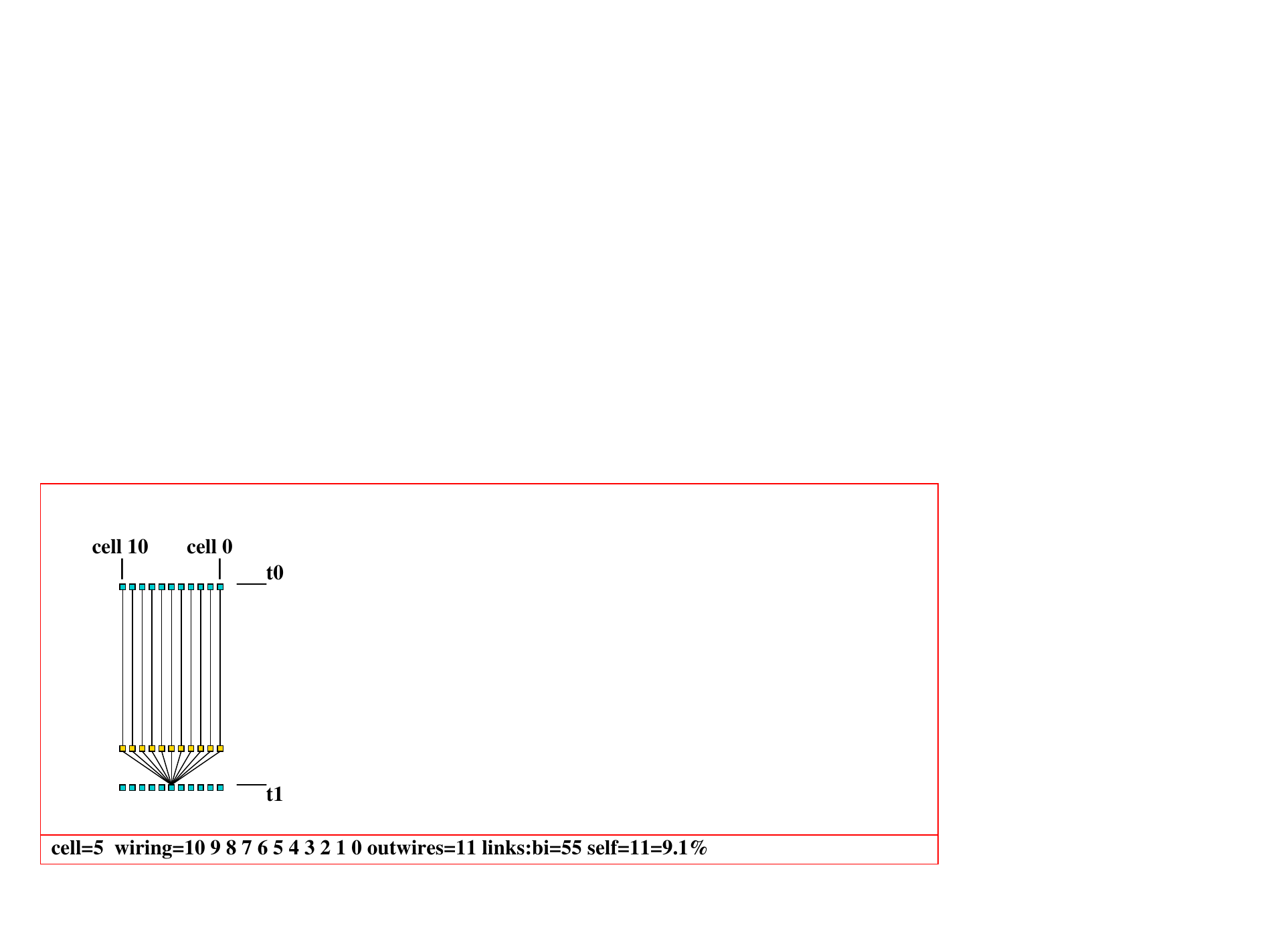}
\raisebox{3ex}{---continues}
\end{minipage}\\[1ex]
\begin{minipage}[c]{.15\linewidth}
\vspace{-8ex}
\begin{flushright}
\textsf{even-$k$\\[-.5ex]
2 to 10\\[-.5ex]
skewed right}
\end{flushright}
\end{minipage}
\hfill
\begin{minipage}[t]{.9\linewidth}
\includegraphics[height=.105\linewidth, bb=80 99 101 143, clip=]{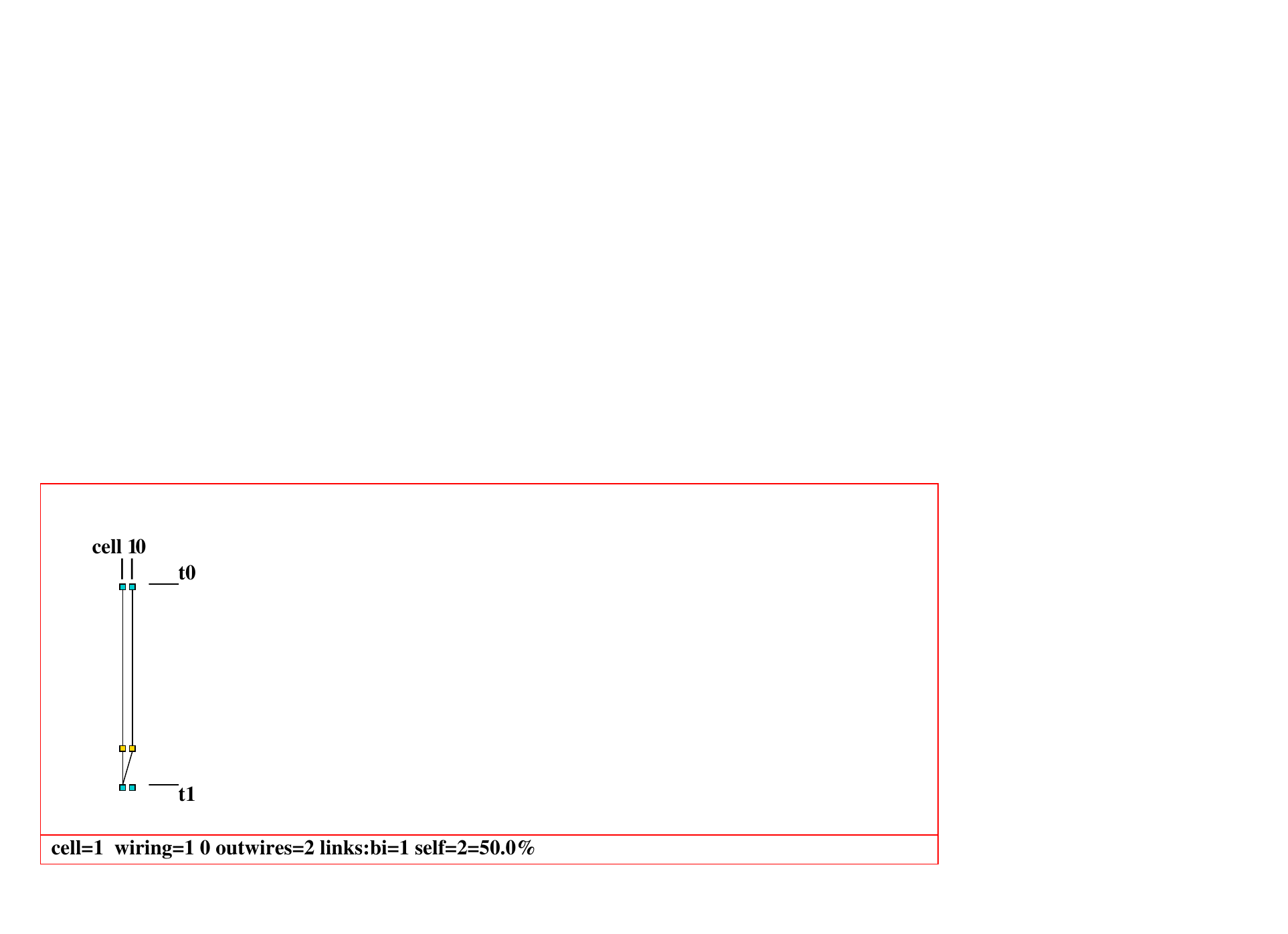}
\includegraphics[height=.105\linewidth, bb=80 99 115 143, clip=]{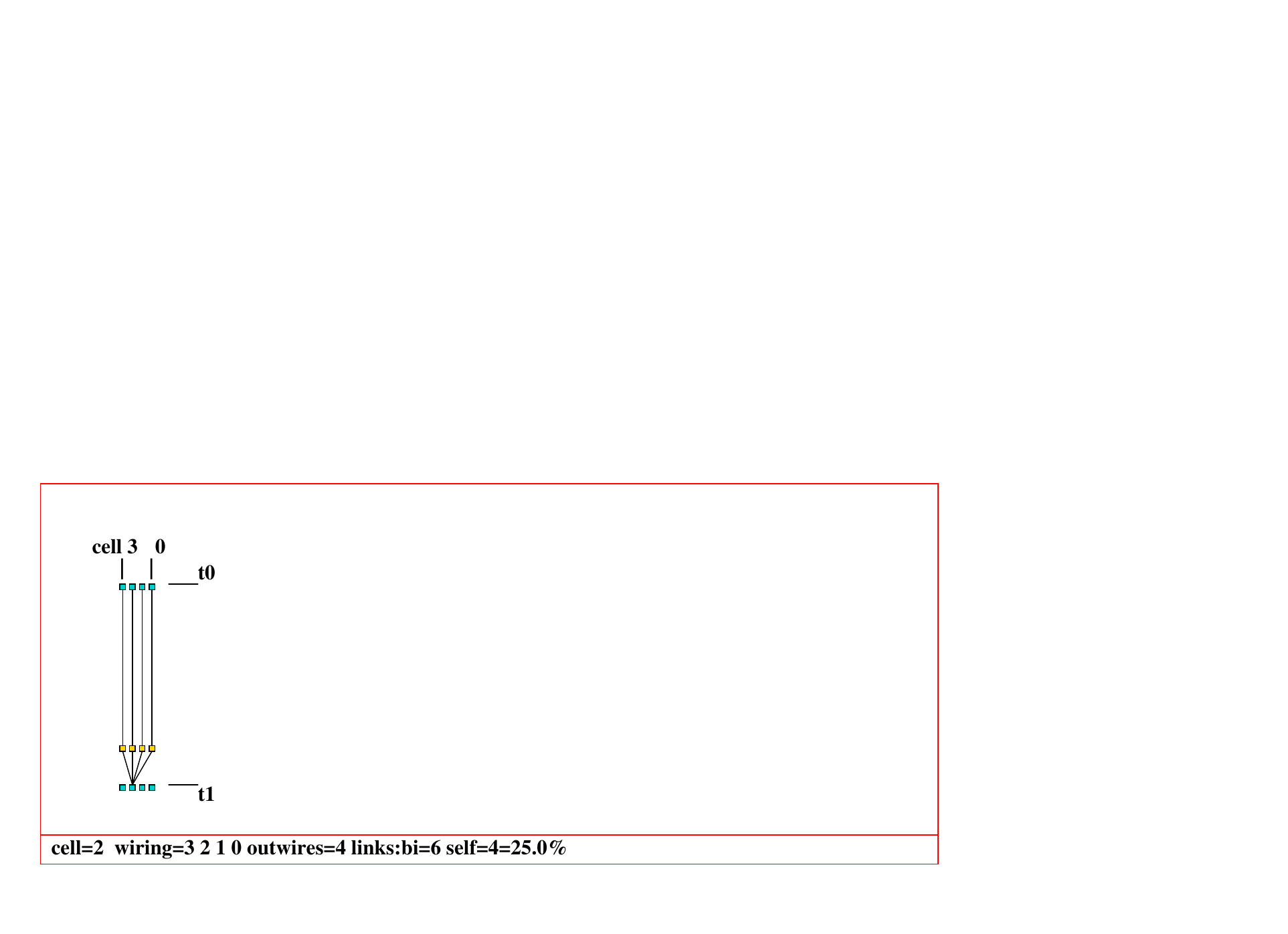}
\includegraphics[height=.105\linewidth, bb=80 99 129 143, clip=]{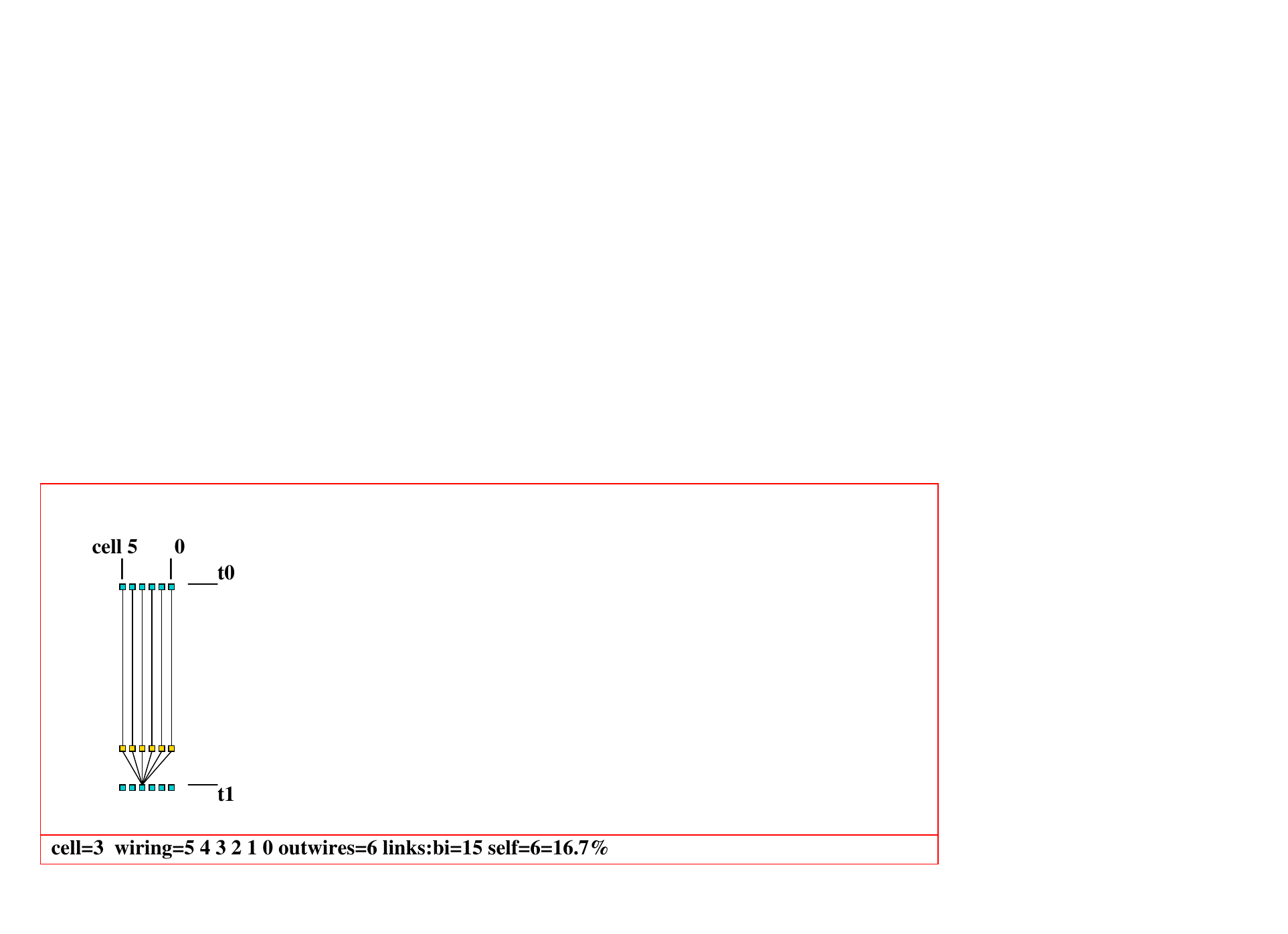}
\includegraphics[height=.105\linewidth, bb=80 99 144 143, clip=]{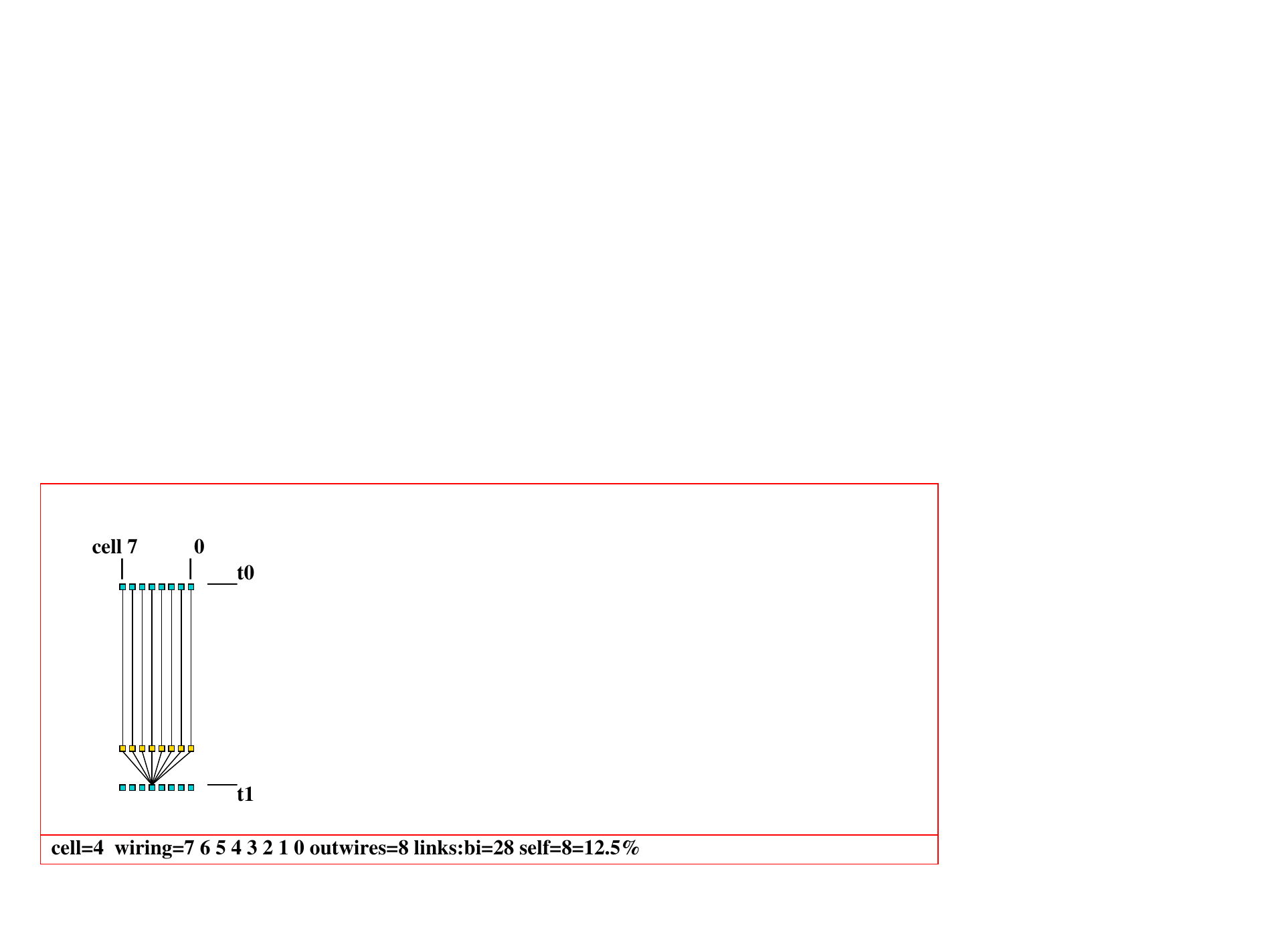}
\includegraphics[height=.105\linewidth, bb=80 99 159 143, clip=]{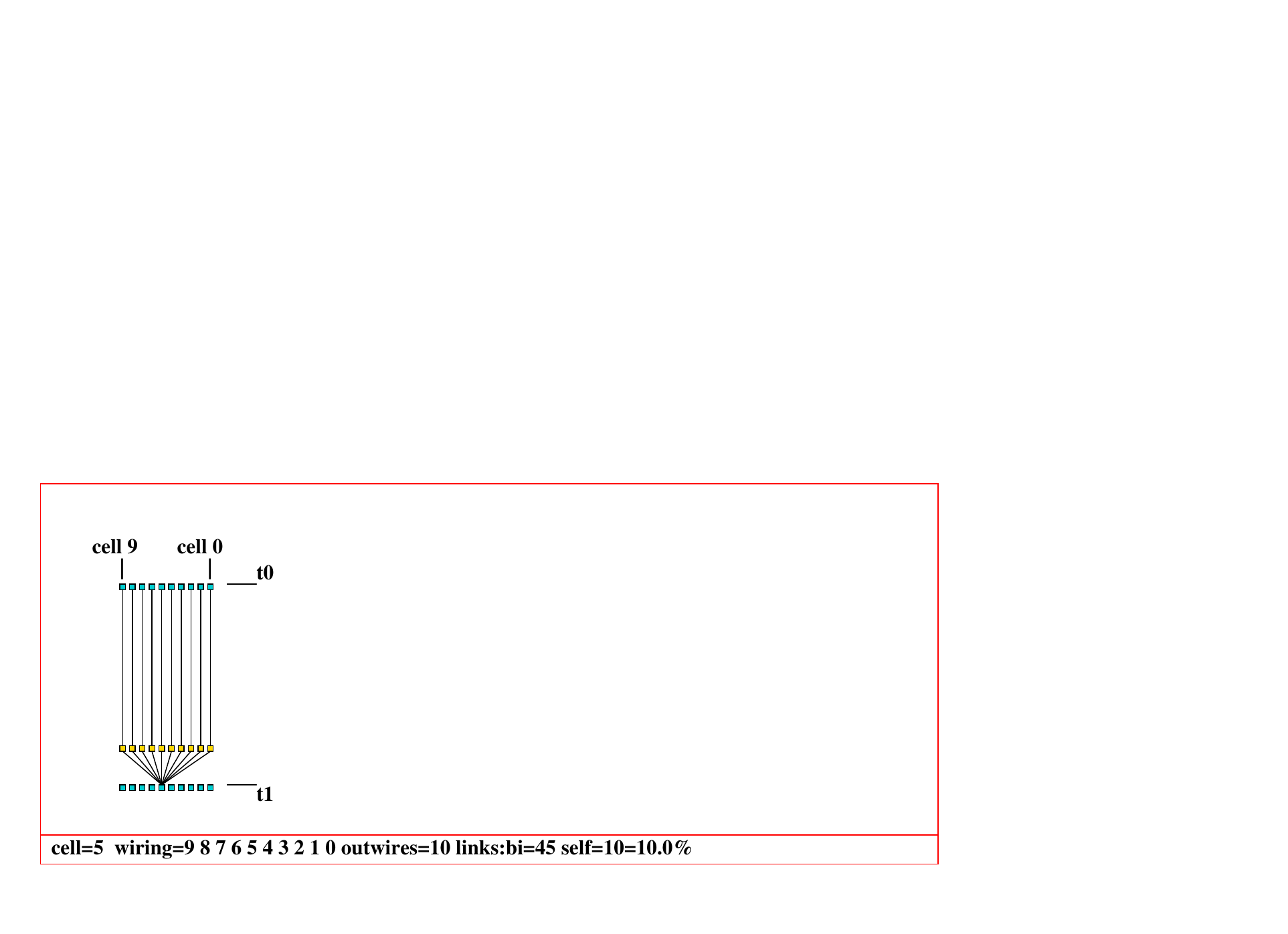}
\raisebox{3ex}{---continues}
\end{minipage}
\end{minipage}
\vspace{-2ex}
\caption[1d n-templates] 
{\textsf{\normalsize 1d n-templates, indexed $k$-1 to 0 from left to right,
 as defined in DDLab for odd and even $k$\cite[EDD:10.1.2]{EDD}. 
Even-$k$  n-templates are asymmetric, skewed to the right.
However, they still support iso-rules as shown in figure~\ref{1d v8k2 space-time pattern}
where the symmetry is preserved by shifting successive time-steps by $1/2$ cell-space.
}}
\label{1d n-templates}
\end{figure}

\begin{figure}[htb]
\textsf{\footnotesize
\begin{center}
\begin{minipage}[c]{1\linewidth}
\begin{minipage}[c]{.1\linewidth} 
\includegraphics[width=1\linewidth,bb=135 172 206 241, clip=]{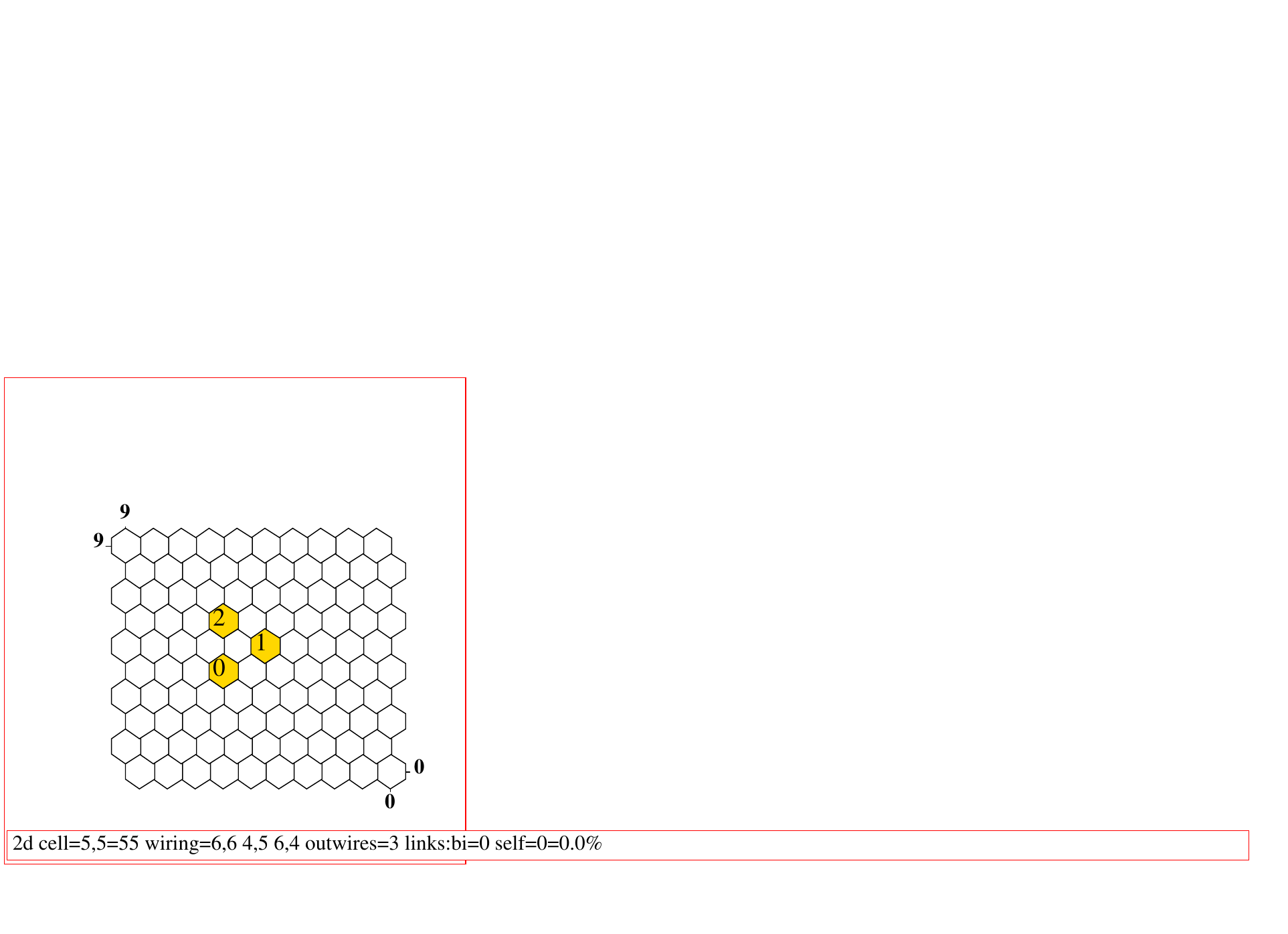}\\[-5ex]
\begin{center}(a)$k$=3\end{center}
\end{minipage}
\hfill
\begin{minipage}[c]{.1\linewidth} 
\includegraphics[width=1\linewidth,bb=135 172 206 241, clip=]{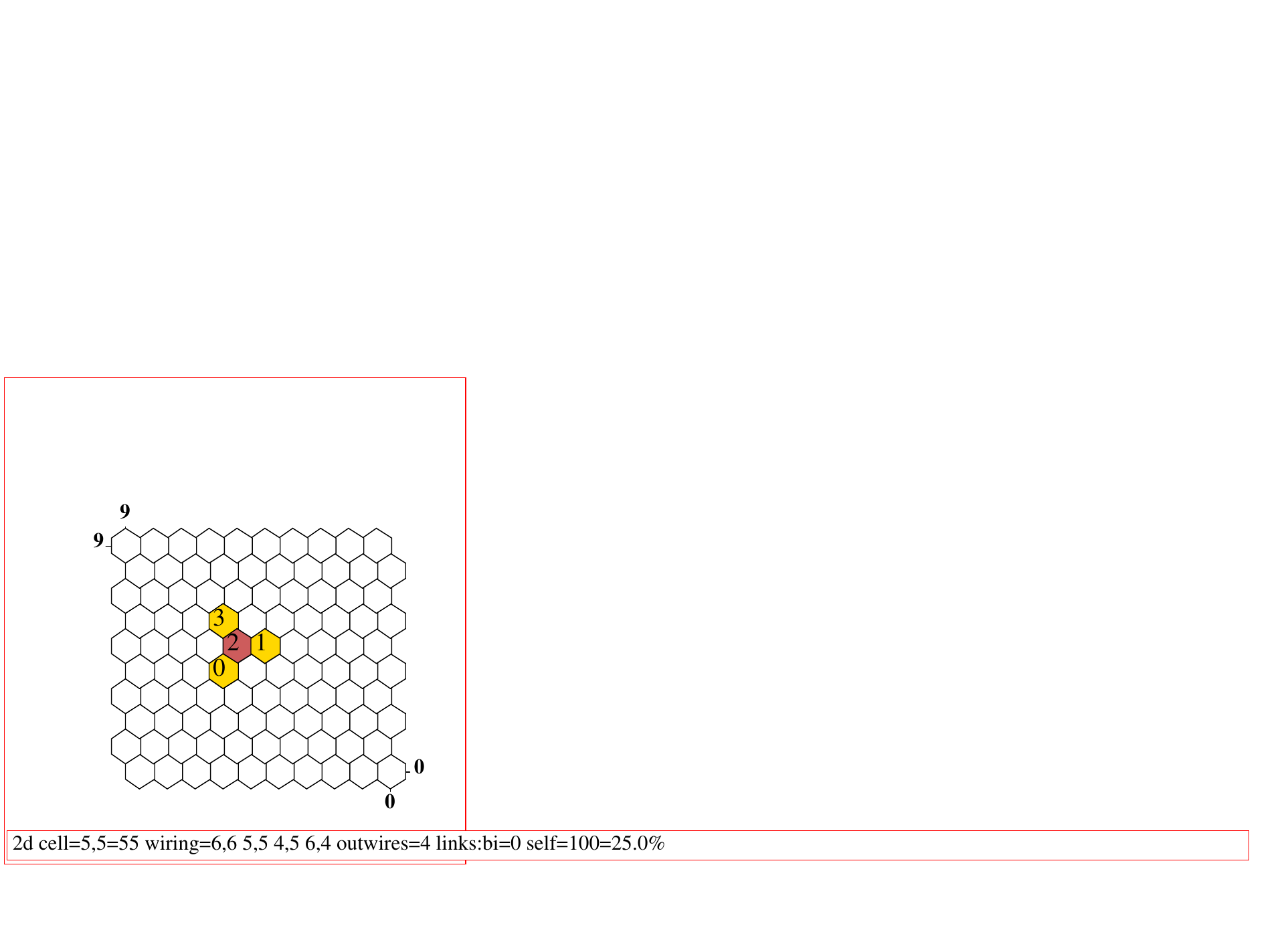}\\[-5ex]
\begin{center}(b)$k$=4t\end{center}
\end{minipage}
\hfill
\begin{minipage}[c]{.1\linewidth} 
\includegraphics[width=1\linewidth,bb=130 178 210 256, clip=]{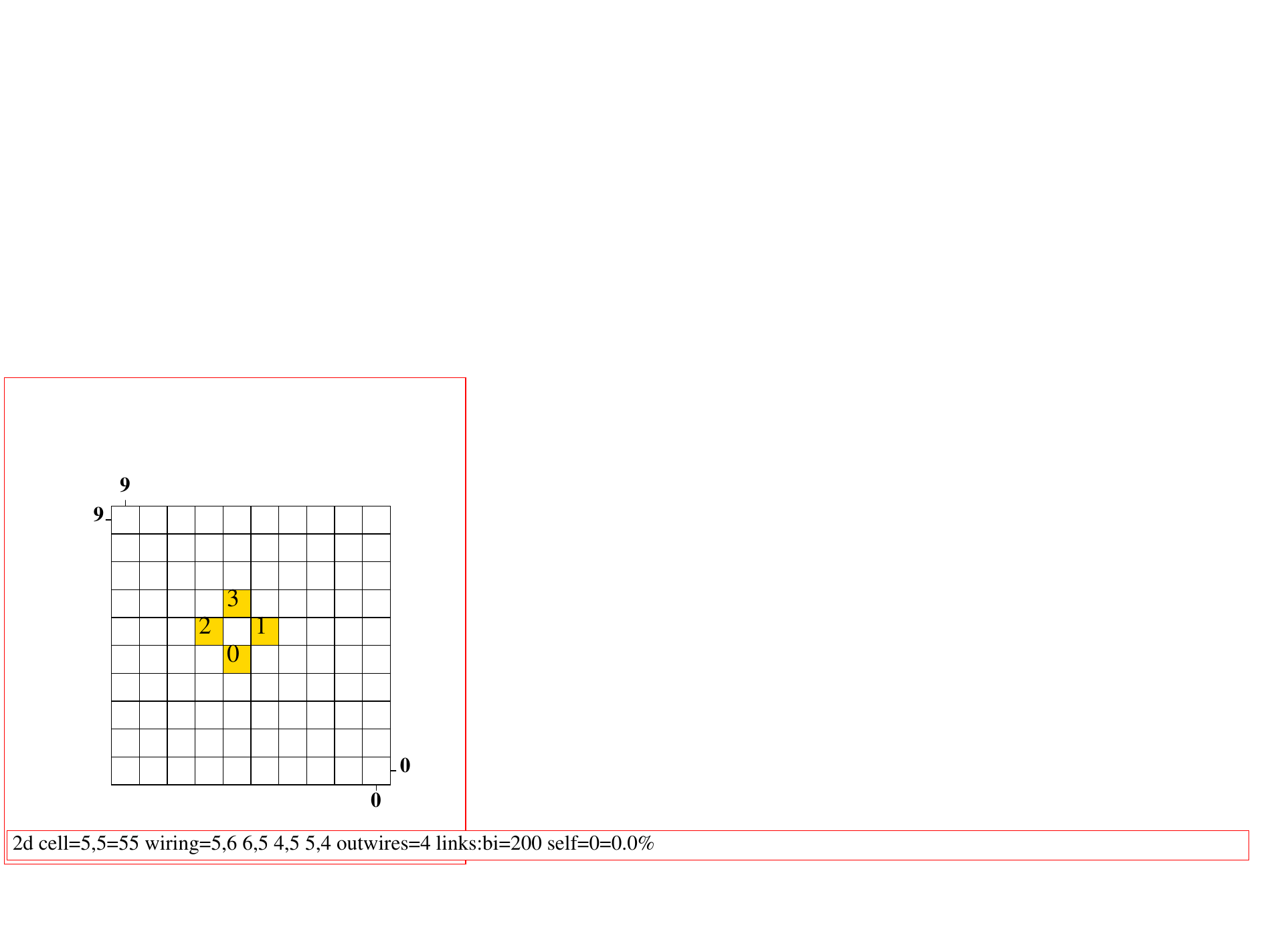}\\[-5ex]
\begin{center}(c)$k$=4s\end{center}
\end{minipage}
\hfill
\begin{minipage}[c]{.1\linewidth} 
\includegraphics[width=1\linewidth,bb=130 178 210 256, clip=]{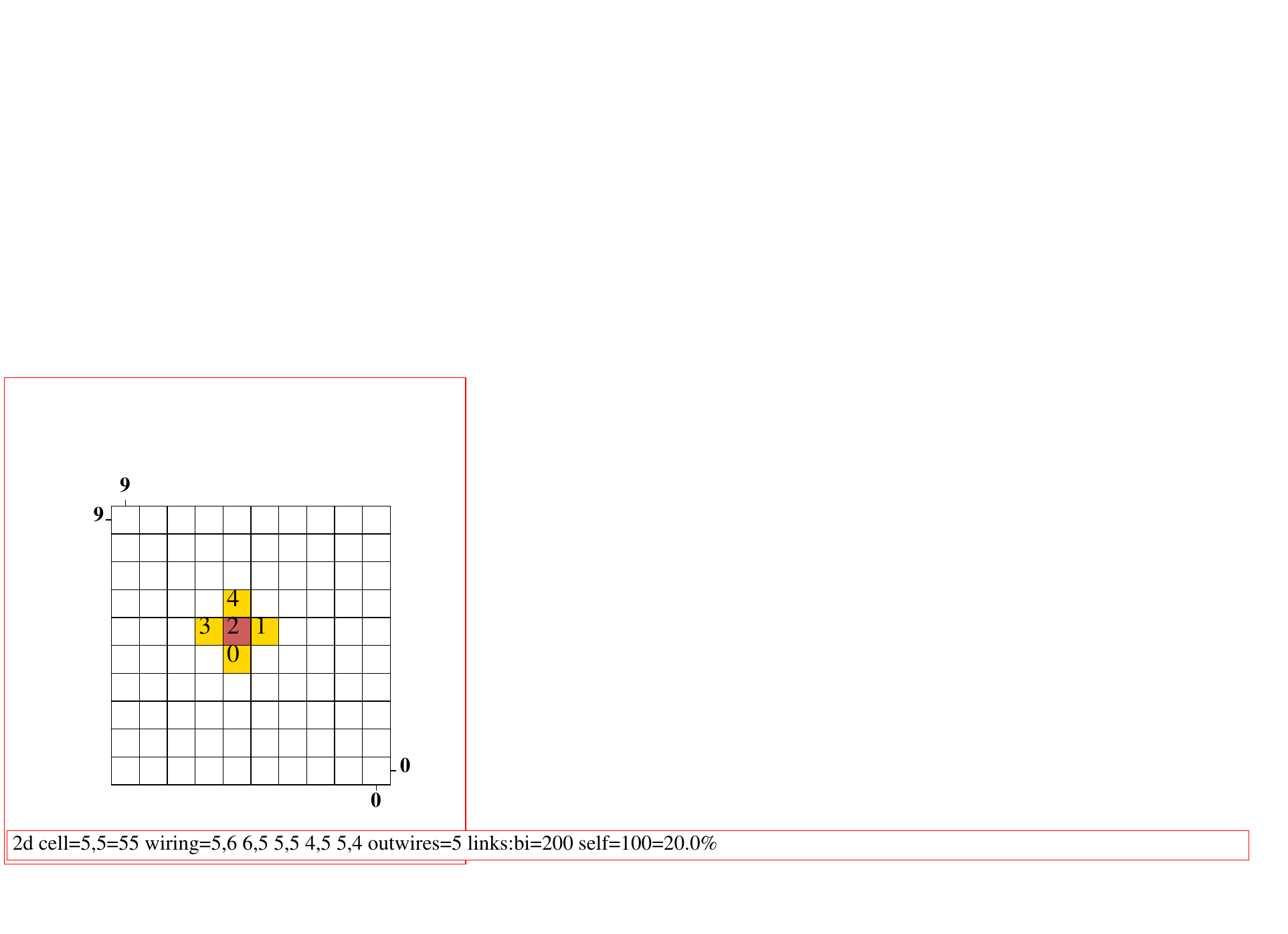}\\[-5ex]
\begin{center}(d)$k$=5\end{center}
\end{minipage}
\hfill
\begin{minipage}[c]{.1\linewidth} 
\includegraphics[width=1\linewidth,bb=135 172 206 241, clip=]{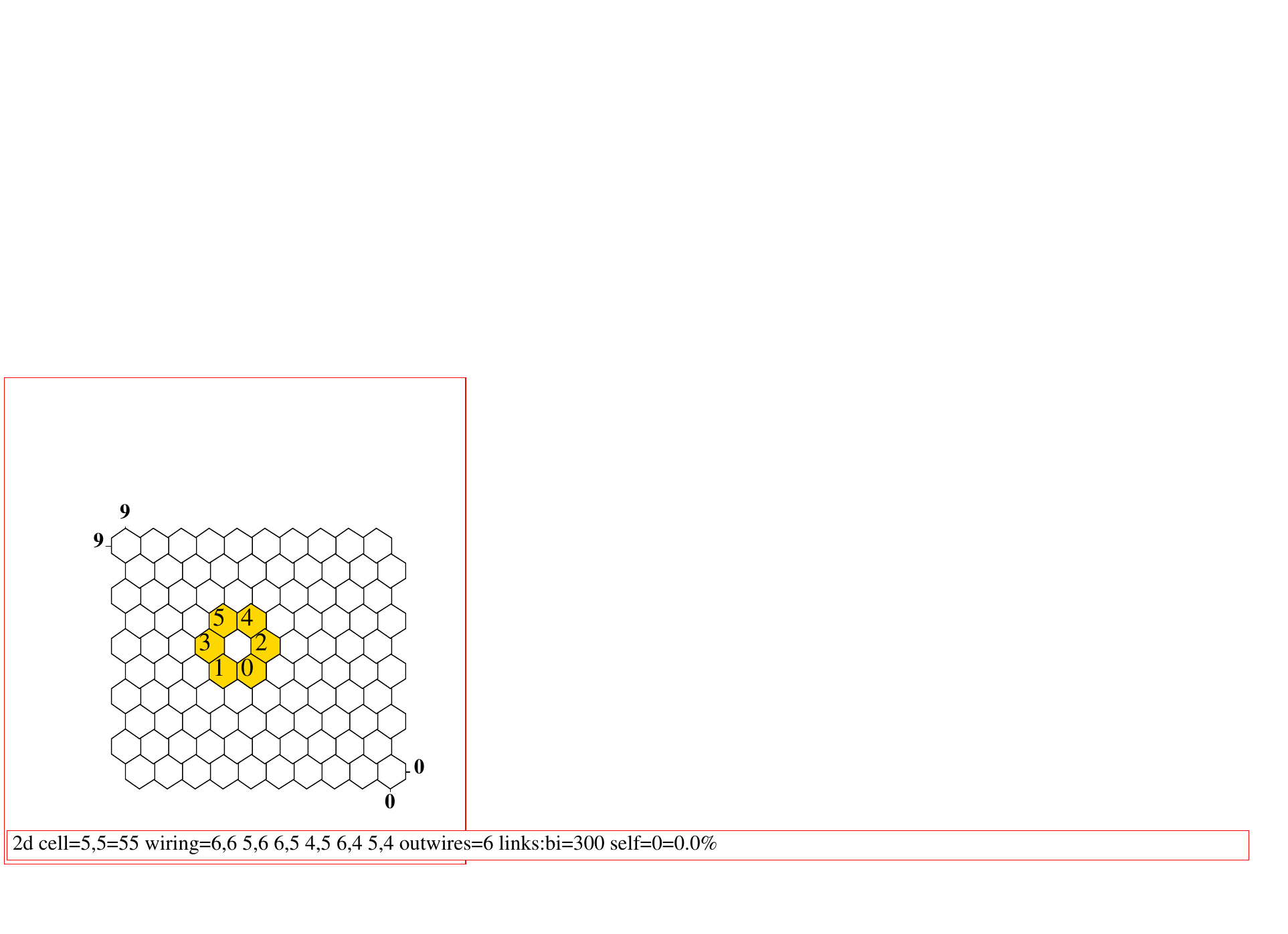}\\[-5ex]
\begin{center}(e)$k$=6\end{center}
\end{minipage}
\hfill
\begin{minipage}[c]{.1\linewidth} 
\includegraphics[width=1\linewidth,bb=135 172 206 241, clip=]{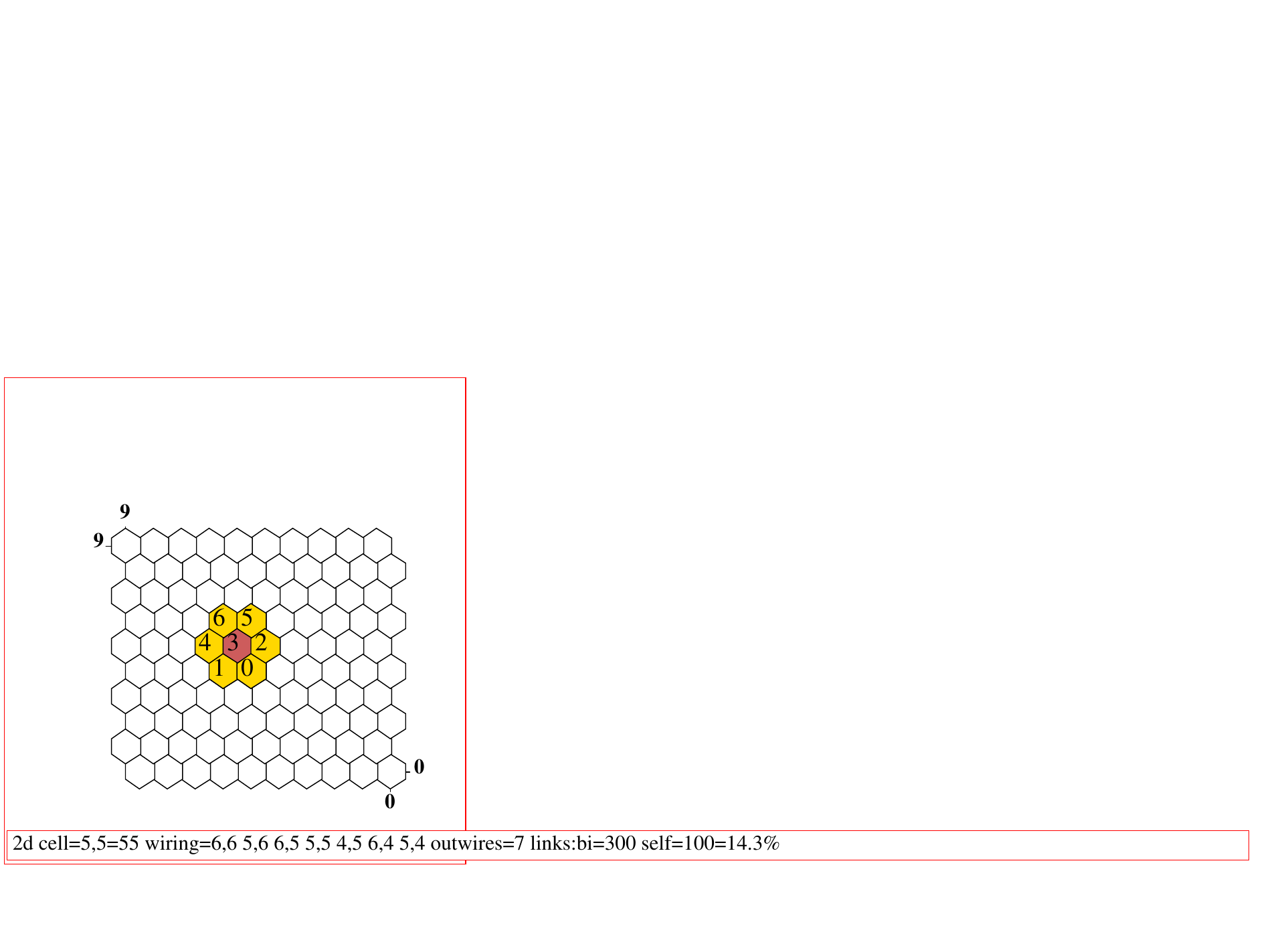}\\[-5ex]
\begin{center}(f)$k$=7\end{center}
\end{minipage}
\hfill
\begin{minipage}[c]{.1\linewidth} 
\includegraphics[width=1\linewidth,bb=113 160 197 244, clip=]{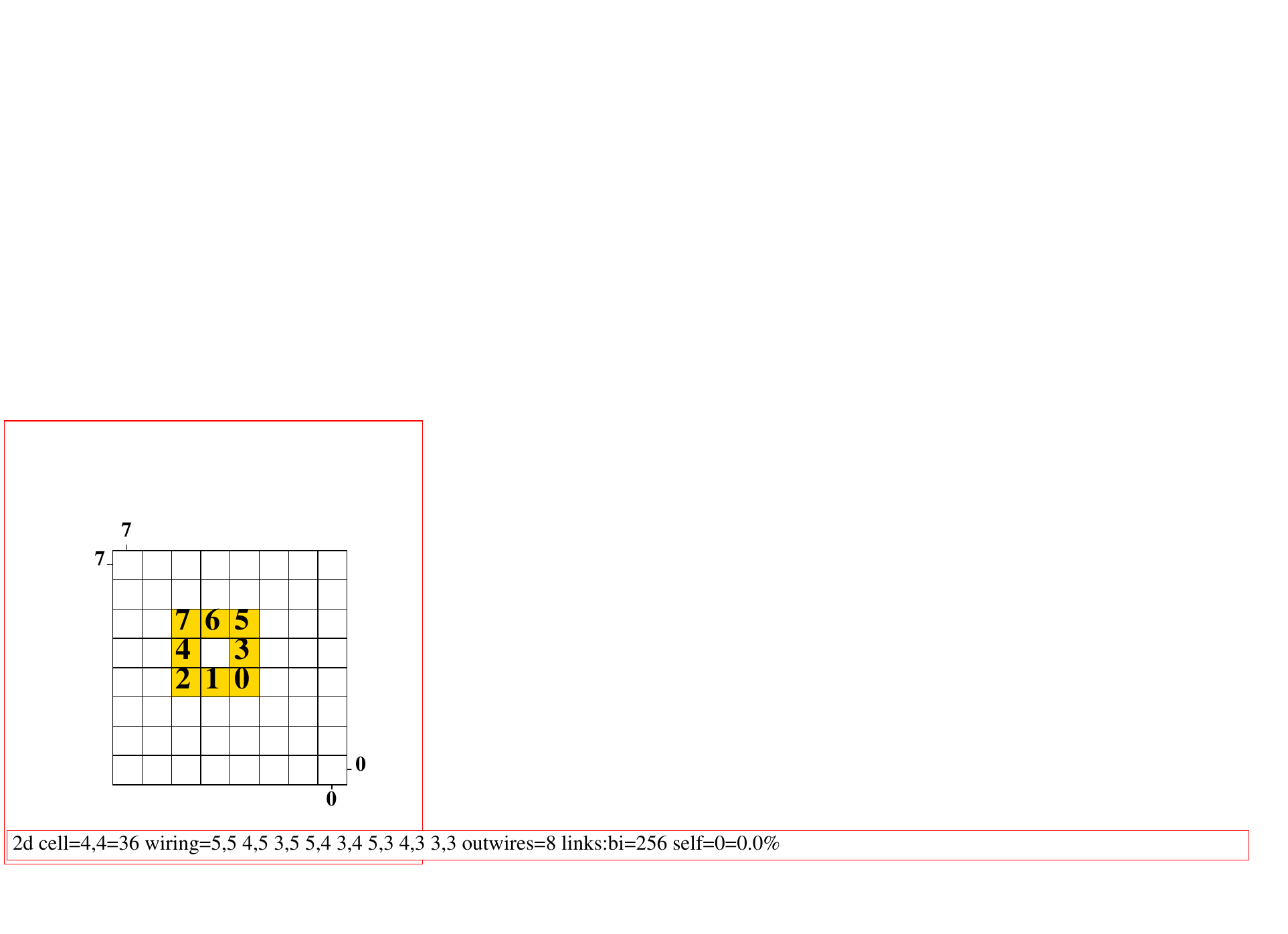}\\[-5ex]
\begin{center}(h)$k$=8\end{center}
\end{minipage}
\hfill
\begin{minipage}[c]{.1\linewidth} 
\includegraphics[width=1\linewidth,bb=113 160 197 244, clip=]{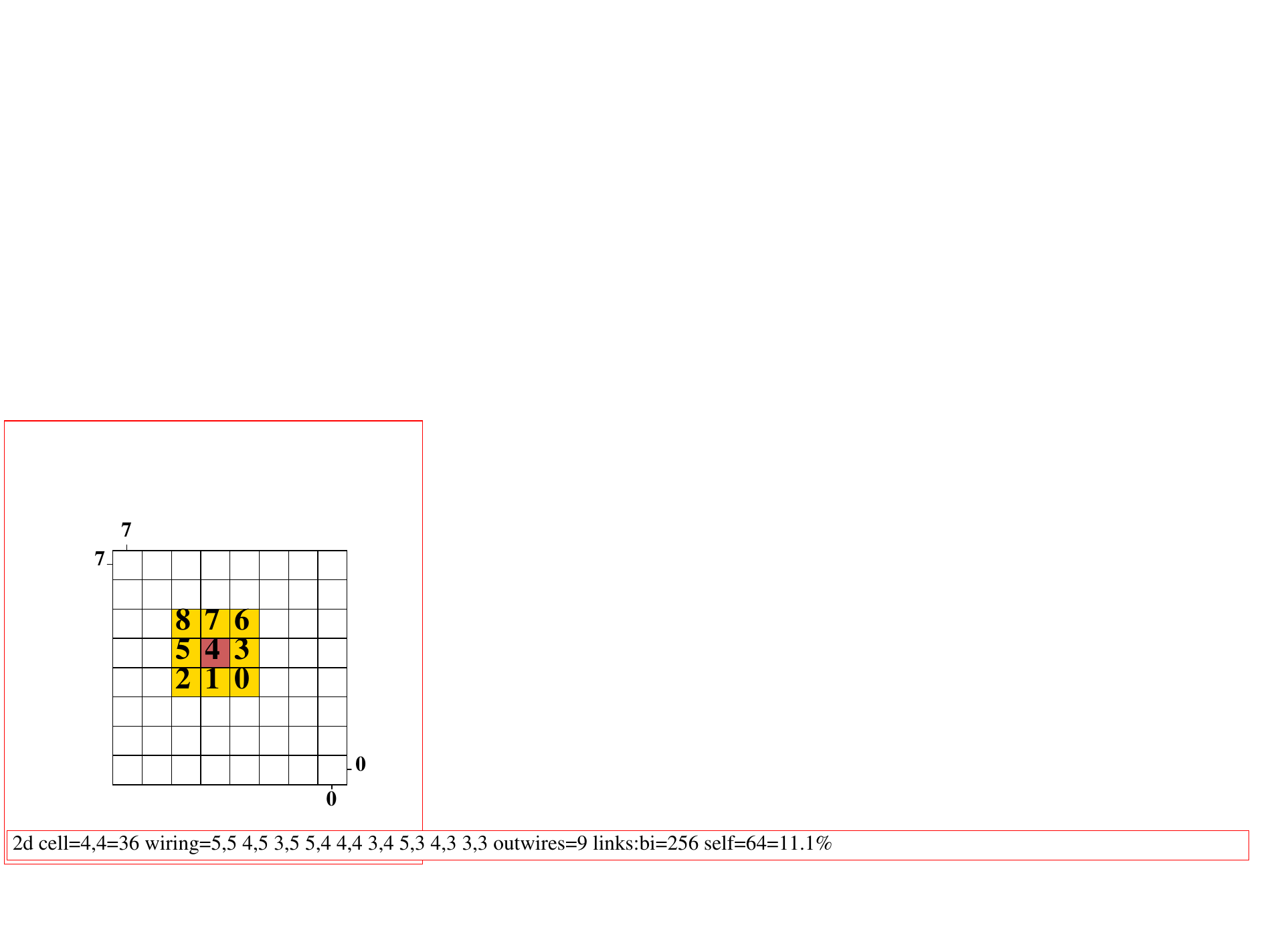}\\[-5ex]
\begin{center}(i)$k$=9\end{center}
\end{minipage}
\end{minipage}
\end{center}
}
\vspace{-2ex}
\caption[2D n-templates]
{\textsf{2D n-templates ($k$=3 to 9) as defined and indexed in 
DDLab\cite[EDD:10]{EDD}, hexagonal or square. $k$=4 n-templates can be either. 
To achieve periodic boundary conditions
hexagonal n-templates require even lattice dimensions.
The target cell is central even if not part of the n-template.
For 3-value CA, glider rules are readily found for these n-templates
as in figure~\ref{2d v3 Glider examples}.}}
\label{2d n-templates}
\end{figure}

\begin{figure}[htb]
\textsf{\footnotesize
\begin{center}
\begin{minipage}[c]{.4\linewidth}
\begin{minipage}[c]{.4\linewidth}
\includegraphics[width=1\linewidth,bb=191 106 299 213, clip=]{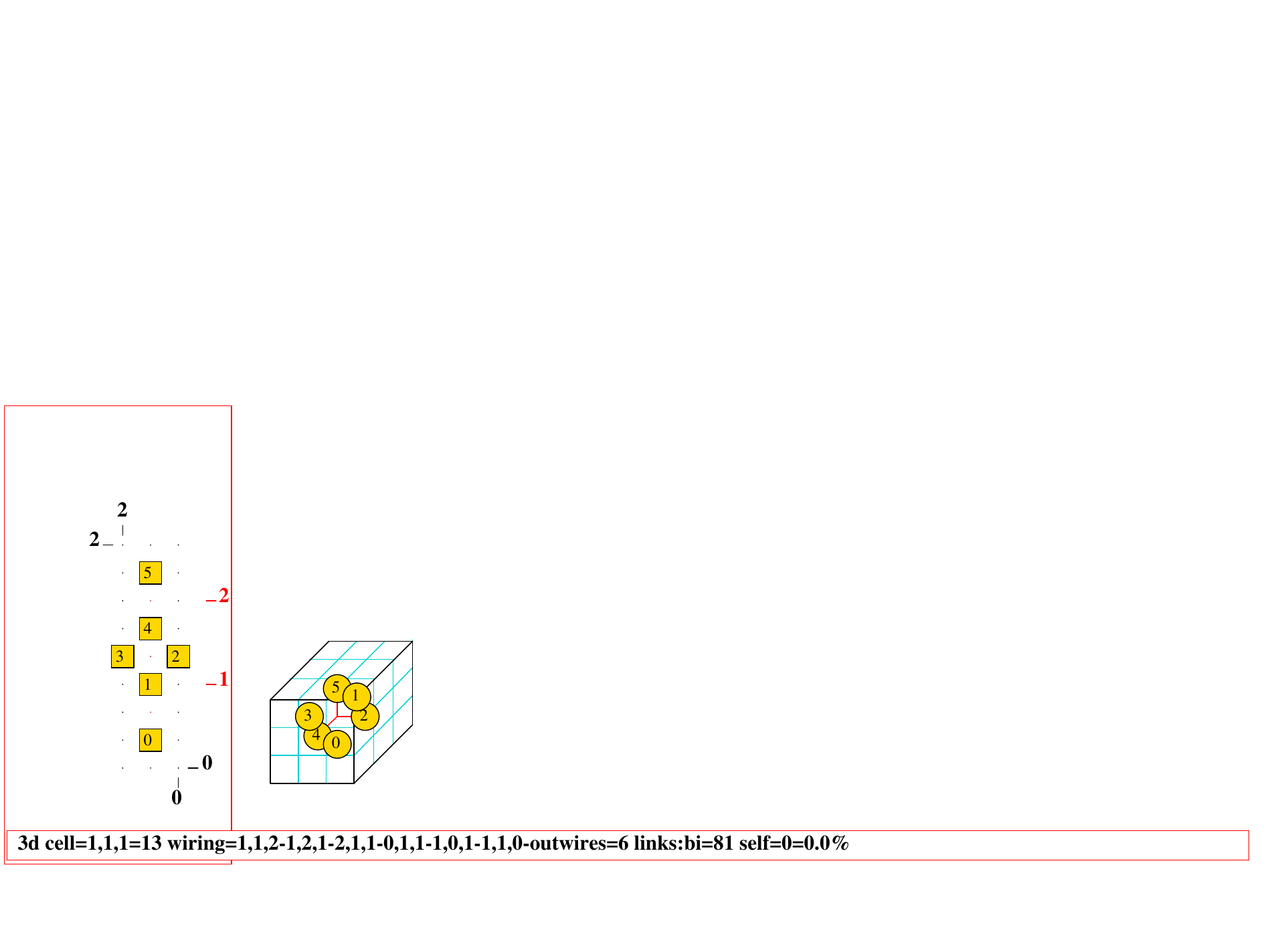}\\[-5ex]
\begin{center}(a)$k$=6\end{center}
\end{minipage}
\hfill
\begin{minipage}[c]{.4\linewidth} 
\includegraphics[width=1\linewidth,bb=191 106 299 213, clip=]{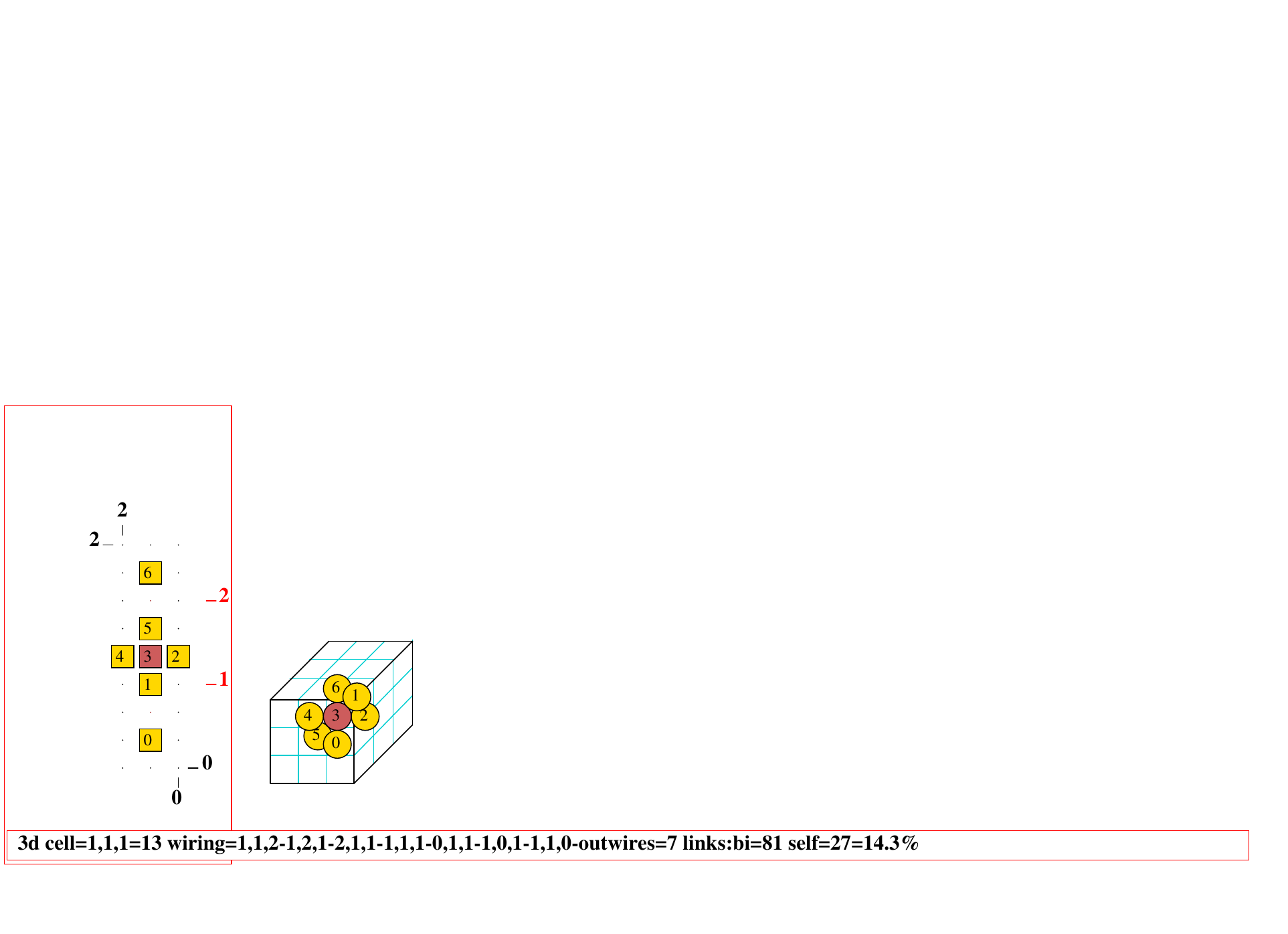}\\[-5ex]
\begin{center}(b)$k$=7\end{center}
\end{minipage}
\end{minipage}
\end{center}
}
\vspace{-3ex}
\caption[3d n-templates]
{\textsf{3d n-templates ($k$=6 or 7) as defined and indexed in 
DDLab\cite[EDD:10]{EDD}. The target cell is central even if not part of the n-template.
}}
\label{3d n-templates}
\vspace{-1ex}
\end{figure}

\noindent The lattice geometry of a CA depends on its
n-template, and there are a wide range of pre-defined n-templates in
DDLab\cite[EDD:10]{EDD}. 
In figures~\ref{1d n-templates}, \ref{2d n-templates} 
and \ref{3d n-templates} and we present those pre-defined n-templates
where iso-groups and iso-rules are computed, and which themselves
have a symmetric geometry.

\enlargethispage{5ex}
A CA ``target'' cell updates according to the values within its
n-template.  As can be seen in 
figures~\ref{2d n-templates} and \ref{3d n-templates},
the target cell in some cases
is not a member of the n-template.  The n-template is homogeneous
throughout the network so requires periodic boundary conditions
where each lattice boundary wraps around to its opposite boundary
resulting in a ring of cells in 1d, a torus in 2d, and
a 3-torus in 3d. However, null boundary condition can also be
imposed\cite[EDD:31.3]{EDD}.

\section{rcode, iso-groups and iso-rules}
\label{rcode, iso-groups and iso-rules}
CA rules can be divided and defined according to a number of
(possibly overlapping)
types\cite[EED:13]{EDD}. These include the full rule-table (rcode),
k-totalistic (kcode), t-totalistic (tcode), outer-totalistic,
reaction-diffusion, survival/birth, and of course iso-rules.

The most general rule type, rcode, can implement any logic including
all the types listed above, and forms the basis for extracting
iso-rules.
Rcode is a list of the outputs of all $v^k$ possible
neighborhoods depending only the value-range $v$ and neighborhoods
size $k$ giving a rule-space of $v^{v^k}$, and is independent of
n-template geometry.  The list order must be specified, and we follow
Wolfram's classical convention\cite{wolfram83,wolfram2002}; a descending order of
neighborhood binary (or \mbox{$v$-ary} for $v$$>$2) values from left to right
which is also the rcode index, as in this example for binary $k$=3
where the decimal equivalent of the rcode string gives the
``Elementary Rule''\footnote{There are $2^{2^3}$=256 $v2k3$ ``elementary'' rules
consisting of 88 equivalents, in 48 rule clusters. Of these
64 rules, 36 equivalents, in 20 clusters, are symmetric\cite{Wuensche92}
so 1d isotropic.} number.

\fontsize{8pt}{8pt}
\begin{verbatim}
     7    6    5    4    3    2    1    0   - binary value and rcode index
    111  110  101  100  011  010  001  000  - k=3 neighborhoods
     0    0    1    1    1    1    0    0   - output string = rcode 60 in decimal
\end{verbatim}
\normalsize

In DDLab the neighborhoods are displayed vertically for compactness with
a $k$ index ($k$-1 to 0, top down), making a so called ``neighborhood
matrix'', shown here for binary $k$=3, and for $k$=5 where rcode is better
expressed in hexadecimal rather than decimal,
\clearpage

\fontsize{8pt}{8pt}
\begin{verbatim}
rule index - 7......0                   31...... ........ ........ .......0 
             :      :                   :                                 :
         2 - 11110000               4 - 11111111 11111111 00000000 00000000
 k-index 1 - 11001100               3 - 11111111 00000000 11111111 00000000
         0 - 10101010       k-index 2 - 11110000 11110000 11110000 11110000
             --------               1 - 11001100 11001100 11001100 11001100
 rcode 193 - 11000001               0 - 10101010 10101010 10101010 10101010
                                        -------- -------- -------- --------
               rcode (dec) 4276676736 - 11111110 11101000 11101000 10000000 
                     (hex) fee8e880 (majority rule)
\end{verbatim}
\normalsize

\begin{figure}[htb]
\textsf{\footnotesize
\begin{center}
\begin{minipage}[c]{1\linewidth}
\begin{minipage}[c]{1\linewidth}
\includegraphics[width=.5\linewidth,bb=43 401 299 469 469, clip=]{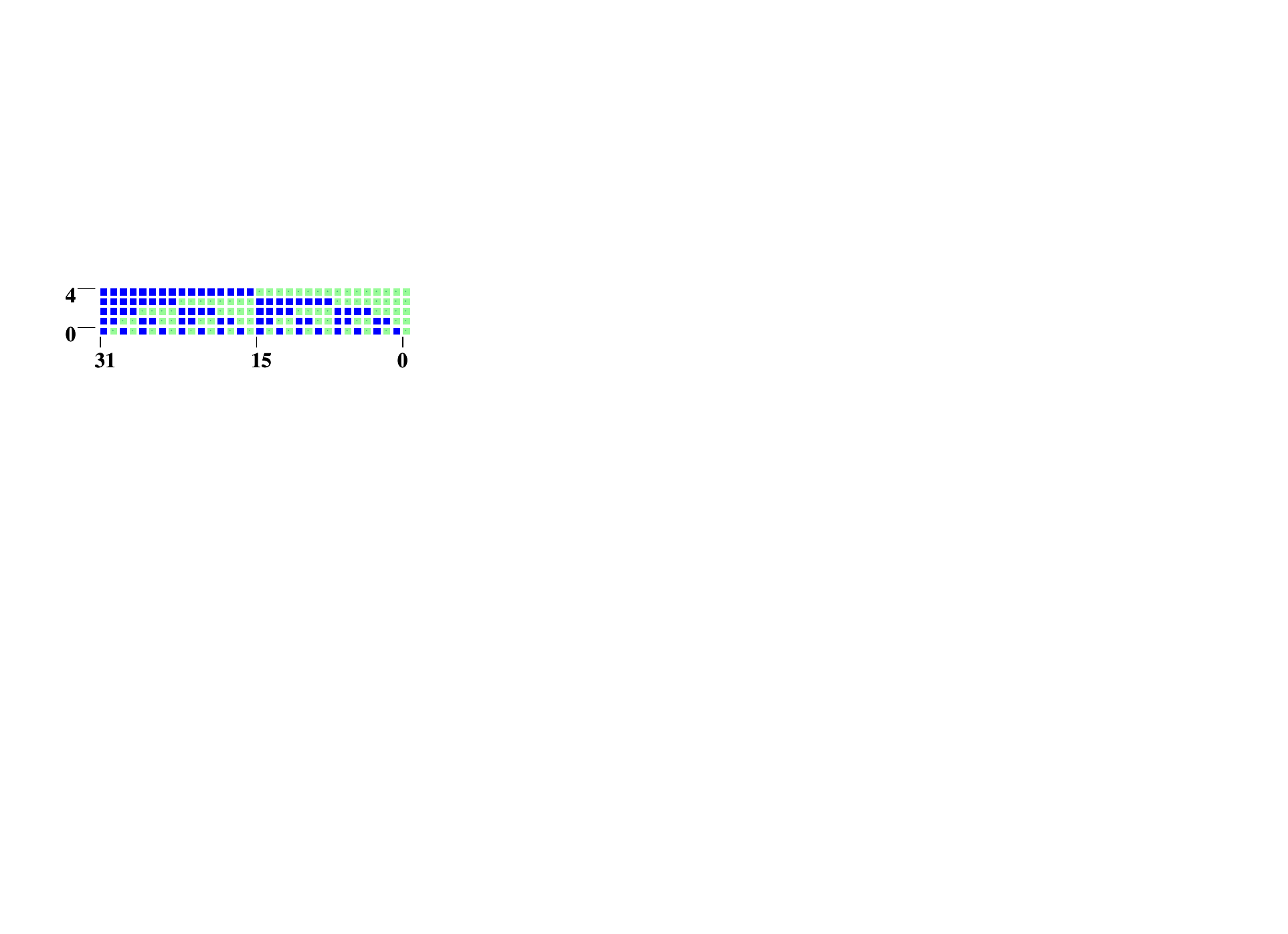}
\raisebox{8ex}{(a) $v2k5$, $2^5$=32, complete matrix}
\end{minipage}\\[1ex]
\begin{minipage}[c]{1\linewidth}
\includegraphics[width=1\linewidth,bb=40 417 1102 470, clip=]{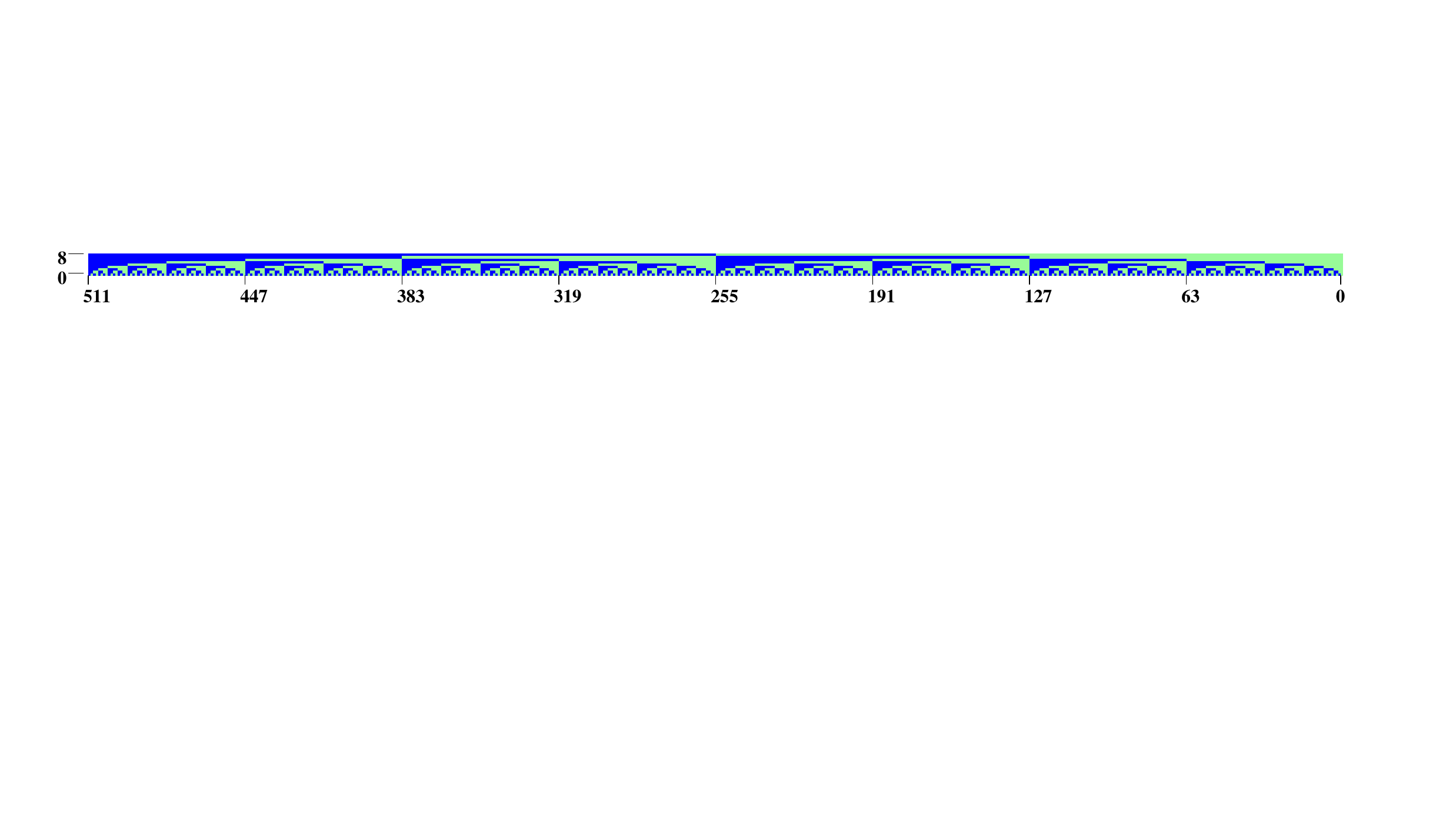}\\[-6ex]
\begin{center}(b) $v2k9$, $2^9$=512, complete matrix\end{center}
\end{minipage}\\[2ex]
\begin{minipage}[c]{1\linewidth}
\includegraphics[width=1\linewidth,bb=40 414 598 470, clip=]{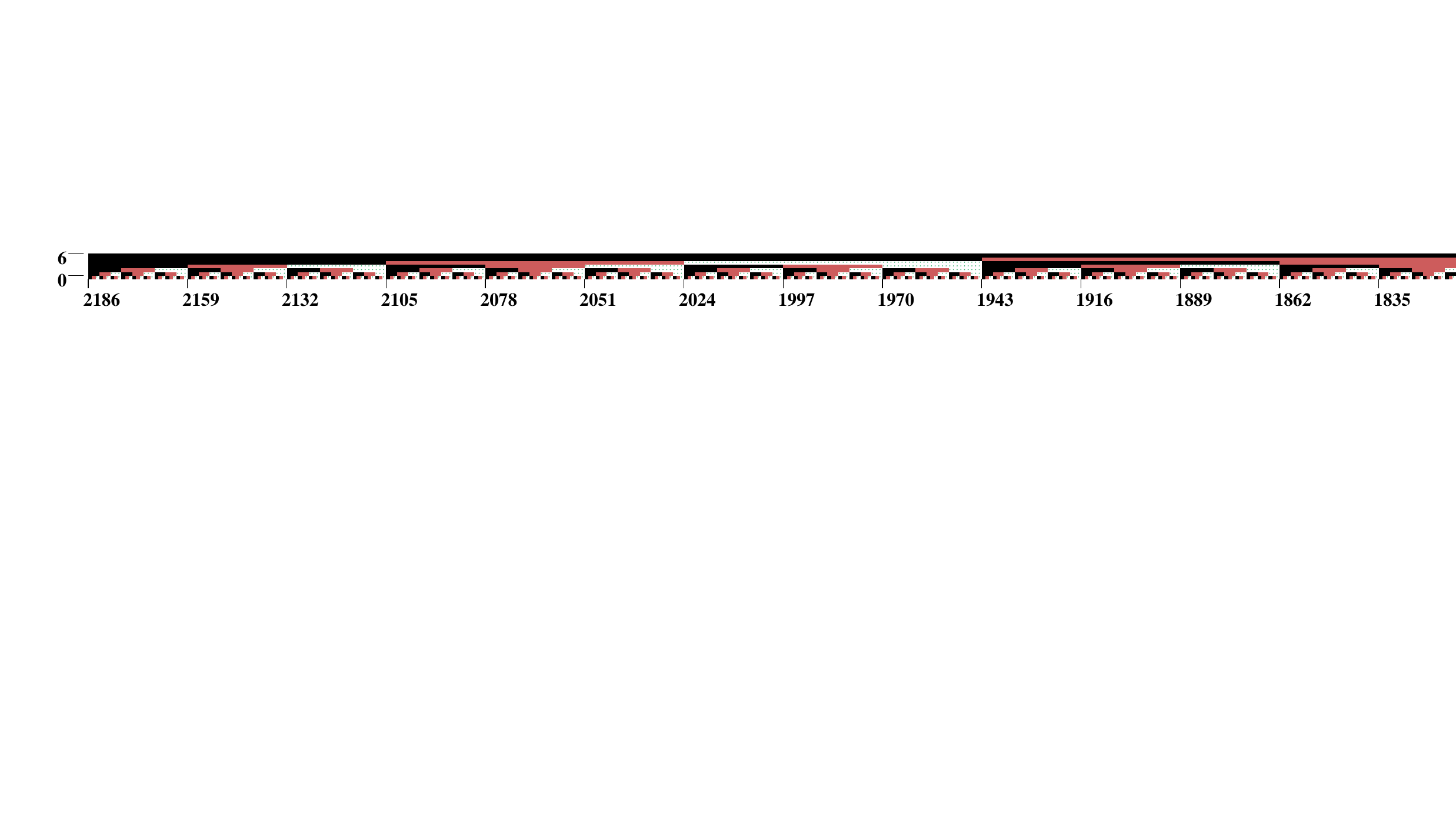}\\[-6ex]
\begin{center}(c)  $v3k7$, $3^7$=2187, left part only\end{center}
\end{minipage}\\[2ex]
\begin{minipage}[c]{1\linewidth}
\includegraphics[width=1\linewidth,bb=40 414 598 470, clip=]{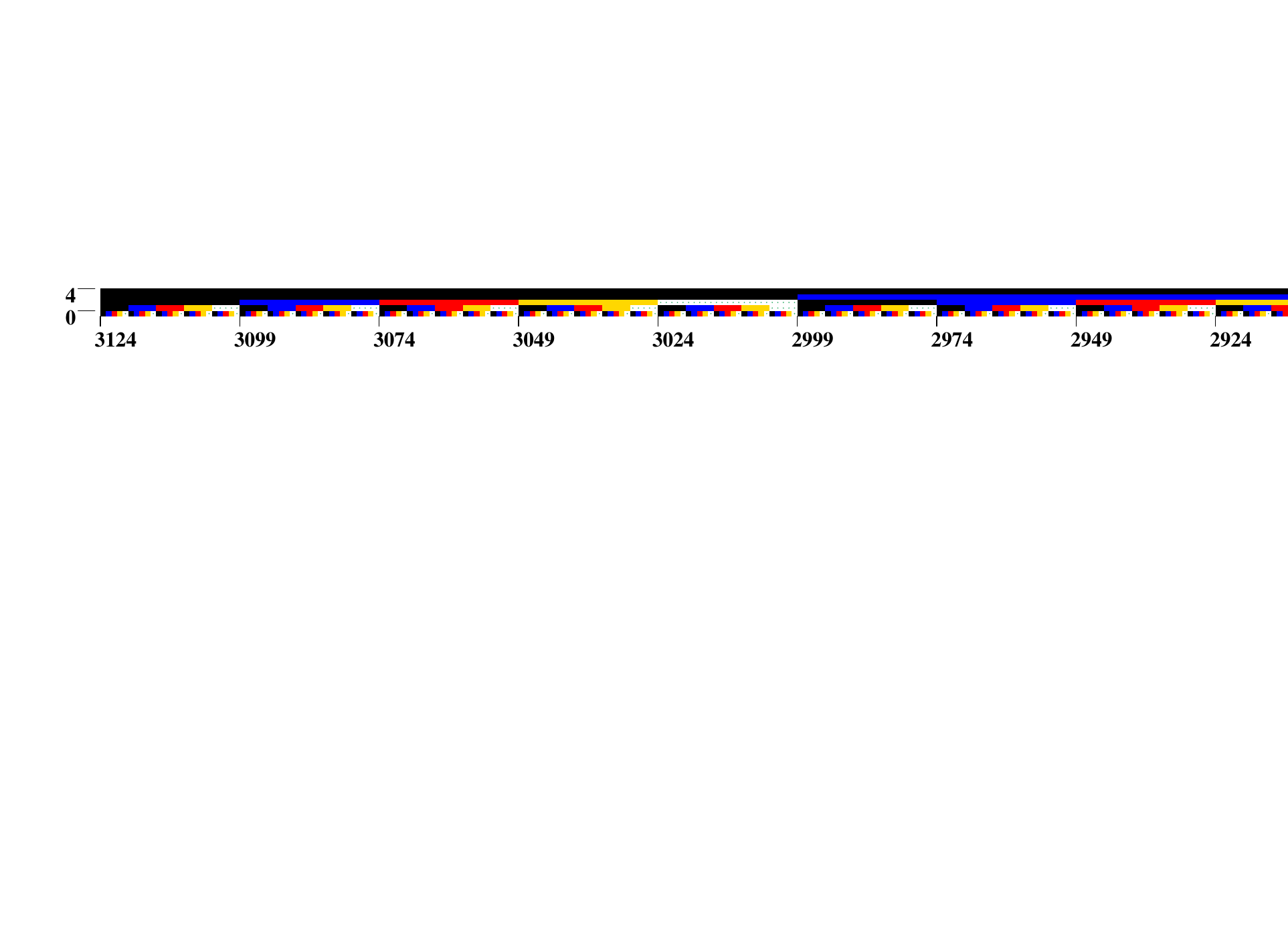}\\[-6ex]
\begin{center}(d) $v5k5$, $5^5$=3125, left part only\end{center}
\end{minipage}
\end{minipage}
\end{center}
}
\vspace{-3ex}
\caption[neighborhood matrices]
{\textsf{Examples of various $v$,$k$ matrices showing rcode size.
}}
\label{neighborhood matrices}
\vspace{-2ex}
\end{figure}

As a reference, the neighborhood matrix is displayed graphically prior
to selecting, editing and transforming the rcode. The same matrix
principles apply for any values of $v$ and $k$, as in the examples in
figure~\ref{neighborhood matrices}.

\enlargethispage{3ex}
\begin{figure}[htb]
\textsf{\footnotesize
\begin{center}
\begin{minipage}[c]{1\linewidth}
\begin{minipage}[c]{.49\linewidth}
\includegraphics[width=1\linewidth,bb=11 17 340 36, clip=]{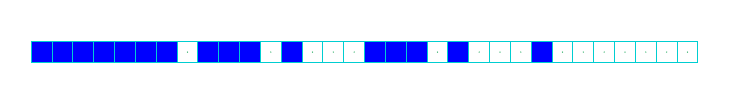}\\
\includegraphics[width=.40\linewidth,bb=11 17 140 36, clip=]{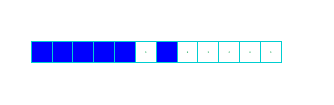}\\[-6ex]
\begin{center}(a) $v2k5$  rcode(32), 2d square iso-rule(12)\end{center}
\end{minipage}
\hfill
\begin{minipage}[c]{.49\linewidth}
\includegraphics[width=1\linewidth,bb=4 7 330 54, clip=]{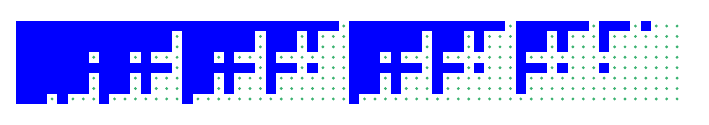}\\
\includegraphics[width=1\linewidth,bb=3 7  333 24, clip=]{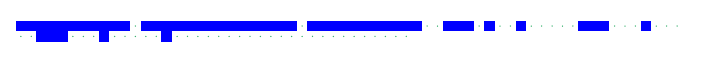}\\[-6ex]
\begin{center}(b) $v2k9$ rcode(512), 2d square iso-rule(102)\end{center}
\end{minipage}\\[2ex]
\begin{minipage}[c]{1\linewidth}
\includegraphics[width=1\linewidth]{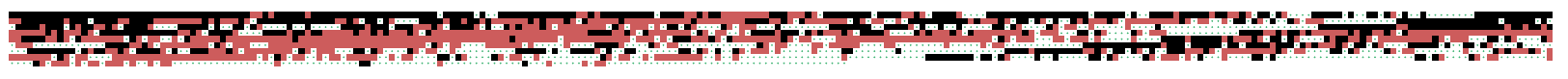}\\
\includegraphics[width=1\linewidth]{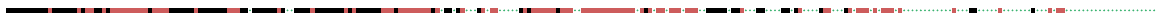}\\[-6ex]
\begin{center}(c) $v3k7$ rcode(2187), 2d hex iso-rule(276)\end{center}
\end{minipage}\\[2ex]
\begin{minipage}[c]{1\linewidth}
\includegraphics[width=1\linewidth]{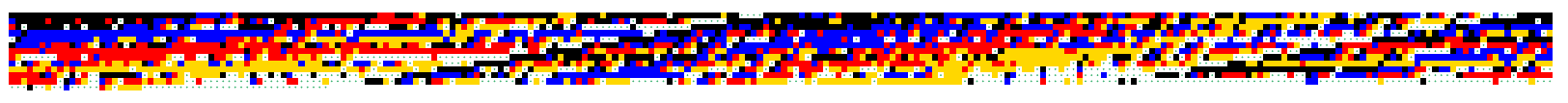}\\
\includegraphics[width=1\linewidth]{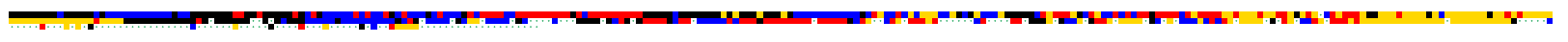}\\[-5ex]
\begin{center}(d)  $v5k5$ rcode(3225), 2d square iso-rule(600)\end{center}
\end{minipage}
\end{minipage}
\end{center}
}
\vspace{-3ex}
\caption[majority rules]
{\textsf{Examples of majority (voting) rules\cite[EDD:16.7]{EDD} for 
$v$,$k$ in figure~\ref{neighborhood matrices}
presented as bit/value strings (and sizes) as rcode, and transformed to 2d iso-rules.
Only binary majority rcode with odd $k$ is isotropic by default when initially selected.
}}
\label{majority rules}
\end{figure} 
\clearpage

The examples in figure~\ref{majority rules} show majority (voting)
rcode\cite[EDD:16.7]{EDD} selected in DDLab, where the majority value
in each neighborhood becomes its output. In case of a tie, for $v$=2
the central cell wins --- the rcode is isotropic by default.
For $v$$>$2, or an empty central cell, one of
the majority values is picked at random --- probably not isotropic. 
However, the transformation to an iso-rule also induces
isotropy in the original rcode. The iso-rule string is much shorter
than the rcode-string as can be seen in table~\ref{iso-rule size
  table}.  Graphical string presentations can be rescaled, adjusted
between single or multiple rows, and allow various functions and
manipulations with the mouse and keyboard \cite[EDD:16.4]{EDD}.

\begin{figure}[htb]
\textsf{\small
\begin{minipage}[c]{1\linewidth}
\includegraphics[width=.95\linewidth,bb=7 568 459 666, clip=]{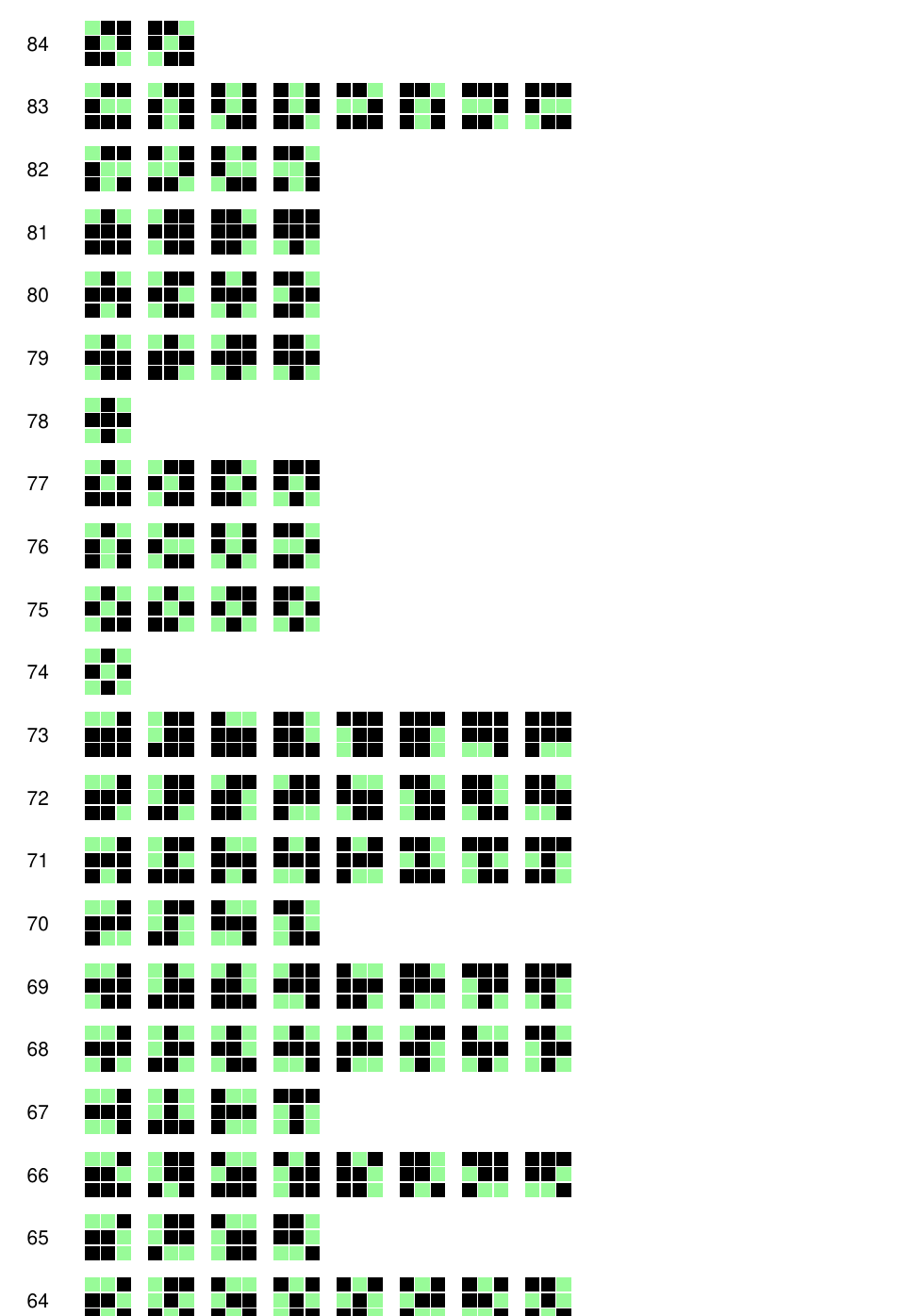}\\
(a) 3 successive $v2k9$ 2d square (Moore) iso-groups for iso-indeces as shown (max=101)\\[.5ex]
\end{minipage}
\begin{minipage}[c]{1\linewidth}
\includegraphics[width=.95\linewidth,bb=7 556 459 664, clip=]{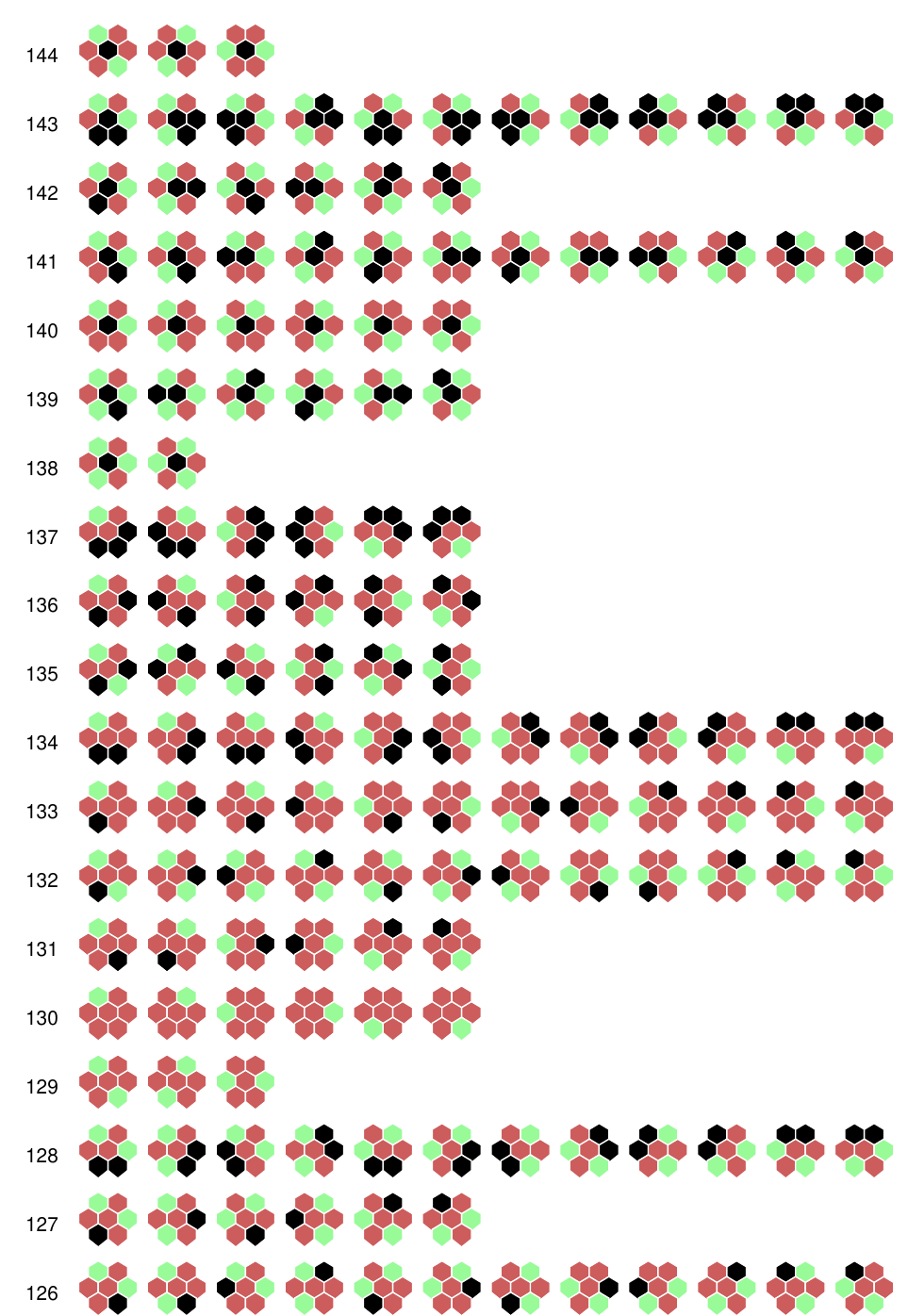}\\
(b) 3 successive $v3k7$ 2d hex iso-groups for iso-indeces as shown (max=275)\\[.5ex]
\end{minipage}
\begin{minipage}[c]{1\linewidth}
\includegraphics[width=1\linewidth,bb=7 488 463 650, clip=]{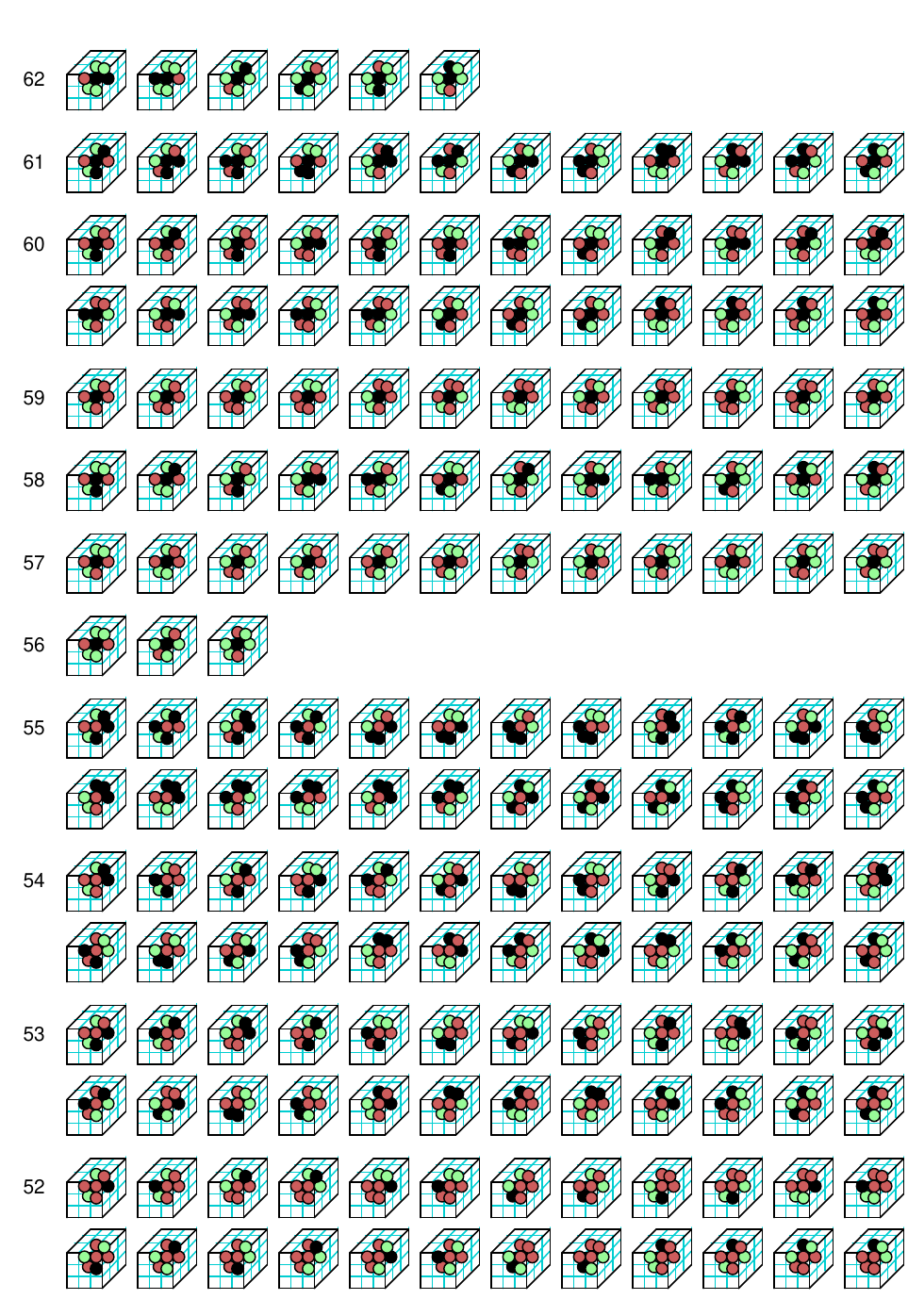}\\
(c) 3 successive $v3k7$ 3d iso-groups for iso-indeces as shown (max=171)\\[.5ex]
\end{minipage}
}\vspace{-4ex}
\caption[graphic iso-groups]
{\textsf{Examples of 3 successive iso-groups shown graphically; their sizes
depend on internal symmetries.
(a) $v2k9$ has 102 iso-groups, (b) and (c) are both $v2k7$ but 
(b) is 2d hex with 276 iso-groups, and (c) is 3d with 172 iso-groups. 
}}
\label{graphic iso-groups}
\end{figure}

\begin{figure}[htb]
\begin{center}
\textsf{\small
\begin{minipage}[c]{.95\linewidth}
\includegraphics[width=1\linewidth,bb=40 518 500 651, clip=]{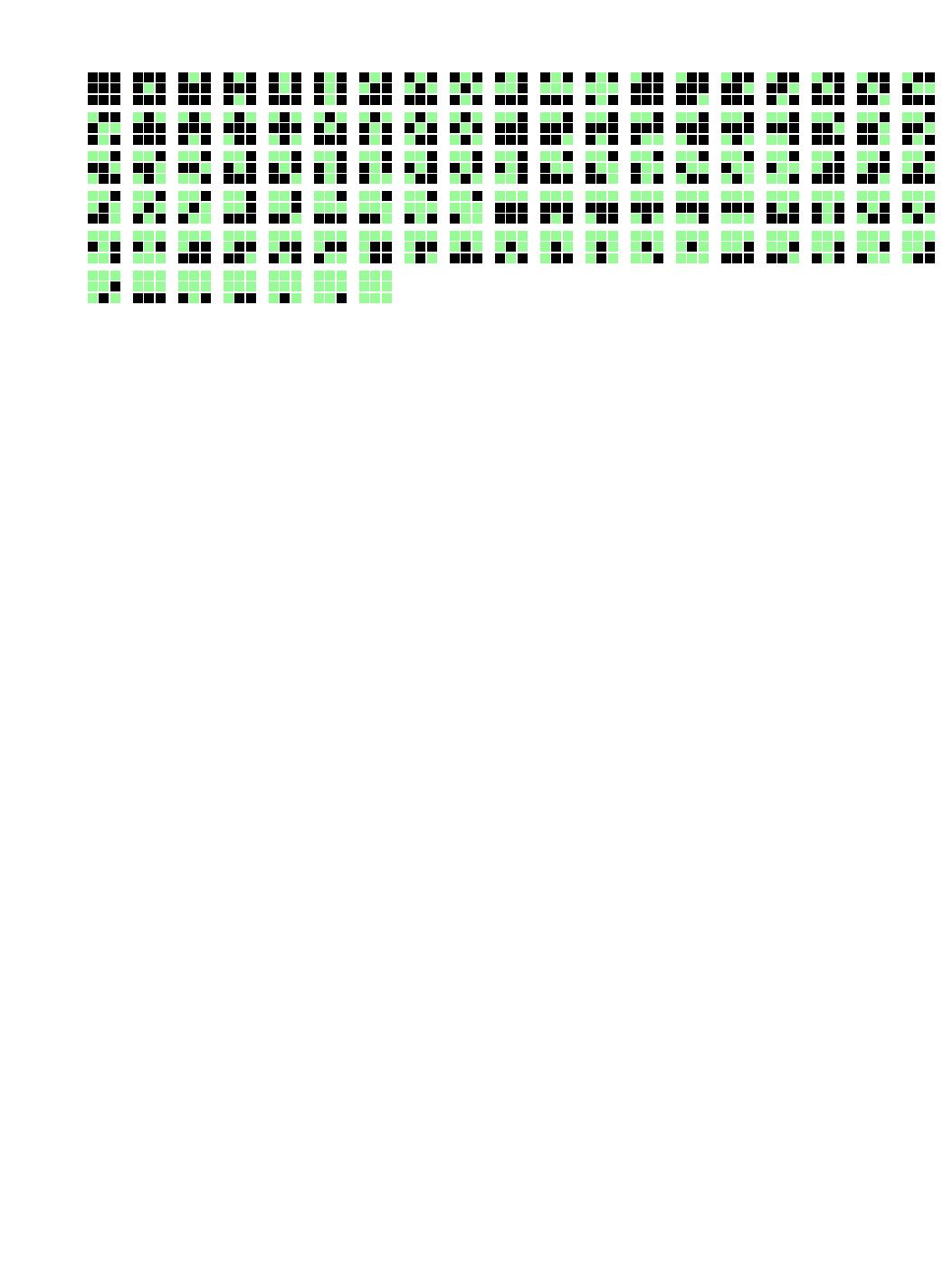}\\
(a) 102 $v2k9$ 2d square (Moore) neighborhood iso-group prototypes\\[.5ex]
\end{minipage}
\begin{minipage}[c]{.95\linewidth}
\includegraphics[width=1\linewidth,bb=4 377 452 636, clip=]{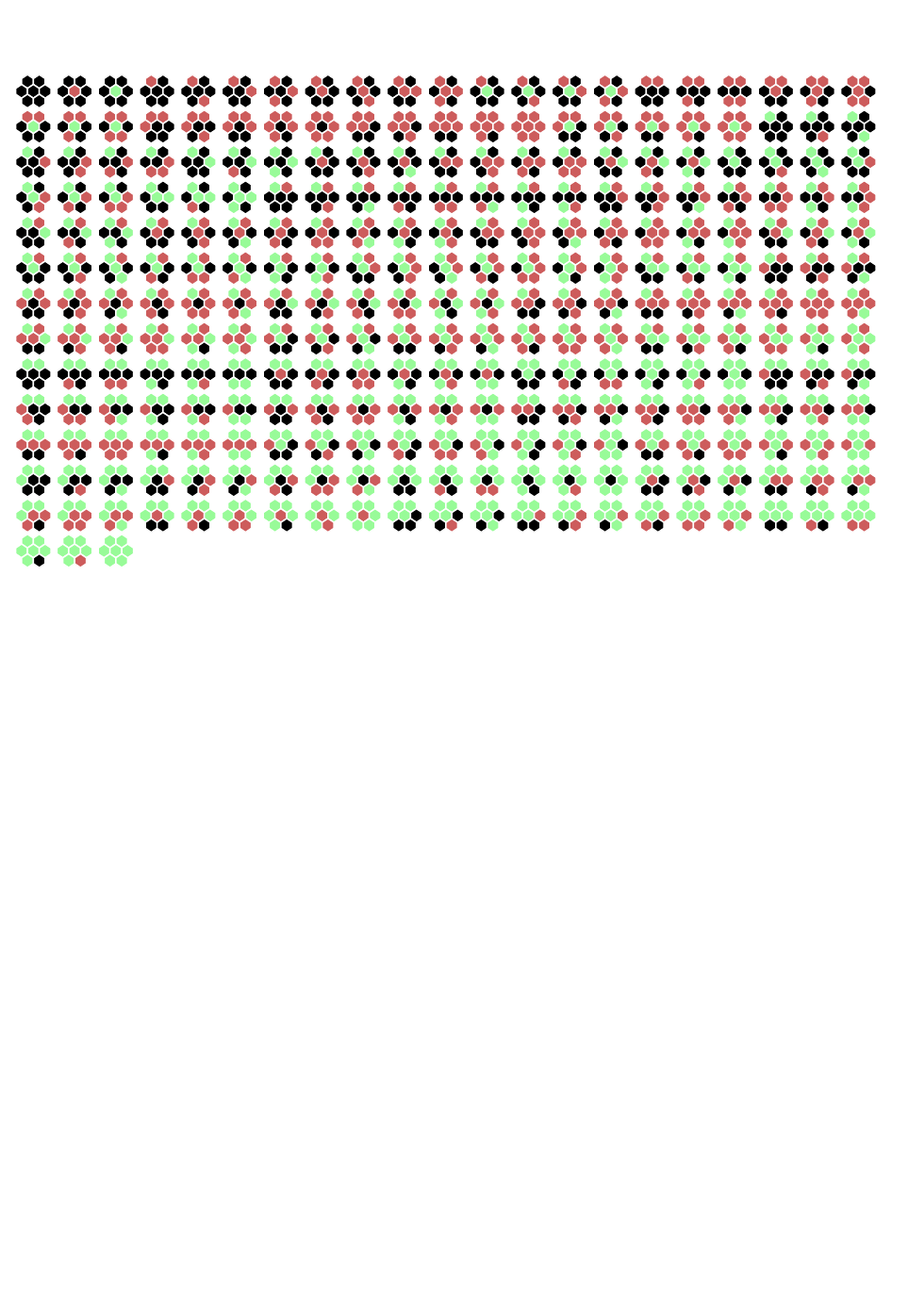}\\
(b) 276 $v3k7$ 2d hex neighborhood iso-group prototypes\\[.5ex]
\end{minipage}
\begin{minipage}[c]{.95\linewidth}
\includegraphics[width=1\linewidth,bb=0 440 460 660, clip=]{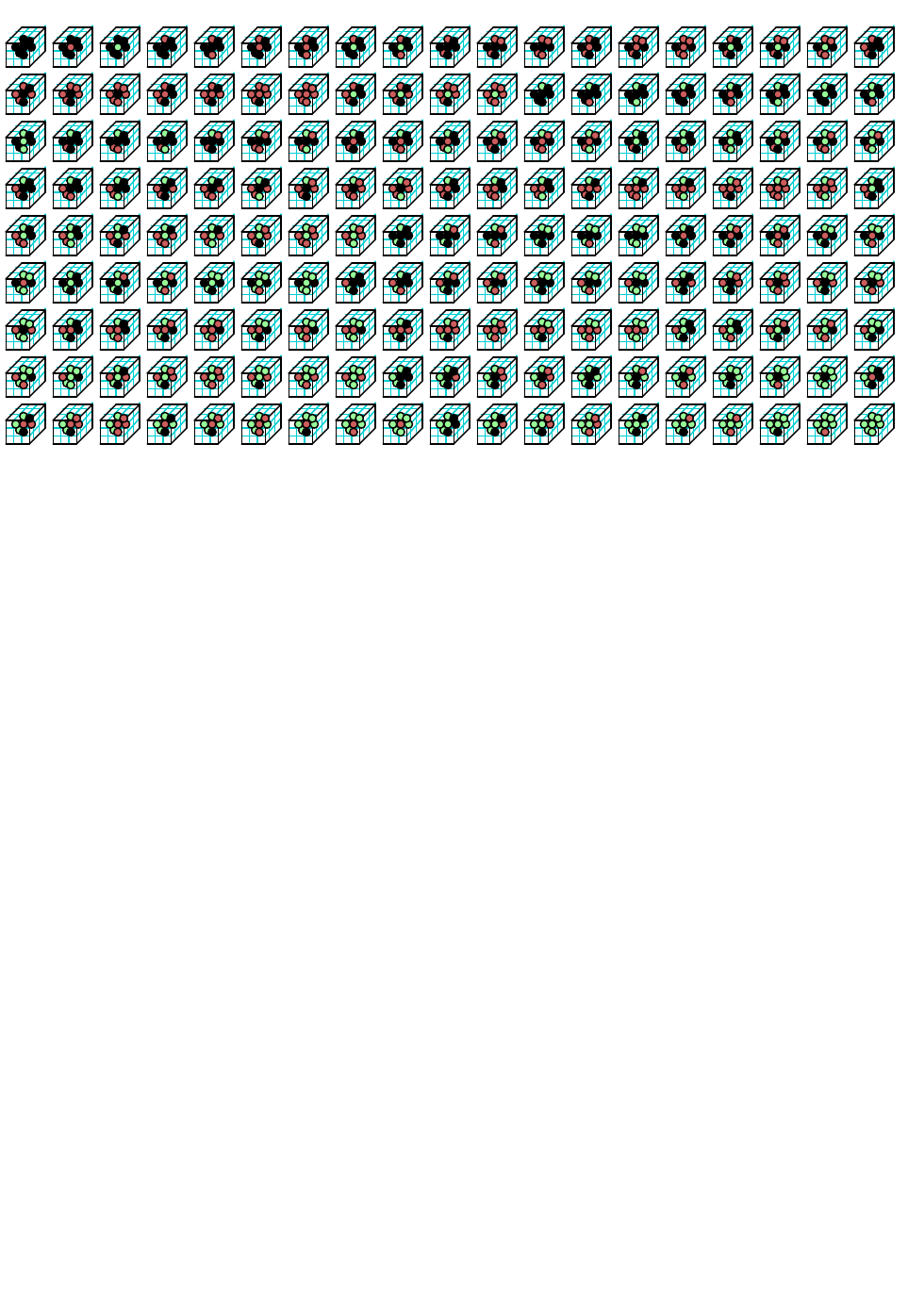}\\
(c) 172 $v3k7$ 3d neighborhood iso-group prototypes\\[.5ex]
\end{minipage}
}\end{center}\vspace{-4ex}
\caption[complete iso-rule neighbourhoods]
{\textsf{Examples of the complete set of iso-rule n-template prototypes
shown graphically in descending order of their decimal equivalents from the top-left.
(a) $v2k9$ has 102 prototypes, (b) and (c) are both $v2k7$ but 
(b) is 2d hex with 276 prototypes, and (c) is 3d with 172 prototypes. 
}}
\label{complete iso-rule neighbourhoods}
\end{figure}

An rcode is transformed to the iso-rule and its graphical string with
a keypress \cite[EDD:16.10.4]{EDD} and further options 
will show iso-groups graphically with accompanying details
(figure~\ref{graphic iso-groups}) or the the complete graphic
of the iso-rule prototype neighborhoods 
(figure~\ref{complete iso-rule neighbourhoods}) in a separate window.

\clearpage

\clearpage

\subsection{iso-rule advantages}
\label{iso-rule advantages}

The iso-rule is arguably an improvement on previous isotropic CA
notations, for example by Sapin\cite{Sapin2004,Sapin2010} and
Hensel\footnote{The Hensel notation, which only applies to a binary
  Moore neighborhood (v2k9), is incorporated in
  DDLab\cite[EDD:16.10.9]{EDD} so that files for rules (and initial
  states) can be interchanged\cite{DDLab-isotropic-rule-update} with
  ``Golly'' software\cite{Golly} used in the game-of-Life
  community\cite{ConwayLife-forum}.}, because the iso-rule is a simple
lookup-table in a conventional order and is general, applying to a
range of n-template sizes $k$ in 1d, square or hex 2d, and 3d, and
extending beyond binary to a range of values $v$.  The iso-rule can be
computed \underline{\it down} by reducing a full CA lookup-table or
\underline{\it up} by enhancing iso-subsets --- totalistic,
reaction-diffusion and survival/birth --- so provides an intermediate
granularity for mutation, bias, manipulation, or in a search by
genetic algorithm\cite{Sapin2004}; isotropy is conserved whatever
changes are made to an iso-rule.

In DDLab, the iso-rule provides the basis for input-frequency/entropy,
filtering and mutation in the same way as conventional full or
totalistic CA lookup-tables: rcode, tcode or kcode.  For
glider/eater/glider-gun iso-rules, the input-frequency histogram (IFH)
identifies both the critical and neutral iso-groups underlying
dynamics, and provides methods to explore rule-space that is
genetically close to significant iso-rules --- the IFH mutation/filter
game\cite[EDD:16.10.8]{EDD}.  The methods for automatically
classifying and examining rule-space based on input-entropy
variability\cite{Wuensche99,Wuensche05,Gomez2015} can be applied to
iso-rules. These ideas are developed further in the paper.

\subsection{iso-rule sizes}
\label{iso-rule sizes}

\enlargethispage{4ex}
The sizes of the iso-rule tables for the n-templates in 
figures~\ref{1d n-templates}, \ref{2d n-templates}
and \ref{3d n-templates}
depend on $v$, $k$, and the internal symmetries of the n-template so are difficult to calculate
analytically. The tables below give iso-group sizes computed algorithmically 
in DDLab.\\[-4ex]

\begin{table}[htb]
\fontsize{8pt}{7pt}
\begin{center}  
\begin{BVerbatim}
                         1d-k
        2   3    4    5    6     7    8    9   10 
        -----------------------------------------
   2 |  3   6   10   20   36    72  136  272  528
   3 |  6  18   45  135  378  1134 3321 9963
|  4 | 10  40  136  544 2080  8320
v  5 | 15  75  325 1625 7875
|  6 | 21 126  666 3996
   7 | 28 196 1255 8575
   8 | 36 288 2080    

             2d hex-k                      2d square-k                3d-k
         3    4    6    7              4    5    8    9              6    7
        -----------------             -----------------             -------
   2 |   4    8   13   26        2 |   6   12   51  102     |  2 |  10   20 
   3 |  10   30   92  276        3 |  21   63  954 2862     v  3 |  57  171
|  4 |  20   80  430 1720     |  4 |  55  220               |  4 | 240  960
v  5 |  35  175 1505          v  5 | 120  600                  5 | 800
|  6 |  56  336               |  6 | 231 1386                
   7 |  74  588                  7 | 406                    
   8 | 120  960                  8 | 666                     
\end{BVerbatim}
\end{center}
\normalsize
\vspace{-4ex}
\caption[iso-rule sizes]
{\textsf{Sizes of iso-rule lookup-tables (number of iso-groups) 
for 1d, 2d~hex, 2d~square and 3d n-templates in
figures~\ref{1d n-templates}, \ref{2d n-templates},
and \ref{3d n-templates},
for $vk$ within non-extended limits\cite[EDD:7.1.1]{EDD} in DDLab.
}}
\label{iso-rule size table}
\end{table} 

\section{iso-subsets expressed as iso-rules}
\label{iso-subsets expressed as iso-rules}

Significant CA iso-subsets, rule types that are
isotropic by default, include \mbox{k-totalistic} (kcode), t-totalistic
(tcode), outer totalistic, reaction/diffusion, and  survival/birth
rules. These iso-subsets have their specific concise definitions and
rule-spaces which allow interesting coarse-grained mutations. However,
by transforming the iso-subsets to equivalent iso-rules,
finer-grained mutations become possible in a search for
a wider range of significant rule families. Below, we define these
iso-subsets and their rule-spaces\footnote{DDLab has three modes, SEED, FIELD and 
TFO\cite[EDD:6.1]{EDD}.
TFO-mode (Totalistic Forwards-Only) has
advantages for these iso-subsets in the scope of $v$ and $k$, but
SEED-mode is required for automatic redefinition as iso-rules.}.

\subsection{k-totalistic rules (kcode)}
\label{k-totalistic rules (kcode}

\vspace{-3ex}
\begin{figure}[htb]
\textsf{\footnotesize
\begin{center}
\begin{minipage}[c]{.8\linewidth}
\begin{minipage}[c]{.43\linewidth}
\fbox{\includegraphics[width=1\linewidth,bb=35 14 421 373, clip=]{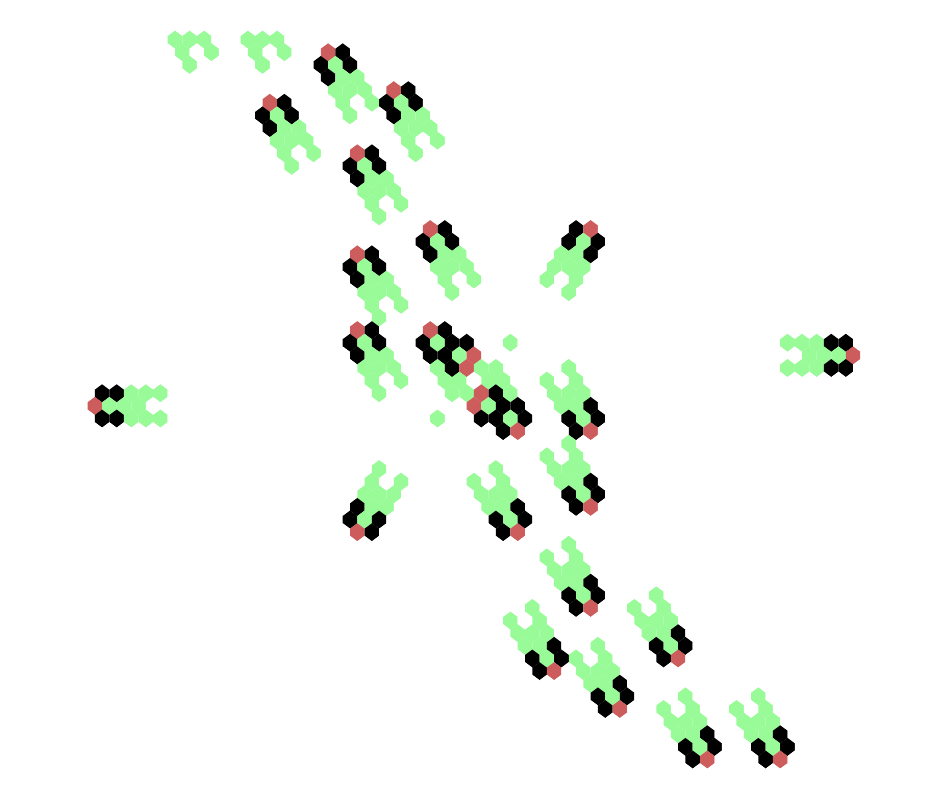}}\\[-4ex]
\begin{center}(a) hex 2d glider-gun \end{center}
\end{minipage}
\hfill
\begin{minipage}[c]{.43\linewidth}
\includegraphics[width=1\linewidth,bb=13 3 495 484, clip=]{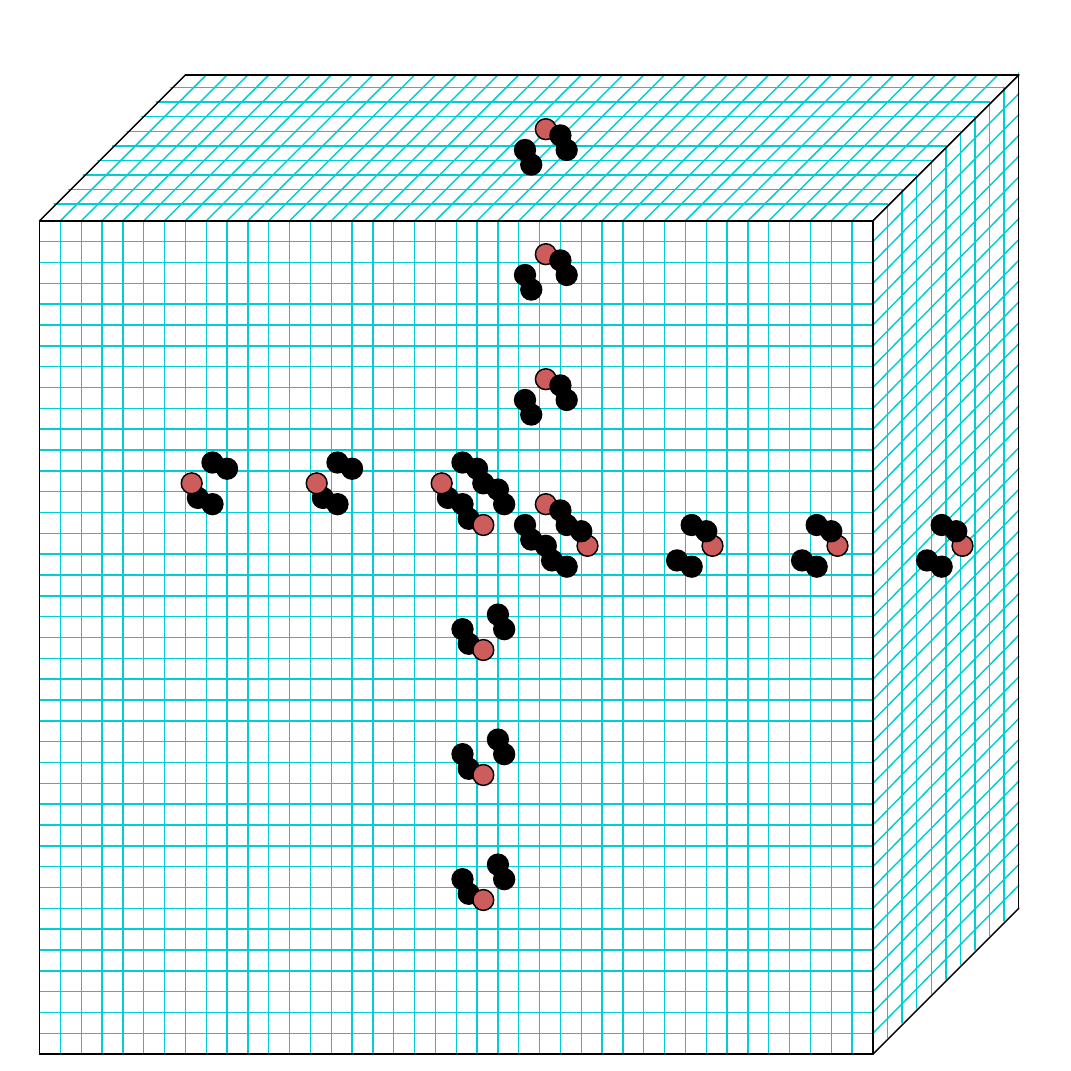}\\[-5ex]
\begin{center}(b) 3d glider-gun\end{center}
\end{minipage}
\end{minipage}
\end{center}
}
\vspace{-3ex}
\caption[beehive rule glider-gun]
{\textsf{Beehive rule glider-guns $v3k6$ for (a) hex 2d shooting gliders in 6 directions,
and (b) 3d 40$\times$10$\times$40 shooting gliders in 4 directions.
The 3d glider-gun emerges spontaneously, but localised structures 
(such as eaters) are absent.}}
\label{beehive rule glider-gun}
\vspace{-1ex}
\end{figure}

\noindent Kcode rules are defined by a list of the outputs for all possible
combinations of value-frequencies in the neighborhood. Each
combination is represented by a string of length $v$, shown
vertically from $v$-1 down in this example for the  
$v3k6$ Beehive rule\cite{Wuensche05,Adamatzky-beehive,beehive-webpage}) 
with frequencies of values 2 to 0, which must add up to $k$, so the last row of
frequencies is redundant and could be omitted.

\fontsize{8pt}{8pt}
\begin{center}  
\begin{BVerbatim}
               27..........................0 <--k=6 kcode index
                |                          |
           > 2: 6554443333222221111110000000  < frequency strings       6    0
v=3 values > 1: 0102103210432105432106543210  < of 2s, 1s, 0s,     from 0 to 0  
           > 0: 0010120123012340123450123456  < shown vertically        0    6
                ||||||||||||||||||||||||||||
                0022000220022001122200021210 <--kcode, outputs [0,1,2]
                                   beehive rule (hex) 0a0282816a0264
\end{BVerbatim}
\end{center}
\normalsize

In the spirit of Wolfram's convention\cite{wolfram83},
the ordering of the combinations depend on their
$v$-ary value, with the higher kcode index on the left.  Kcode rules are
independent of the n-template; they are isotropic because the positions of values
in are irrelevant.
 
In the example above for $v3k6$, outputs [0,1,2] are listed in reverse
order of the kcode index, and can be expressed in decimal (if
applicable) or in \mbox{hexadecimal}.  DDLab automatically transforms
kcode into its equivalent rcode, which can then be transformed to
isotropic rcode and the iso-rule according to the n-template. 
The 2d $v3k6$ Beehive rule on a hex lattice also works on
a 3d lattice, and both support a glider-gun 
(figure~\ref{beehive rule glider-gun}). 
The 2d ad 3d iso-rules are different, with lengths 92
an 57 respectively as shown below as a table and in hex,

\fontsize{6.5pt}{6.5pt}
\begin{center}  
\begin{BVerbatim}
(92) 00220200000102000222020022222222222122220222212102222221022000220222022222212000022212022010
(hex) 0a 20 01 20 2a 20 aa aa a9 aa 2a 99 2a a9 28 0a 2a 2a a9 80 2a 62 84

(57) 202200001120002222222221210202200022220222122120021022010
(hex) 02 28 01 60 2a aa a6 48 a0 2a 8a 9a 60 92 84 
\end{BVerbatim}
\end{center}
\normalsize

The size of a kcode-table, $S_k = (v + k - 1)! / (k! \times (v-1)!)$, 
and increases more slowly
than iso-rule size for larger $vk$ as in the table below,\\[-4ex]

\fontsize{8pt}{8pt}
\begin{center}  
\begin{BVerbatim}
                     -- k --
        2   3   4   5    6    7    8     9    10    
        ----------------------------------------   
   2 |  3   4   5   6    7    8    9    10    11    
   3 |  6  10  15  21   28   36   45    55    66 
|  4 | 10  20  35  56   84  120  165   220   286 
v  5 | 15  35  70 126  210  330  495   715  1001 
|  6 | 21  56 126 252  462  792 1287  2002  3003
   7 | 28  84 210 462  924 1716 3003  5005  8008
   8 | 36 120 330 792 1716 3432 6435 11440 19448 
\end{BVerbatim}
\end{center}
\normalsize                                                                                                                           

Kcode where $v$=3 can be expressed as an $ij$-matrix based on the
frequency of 2s and 1s (0s are  given by \mbox{$k-(i+j)$} so are not required).
Such rules can be reinterpreted
as conceptual discrete models of reaction-diffusion systems with inhibitor 
and activator reagents\cite{Adamatzky-beehive,Adamatzky&Wuensche2006}.
Figure~\ref{$v$3 $ij$-matrix} gives examples.

\begin{table}[htb]
\begin{center}
\begin{minipage}[b]{.9\linewidth}
\begin{minipage}[b]{.4\linewidth}
{\footnotesize
$
\begin{array}{ll|llllllll}
&       &       &   &   &   & j  &   &   &\cr
 &       & 0    & 1 & 2 & 3 & 4 & 5 & 6\cr \hline
 &  0    & 0&1&\color{blue}2&\color{red}1&\color{red}2&\color{red}0&\color{red}0 \cr
 &  1    & 0&2&\color{blue}2&2&\color{red}1&\color{red}1& \cr
 &  2    & 0&\color{red}0&\color{blue}2&2&\color{red}0& & \cr
i&  3    & 0&2&\color{blue}2&\color{red}0& & & \cr
 &  4    & \color{red}0&\color{blue}0&\color{blue}2& & & & \cr
 &  5    & \color{blue}2&\color{blue}0& & & & & \cr
 &  6    & 0& & & & & & \cr
\end{array}
$
}\\[1ex]
Beehive rule
\end{minipage}
\hfill
\begin{minipage}[b]{.4\linewidth}
{\footnotesize
$
\begin{array}{cc|cccccccc}
&       &       &   &   &   & j  &   &   &\cr
 &       & 0    & 1 & 2 & 3 & 4 & 5 & 6 & 7\cr \hline
&  0    & 0&1&2&\color{red}1&\color{blue}2&\color{blue}2&\color{red}2&\color{red}2 \cr
 &  1    & 0&2&2&\color{red}1&\color{red}2&\color{red}2&\color{red}2& \cr
 &  2    & 0&0&2&\color{blue}1&\color{red}2&\color{red}2& & \cr
i&  3    & 0&2&2&1&\color{blue}2& & & \cr
 &  4    & 0&0&2&1 & & & & \cr
 &  5    & 0&0&\color{blue}2 & & & & & \cr
 &  6    & 0&0 & & & & & & \cr
 &  7    & \color{blue}0& & & & & & & \cr
\end{array}
$
}
Spiral rule
\end{minipage}
\end{minipage}
\end{center}
\vspace{-3ex}
\caption[$v$3 $ij$-matrix]
{\textsf{The $ij$-matrices of (left) the
$v3k6$ Beehive rule~\cite{Wuensche05,Adamatzky-beehive,beehive-webpage} and (right) the
$v3k7$ Spiral rule~\cite{Wuensche&Adamatzky2006,Adamatzky&Wuensche2006,spiral-webpage}, 
on 2d hex lattices. The output can be read off from the frequency of 2s (rows $i$)
and 1s (columns $j$).
Previous studies (as cited above) have shown that a significant proportion of 
outputs are quasi-neutral --- wildcards --- with little impact on dynamics. 
These are shown in color: red for strong wildcards, blue for weak wildcards.
}}
\label{$v$3 $ij$-matrix}
\end{table} 

\normalsize
\subsection{t-totalistic rules (tcode)}
\label{t-totalistic rules (tcode)}

Tcode rules are defined by a list of outputs for each possible total,
the sum of values in the neighborhood, and are useful for setting
threshold functions. Tcode is a subset of kcode because each total can
include several kcode combinations of value-frequencies, so tcode is
also isotropic, but more directly because any n-template's
rotation/reflection must give the same total.  The size of a
tcode-table $S_t = k(v-1)+1$.  

To set tcode each total is set out in reverse value order, $S_t$-1 to 0,
and is assigned an output [0,1,$\dots$,$v$-1], which is the tcode value-string.
Here is an example for the $v5k5$ majority rule,
 
\fontsize{8pt}{8pt}
\begin{center}  
\begin{BVerbatim}
  20...................0 - all possible totals
   |                   |
   444433332222111100000 - tcode, outputs [0,1,2,3,4]
                           (hex) 49236da491248000
\end{BVerbatim}
\end{center}
\normalsize

Tcode may be expressed in decimal (if applicable) or in
hexadecimal.  DDLab automatically transforms 
tcode into its equivalent rcode, which can then be transformed to 
isotropic rcode and the iso-rule according to the n-template.  
For the majority rule (above) the rule tables for rcode (size 3225), 
and the iso-rule (size 600) for a 2d square n-template,
are shown graphically in figure~\ref{majority rules}(d).

For binary ($v$=2), tcode depends
on just the sum of 1s the neighborhood so tcode and kcode are identical, 
$S_t$=$S_k$, but for $v$$\geq$=3 $S_t$ becomes
progressively smaller than $S_k$.
Here is an example of binary tcode for the $v2k5$ majority rule,
its rcode, and iso-rule on a 2d square n-template.

\fontsize{8pt}{8pt}
\begin{center}  
\begin{BVerbatim}
   5 4 3 2 1 0 - all possible totals
   - - - - - -
   1 1 1 0 0 0 - output = tcode = kcode, 56 in dec, 38 in hex,
                 rcode = 11101000100000011000000100010110 
                 iso-table(12)=111110100000 (hex)0f a0
\end{BVerbatim}
\end{center}
\normalsize

\subsection{outer-totalistic kcode or tcode}
\label{outer-totalistic kcode or tcode}

Outer-totalistic CA require $v$ rules,
one for each possible value of the center cell, so the size of the total string
$S_o$ is $v$$\times$$S_k$ for kcode, or $v$$\times$$S_t$ for tcode.
The method in DDLab works with any $vk$, but makes most sense if the central cell
is empty in the n-template.

Binary ($v$=2) Life-like 2d CA can be defined by two $k$=8 tcode rules, 
with a total table size $S_o$=18, whereas the equivalent $S_i$=102 and $S_r$=512.
For the game-of-Life two $v2k8$ tcodes can be set 
``by hand''\cite[EDD:13.7]{EDD}  for the central cell of 0 and 1 as follows,

\begin{center}
\begin{minipage}[b]{.8\linewidth}
\begin{minipage}[b]{.8\linewidth}
 0: 000001000 - birth: exactly 3 live neighbors\\
 1: 000001100 - survival: 2 or 3 live neighbors
\end{minipage}
\hfill
\begin{minipage}[b]{.07\linewidth}
\raisebox{-.2ex}{\includegraphics[width=1\linewidth,bb=70 106 105 141, clip=]{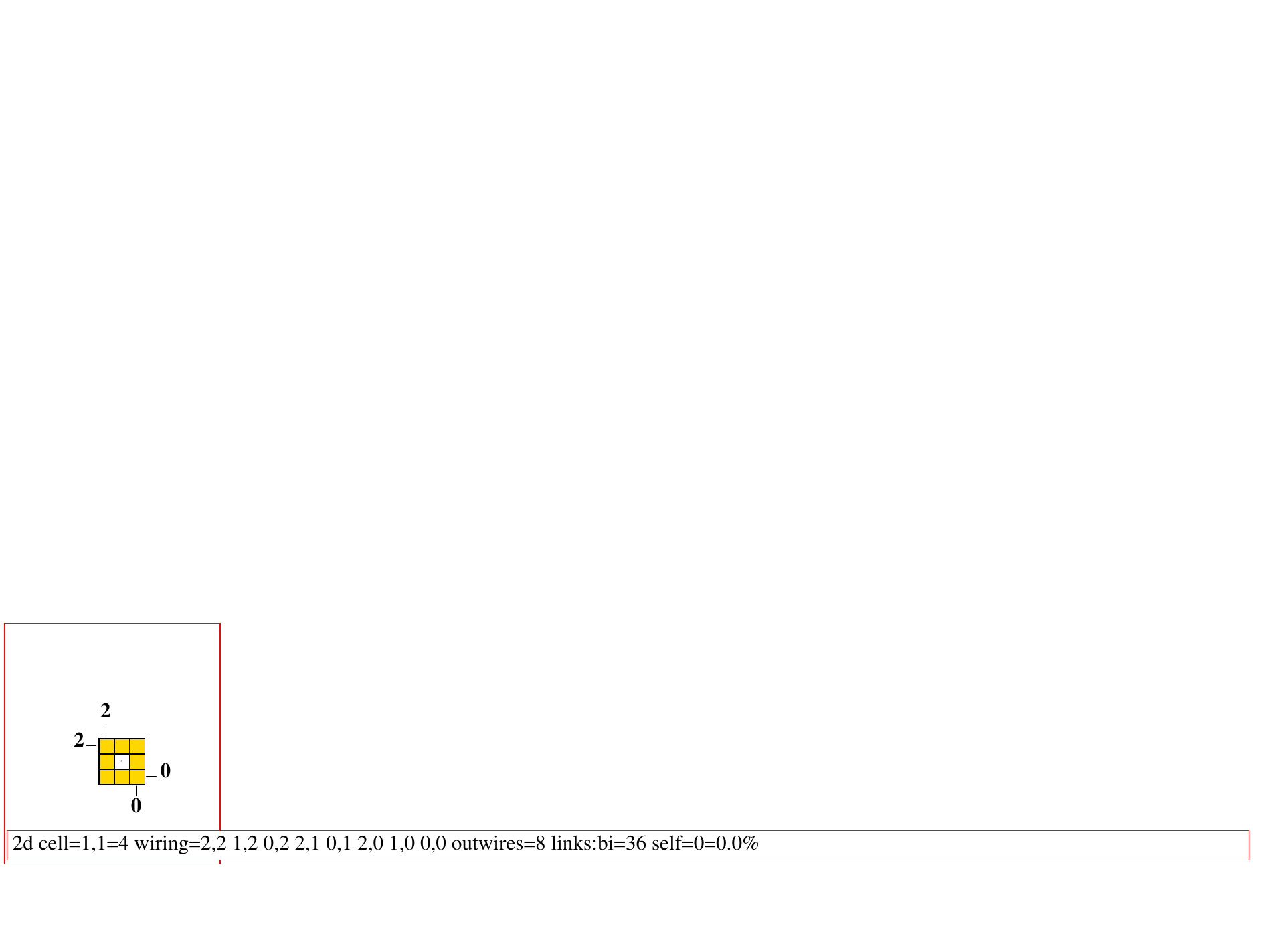}} 
\end{minipage}
\begin{minipage}[t]{.1\linewidth}
\raisebox{1.2ex}{$k$8 n-template} 
\end{minipage}
\end{minipage}
\end{center}

However, DDLab has automatic methods for setting the game-of-Life and
other survival/birth rules (section~\ref{survival/birth rules}) 
as both outer-kcode\cite[EDD:14.2.2]{EDD}
and as full rcode rule-tables\cite[EDD:16.10.7]{EDD} which limits the $vk$ range but 
allows reassigning as equivalent iso-rules.

\subsection{reaction-diffusion rules}
\label{reaction-diffusion rules}

\begin{figure}[htb]
\textsf{\footnotesize
\begin{center}
\begin{minipage}[c]{.9\linewidth}
\begin{minipage}[c]{.46\linewidth}
\includegraphics[width=1\linewidth]{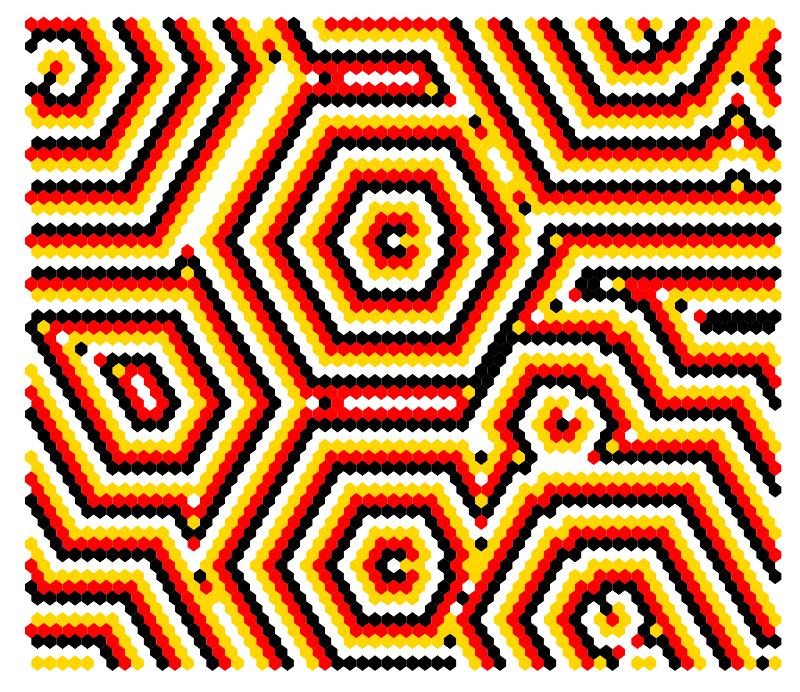}\\[-5ex]
\begin{center}(a) unfiltered \end{center}
\end{minipage}
\hfill
\begin{minipage}[c]{.46\linewidth}
\includegraphics[width=1\linewidth]{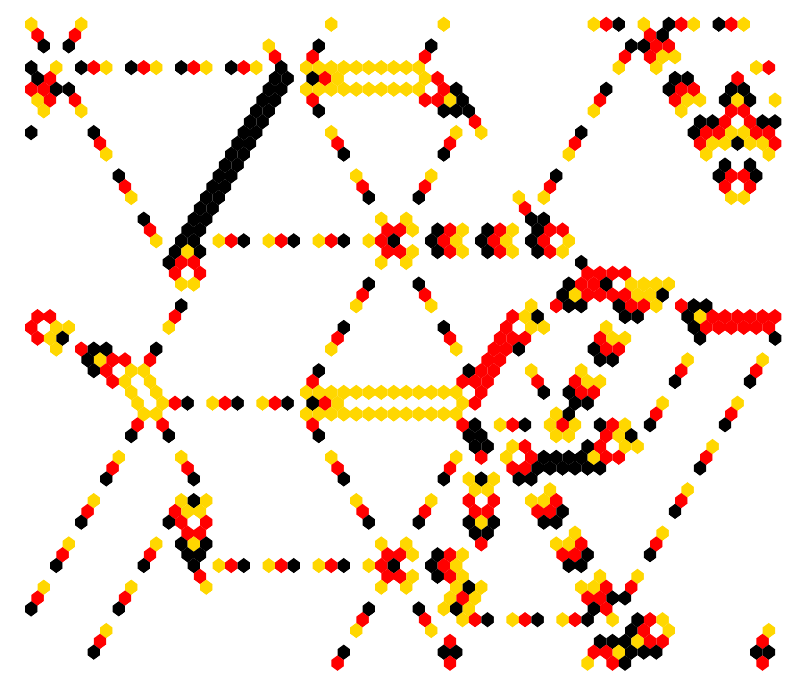}\\[-5ex]
\begin{center}(b) filtered\end{center}
\end{minipage}\\
\includegraphics[width=1\linewidth]{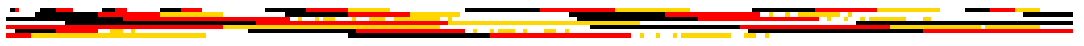}
\end{minipage}
\end{center}
}
\vspace{-3ex}
\caption[$v4k7$ reaction-diffusion CA]
{\textsf{Snapshots of a hex 2d reaction-diffusion CA $v4k7$,
with the iso-rule (size 1720) shown below.
The threshold interval was set 1 to 4.  
The initial state 60$\times$60 has a low density (0.01) of non-zero cells.
(a) the emergent pattern, and (b) the pattern with the 3 most 
frequent iso-groups filtered, showing structures
that resemble glider-guns.}}
\label{reaction-diffusion CA}
\vspace{-1ex}
\end{figure}

Reaction-diffusion or excitable media dynamics~\cite{greenberg}, can
be generated with a type of CA with 3 cell qualities: resting,
excited, and refractory (or substrate, activator, and inhibitor).
The rules are isotropic by default because they are
basically totalistic, depending on just totals of values in the
neighborhood.  There is usually one resting type, one excited type,
and one or more refractory types. In~DDLab these correspond to the
values $v$=0, $v$=1, and $v$$\geq$2, which cycle between each other.
A resting cell (0) remains as is until the number of excited cells in
its neighborhood falls within the threshold interval $t$, whereupon it
becomes excited (1). An excited cell (1) changes to the first
refractory value (2) at the next time-step, then to the next
refractory value (3) and so on, and the final ($v$-1) refractory value
changes back to resting (0), completing the following clockwise cycle,

\begin{center}  
\begin{BVerbatim}
 resting(0)--->if within t threshold interval
          \               \
           \              (1) excited
            \               \
            (v-1)<----(3)<--(2) refractory
\end{BVerbatim}
\end{center}
\normalsize

The variables required to define a reaction-diffusion rule
are $vk$ and the threshold interval within $k$. The number of refractory values
is $v$-2.  In DDLab, reaction-diffusion\cite[EDD:13.8]{EDD} can be set
as rcode which allows transformation to an iso-rule, or as outer-kcode
which allows a greater range of $vk$.

The resulting dynamics, in 2d or 3d, can produce waves, spirals and
related patterns that can resemble the Belousov-Zhabotinsky reaction
in a non-linear chemical medium and other types of excitable media.
Filtering the wave-like patterns by descending frequency of iso-groups
can reveal dynamics reminiscent of glider-guns (figure~\ref{reaction-diffusion CA}).
The filtering method\cite[32.11.5]{EDD} is based on the input-frequency
histogram described in later sections.

As well as the threshold interval, the dynamics are sensitive to the
initial state and its density of non-resting types (non-zero
values)\cite[21.3]{EDD} --- usually low for best spiral-wave results.

\subsection{survival/birth rules}
\label{survival/birth rules}

\begin{figure}[htb]
\begin{center}
  \includegraphics[width=.8\linewidth,bb=8 12 554 87, clip=]{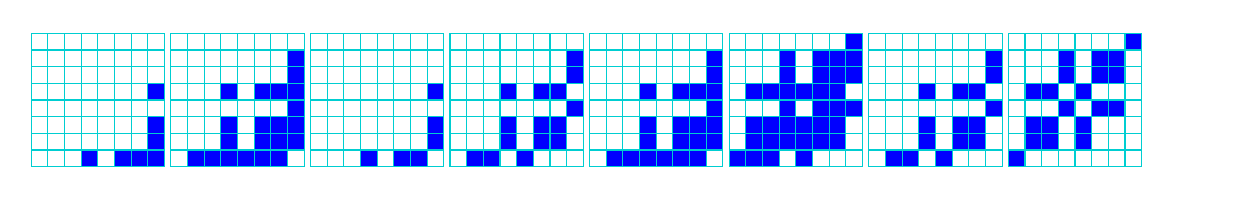}\\
  \includegraphics[width=.8\linewidth,bb=6 12 478 42, clip=]{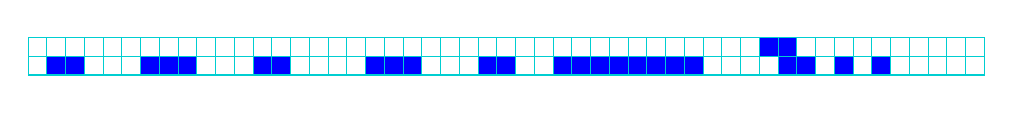}
\end{center}
\vspace{-4ex}
\caption[The game-of-Life rule]   
        {\textsf{Conway's game-of-Life\index{Conway, John} \mbox{(s23/b3)} shown as a 
            512 bit rcode in 8 rows.
            The diagonal symmetry in each 8x8 block is a necessary (but insufficient)
            indication of isotropy
            but a useful visual clue for the general case of isotropic rcode for
            a binary $v2k9$ 2d CA with a Moore neighborhood. Below the rcode is 
            the 102 bit iso-rule ---
            (hex) 00 00 00 00 00 60 03 1c 61 c6 7f 86 a0.
        }}           
       \label{Game-of-Life rule}
\end{figure}

\begin{figure}[htb]
\begin{center}
\includegraphics[width=.45\linewidth,bb=55 34 299 207, clip=]{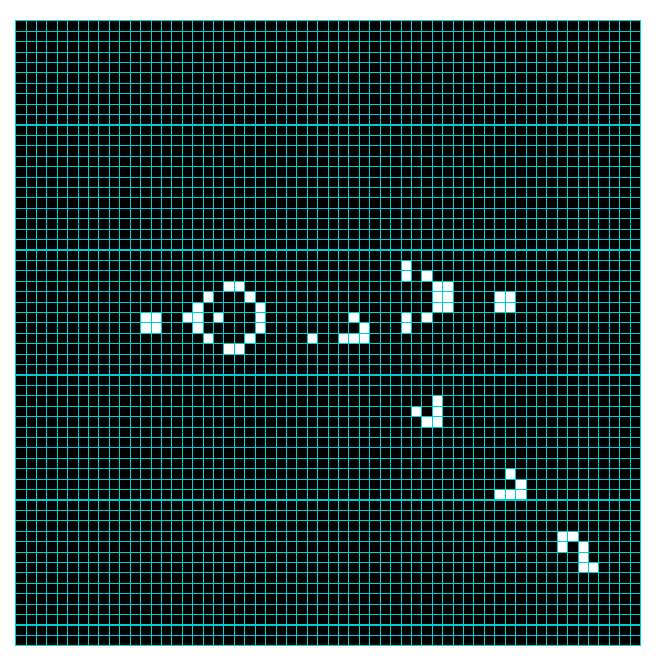}
\end{center}
\vspace{-3ex}
\caption[negative GoL]
{\textsf{The negative game-of-Life with a negative Gosper glider-gun.
    The iso-rule is \raisebox{-.7ex}
    {\includegraphics[width=.6\linewidth,bb=4 7 265 23,clip=]{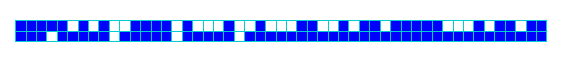}}\\
(hex) 3e a7 a2 46 5b e2 df 7d f7 df ff ff ff.
A negative\cite[EDD:18.5.2]{EDD} iso-rule gives equivalent
dynamics given a complementary initial state, which would reduce the effective
size of iso-rule-space.}}
\label{negative GoL}
\end{figure}

A survival/birth rule, including the game-of-Life, can be set in 
DDLab\cite[EDD:16.10.7]{EDD}. 
The rule is turned into
rcode automatically and can then be transformed into an 
iso-rule (figure \ref{Game-of-Life rule}). The rcode of the iso-rule
can be transformed for a negative universe\footnote{A negative universe also applies
for $v$$>$=3 where black values are exchanged for white.} as in figure~\ref{negative GoL}
by complementing both neighbourhoods and outputs\cite[18.5.2]{EDD}. 

\enlargethispage{2ex}
\clearpage

\begin{figure}[htb]
\begin{minipage}[b]{1\linewidth}
\fbox{\includegraphics[width=.22\linewidth,bb=33 36 378 381, clip=]{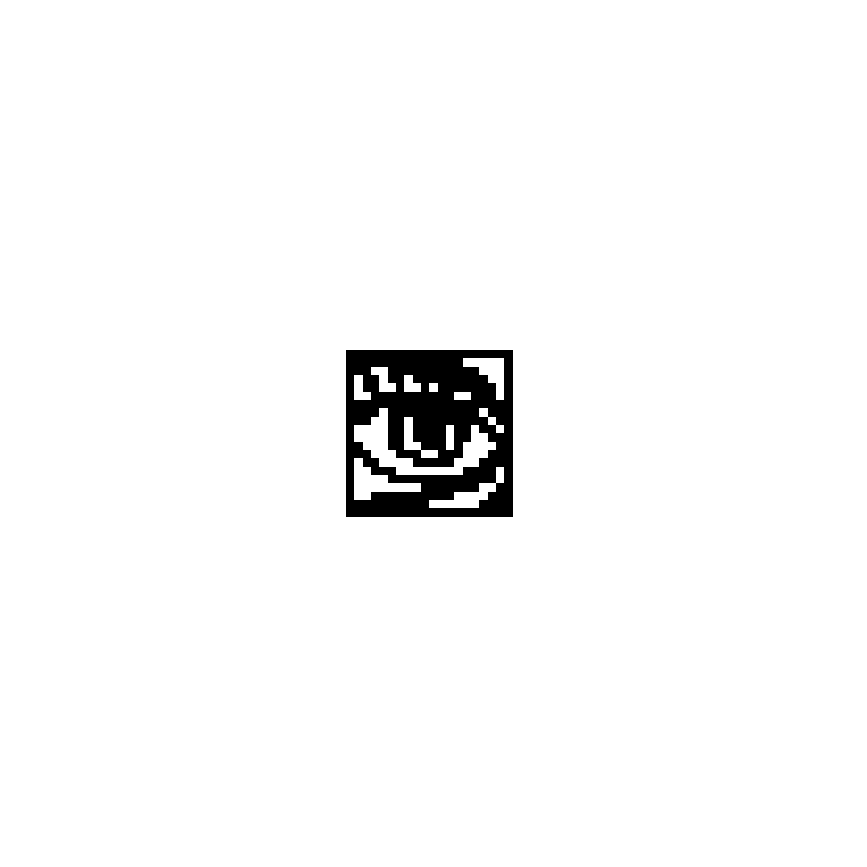}}
\hfill
\fbox{\includegraphics[width=.22\linewidth,bb=33 36 378 381, clip=]{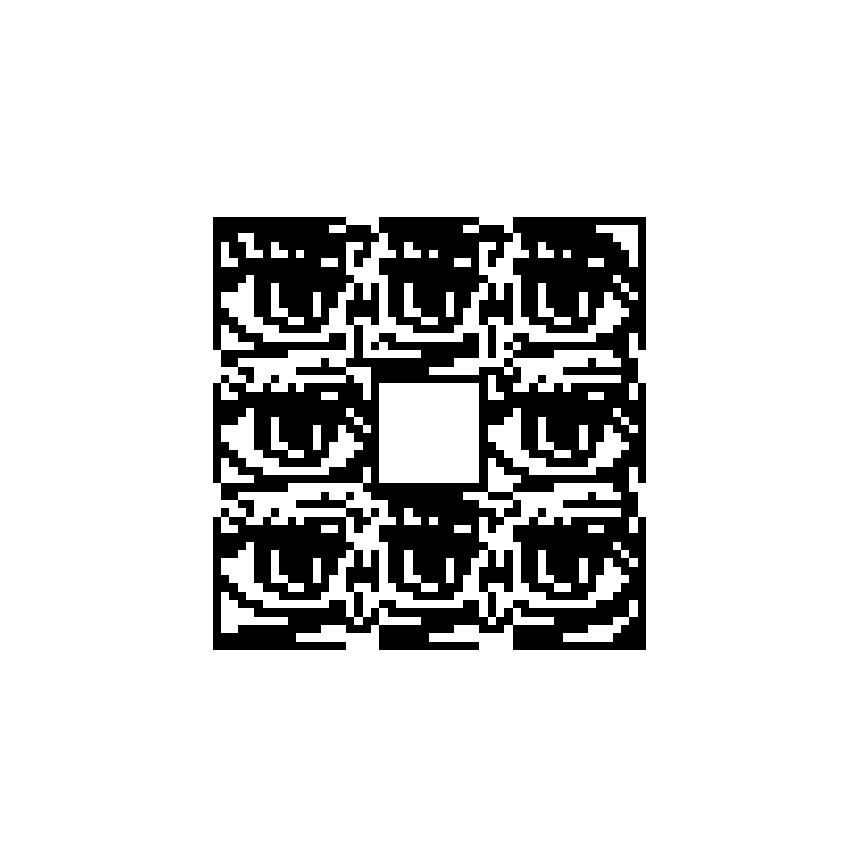}}
\hfill
\fbox{\includegraphics[width=.22\linewidth,bb=33 36 378 381, clip=]{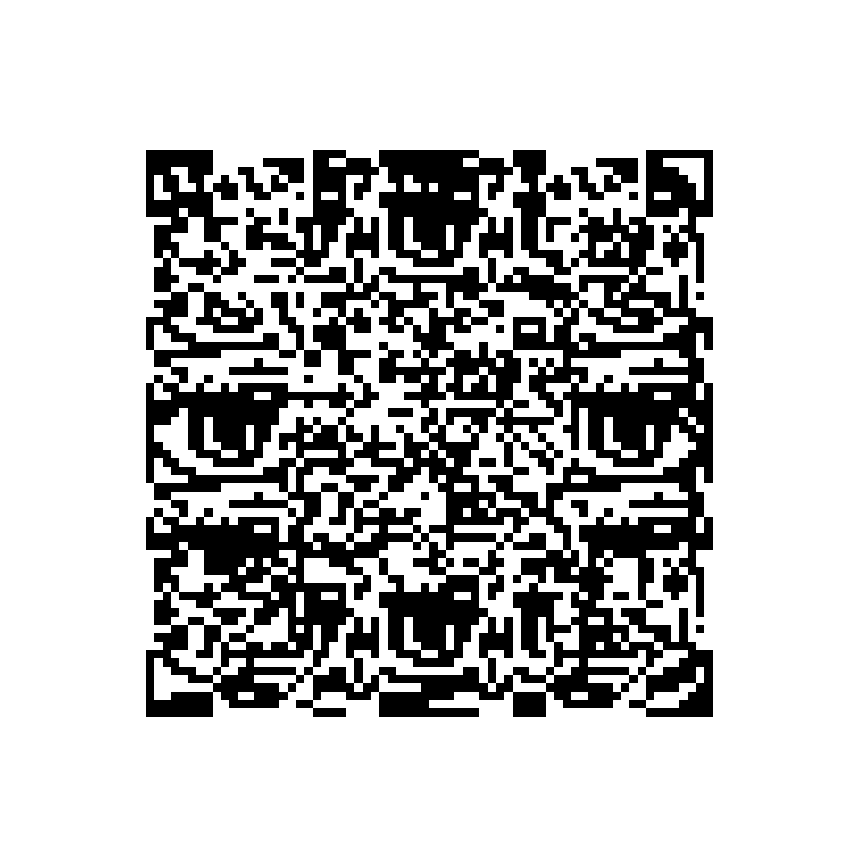}}
\hfill
\fbox{\includegraphics[width=.22\linewidth,bb=33 36 378 381, clip=]{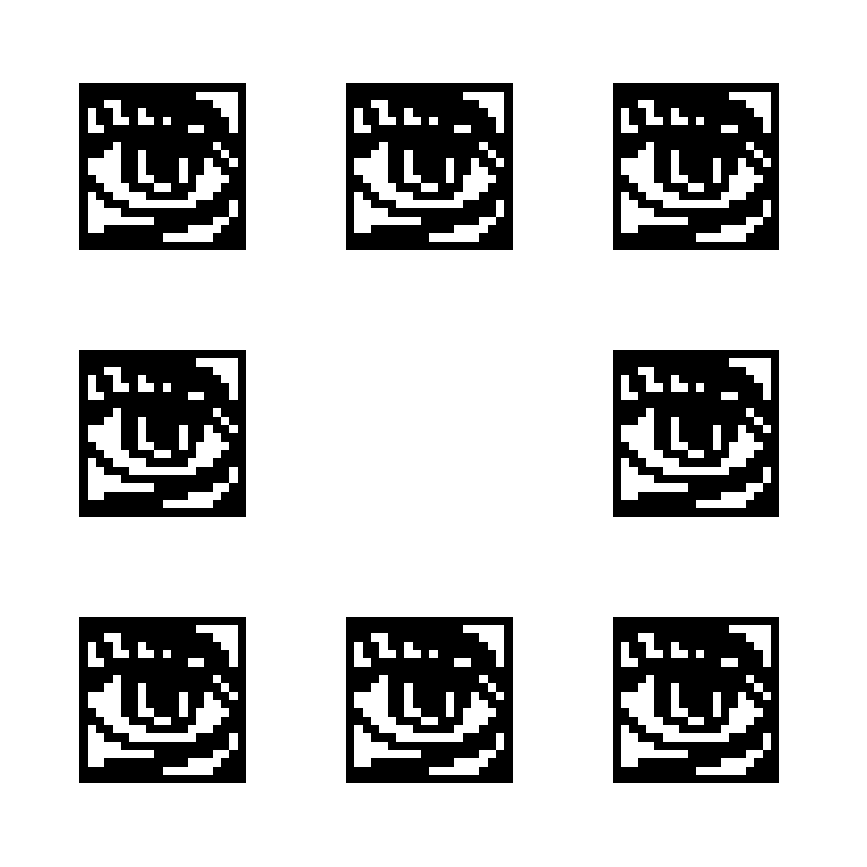}}
\end{minipage}
\vspace{-3ex}
\caption[Fredkin's replicator]
       {\textsf{Fredkin's replicator. An initial pattern (the eye)
         re-emerges from apparent
        disorder as partial or complete multiple copies (time-steps 1, 17, 25, 33).}}
       \label{Fredkin's replicator}
\end{figure}

DDLab defines the classical game-of-Life as (s23/b3) following a
common shorthand for survival/birth\cite{Mirek}, though this is often
reversed to birth/survival (b3/s23)\cite{ConwayLife-forum}.  A cell is
either alive~(1) or dead~(0).  The first part of (s23/b3) defines
the survival of a cell requiring 2 or 3 live neighbors, the second
defines birth, requiring 3 live neighbors, otherwise the cell is dead
by overcrowding or exposure.  Any other survival/birth settings can
be selected in this notation, for example (s1357/b1357) for Fredkin's
replicator (figure~\ref{Fredkin's replicator}).

\vspace{2ex}

\begin{figure}[htb]
\begin{center}
\fbox{\includegraphics[width=.6\linewidth,bb=50 135 335 290, clip=]{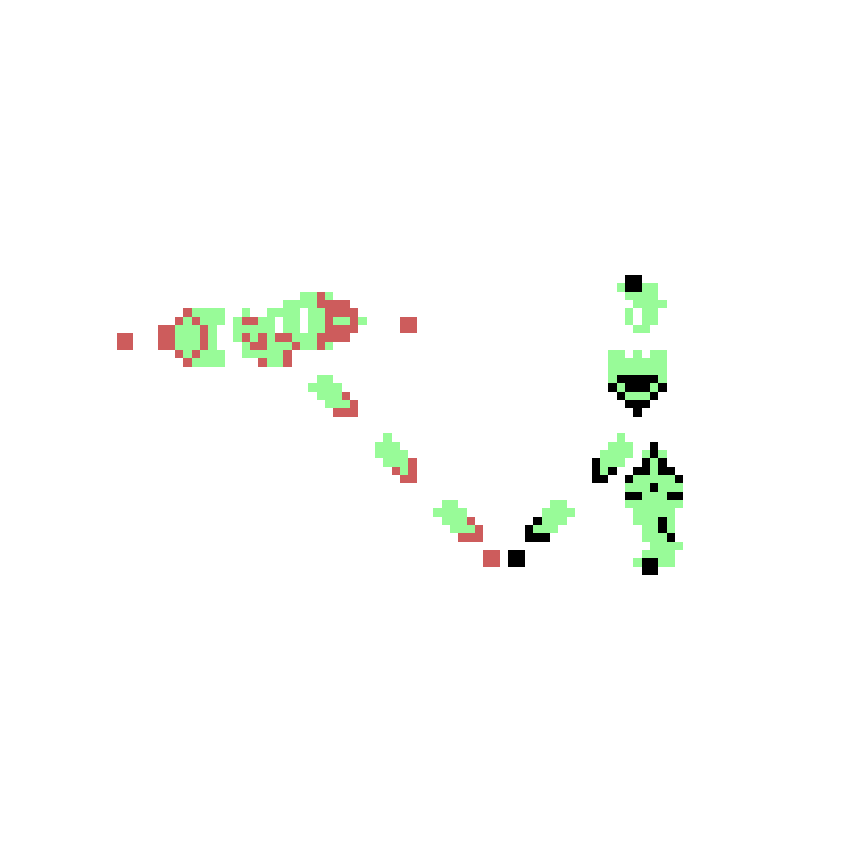}}
\end{center}
\vspace{-3ex}
\caption[Glider guns in $v$=3 Life]
{\textsf{The game-of-Life (s23/b3) applied to a $v$=3 CA.  The
    algorithm in DDLab generates an equivalent rcode and iso-rule
    (size=2862).  The dynamics is the same as binary Life but with 2
    colors + background. This example shows two different color Gosper glider-guns,
    and added green dynamic trails.    
}}
\label{Lguns_v3} 
\end{figure}

The survival/birth option is available for $v$$\geq$2, and any $k$$\geq$5
as well as the $k$=9 neighborhood, for 1d and 3d as well as 2d.  For
$v$$\geq$3 any value $v$$>$0 is considered alive and the algorithm in
DDLab generates an equivalent rcode/iso-code giving dynamics similar
to binary, but including $v$ colors as in figure~\ref{Lguns_v3}.

\section{rule-table size summary}
\label{rule-table size summary}

The sizes of rule-tables, or the amount of information required, $S$,
to define different rule or logic types, with a rule-space of $v^S$
are summarised below,

\begin{list}{$\Box$}{\parsep 0ex \itemsep 0ex 
 \leftmargin 30ex \rightmargin 0ex \labelwidth 30ex \labelsep 1ex} 
    \item[rcode $\dots$] $S_r=v^k$ 
    \item[iso-rule $\dots$] $S_i$: according to $vk$ and n-template as in table~\ref{iso-rule size table}
    \item[kcode $\dots$] $S_k = (v + k - 1)! / (k! \times (v-1)!)$
    \item[tcode $\dots$] $S_t = k(v-1)+1$
    \item[k-outer-totalistic $\dots$] $S_{ok}$ = $v$$\times$$S_k$
    \item[t-outer-totalistic $\dots$] $S_{ot}$ = $v$$\times$$S_t$
    \item[reaction-diffusion $\dots$] $S_{RD}$ $<$ $S_{ot}$, ($v$, $k$ and threshold interval)
    \item[survival/birth $\dots$] $S_{SB}$ $<$ $S_{ot}$, (survival and birth totals)
\end{list}

In general, $S_r$$>$$S_i$$>$$S_k$$>$$S_t$. 
$S_{RD}$ $<$ $S_{ot}$ and $S_{SB}$ $<$ $S_{ot}$ because both reaction-diffusion and survival/birth
logic can be set within t-outer-totalistic rules.

\section{input-frequency histogram (IFH)}
\label{input-frequency histogram}

\enlargethispage{3ex}
\begin{figure}[htb]
\textsf{\small
\begin{minipage}[b]{1\linewidth}
\includegraphics[width=.7\linewidth]{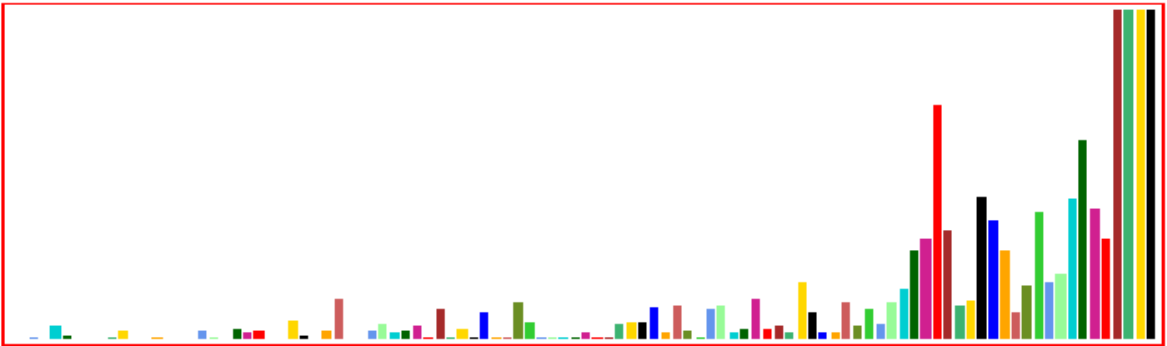}
\phantom{i}\raisebox{1.4ex}{\fbox{\includegraphics[height=.18\linewidth, bb=109 20 282 258, clip=]{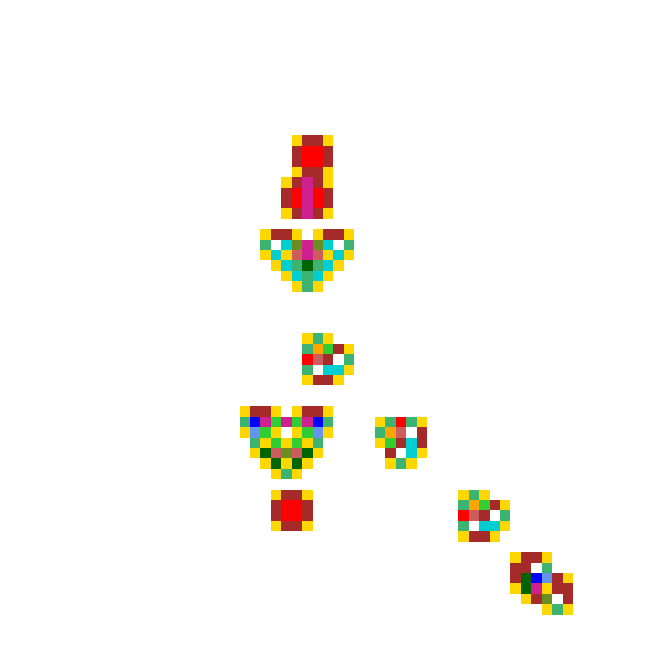}}}
\raisebox{7ex}{Life}
\\[1ex]
\includegraphics[width=.7\linewidth]{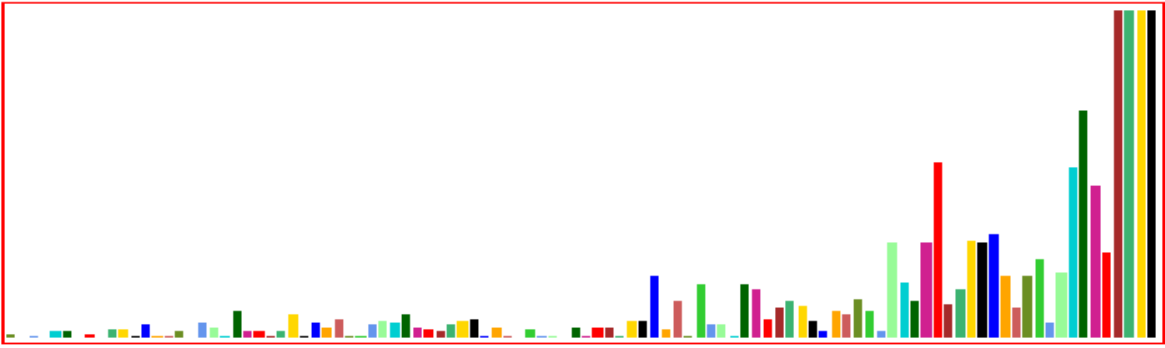}
\raisebox{1.4ex}{\fbox{\includegraphics[height=.18\linewidth, bb=250 55 351 166, clip=]{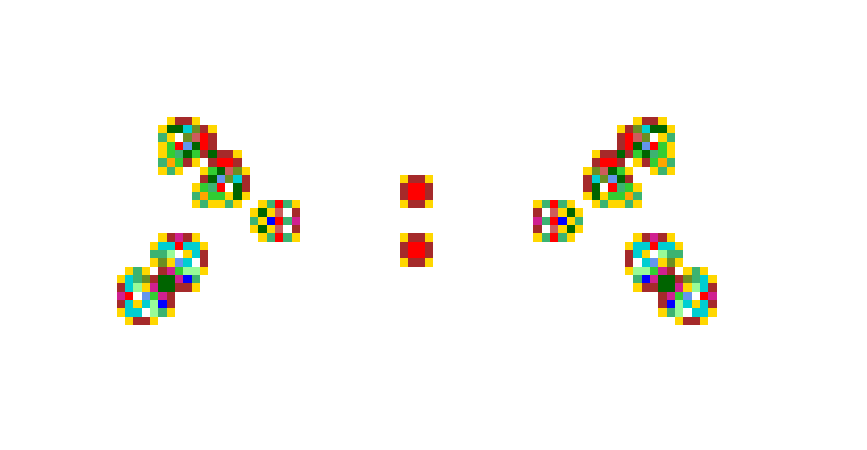}}}
\raisebox{7ex}{Eppstein}
\\[-.3ex]
\includegraphics[width=.7\linewidth]{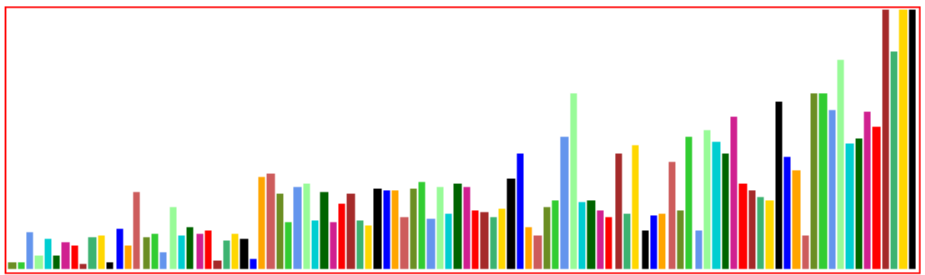}
  \fbox{
    \includegraphics[height=.18\linewidth, bb=25 12 229 253, clip=]{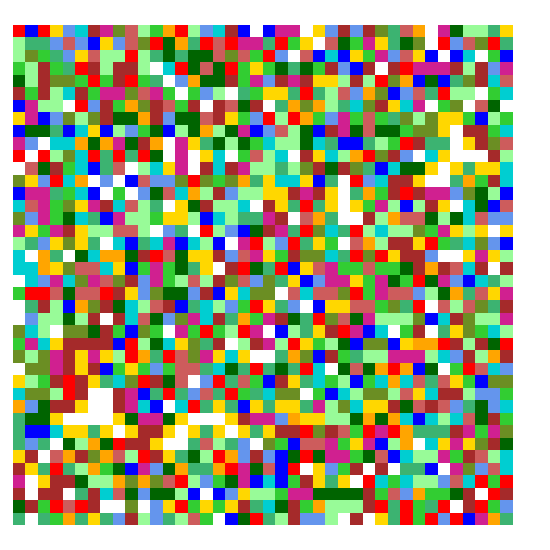}}
\begin{minipage}[t]{.1\linewidth}
\vspace{-14ex}
Eppstein\\random\\initial\\state
\end{minipage}
\\[1ex]
\includegraphics[width=.7\linewidth]{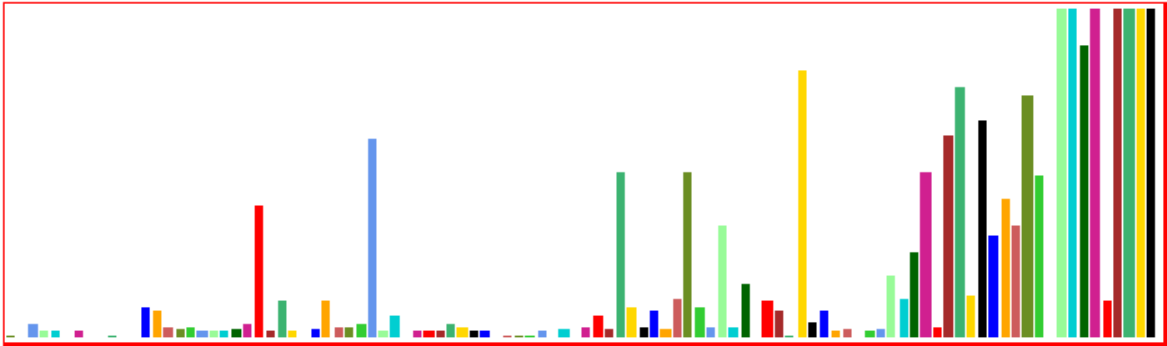}
\raisebox{1.4ex}{\fbox{\includegraphics[height=.18\linewidth, bb=5 22 206 234, clip=]{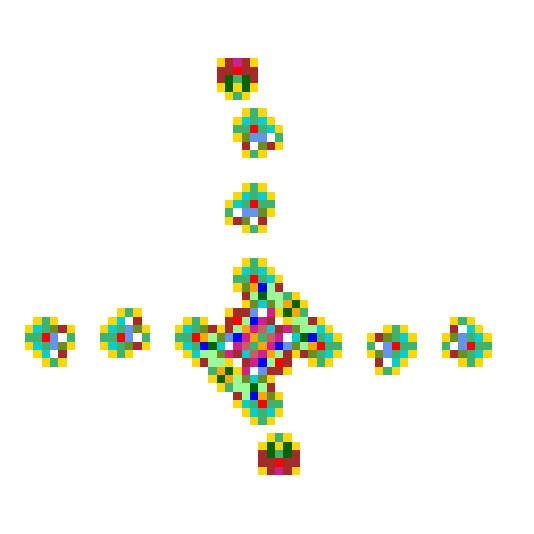}}}
\raisebox{7ex}{SapinR}
\end{minipage}
}
\vspace{-4ex}
\caption[input-frequency histograms for Life, Eppstein, SapinR]
{\textsf{The input-frequency histograms (actual plots) in a moving window
of 100 time-steps, for the glider-guns
of the following rules: Life, Eppstein (shown also from a random initial state), and SapinR,
all from figure~\ref{significant binary iso-rules}(a).
Pattern snapshots are colored to match histogram colors.
Note that apart from its glider-gun, the Eppstein rule is chaotic.
}}
\label{histograms for Life, Eppstein, SapinR}
\end{figure}
\clearpage

\begin{figure}[htb]
\textsf{\small
\begin{minipage}[b]{1\linewidth}
\includegraphics[width=.7\linewidth]{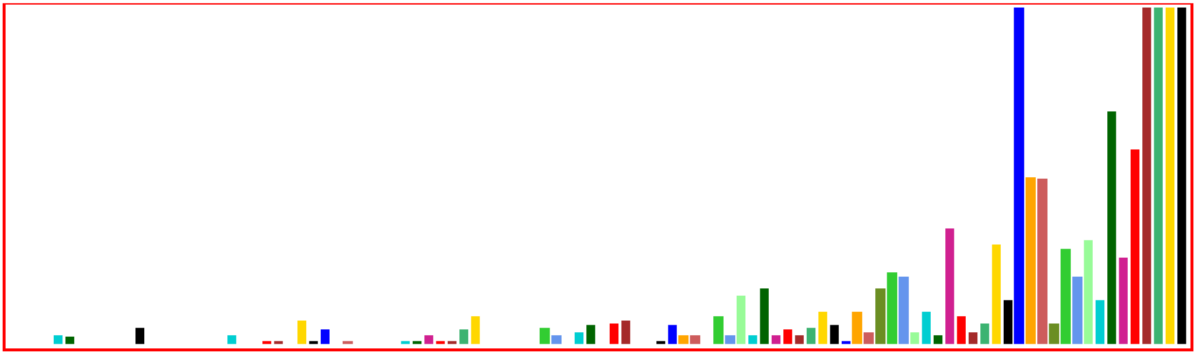} 
\raisebox{1.5ex}{\fbox{\includegraphics[height=.18\linewidth, bb=46 43  198 226, clip=]{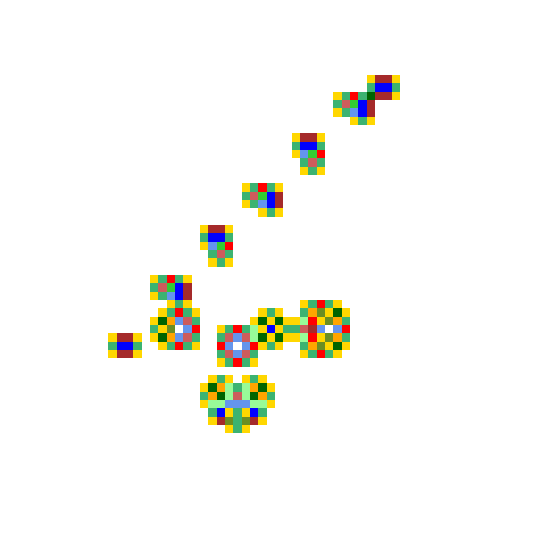}}}
\raisebox{7ex}{Variant}
\\[1ex]
\includegraphics[width=.7\linewidth]{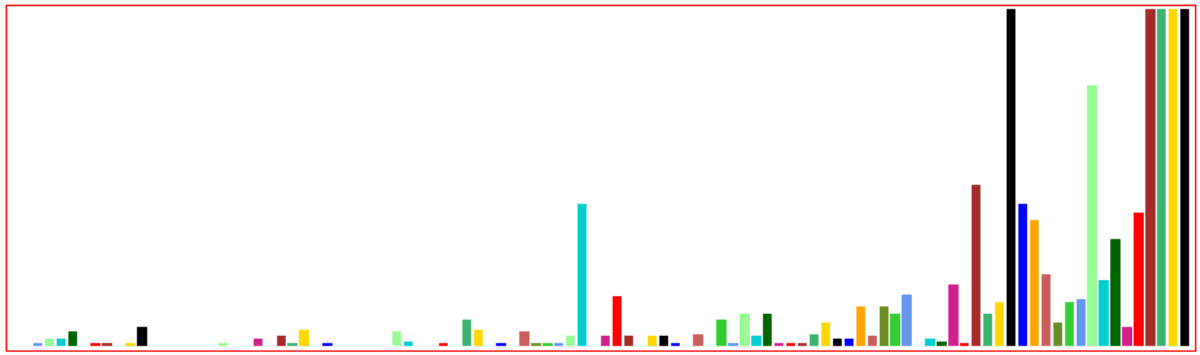} 
\raisebox{1.5ex}{\fbox{\includegraphics[height=.18\linewidth, bb=84 31 260 255, clip=]{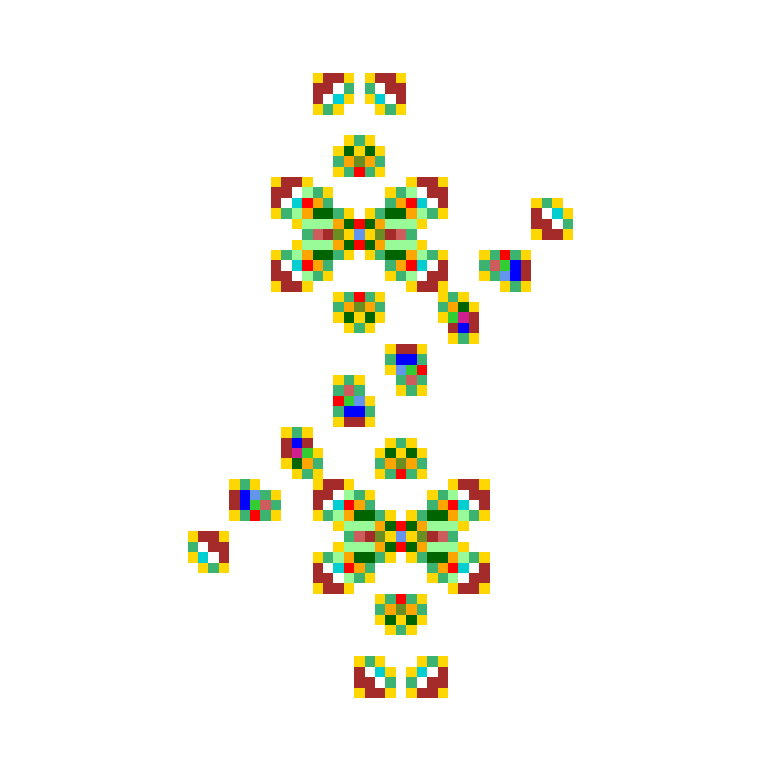}}}
\raisebox{7ex}{Precursor}
\\[1ex]
\includegraphics[width=.7\linewidth]{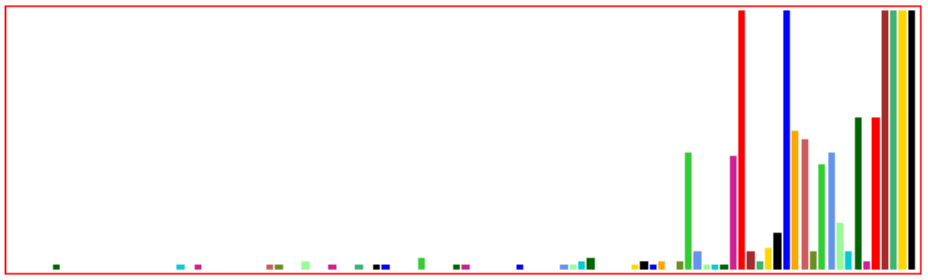} 
\raisebox{1.5ex}{\fbox{\includegraphics[height=.18\linewidth,  bb=45 25 177 190, clip=]{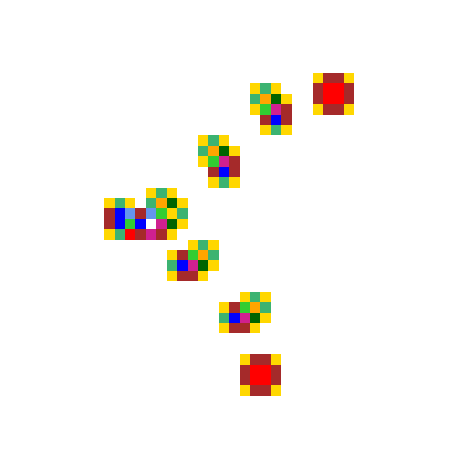}}}
\raisebox{7ex}{Sayab}\\
\includegraphics[width=.7\linewidth]{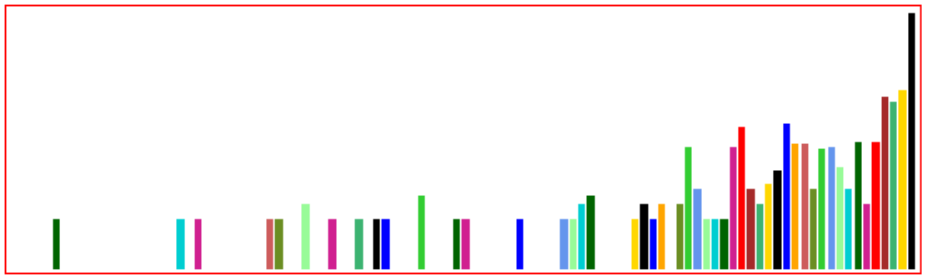}
\phantom{i}
\raisebox{11ex}{\begin{minipage}[t]{.25\linewidth}
{\it left}: the same Sayab histogram as above but on the alternative log$_2$ plot
\end{minipage}}
\hfill
\end{minipage}
}
\vspace{-4ex}
\caption[input-frequency histograms for Variant, Precursor, Sayab]
{\textsf{The IFH (actual plots) in a moving window
of 100 time-steps, for the glider-guns of the following rules:
Variant, Precursor, and Sayab with an alternative log$_2$ plot,
all from figure~\ref{significant binary iso-rules}(b).
Pattern snapshots are colored to match IFH colors.
}}  \label{histograms for Variant, Precursor, Sayab}
\end{figure}

\noindent While iterating CA space-time patterns, DDLab is able to
keep track of the frequency of rule-table lookups in a moving window
of time-steps by means of a dynamic ``input-frequency'' histogram
(IFH)\cite[EDD:31.5]{EDD}.  This applies to any rule type.  For
complex rules the IFH reveals the key inputs (neighborhoods or
iso-groups) that maintain gliders and glider-guns, as well as those
that are rarely or never visited.

Figures~\ref{histograms for Life, Eppstein, SapinR} and
\ref{histograms for Variant, Precursor, Sayab} 
show IFH examples for iso-rules generated by the glider-guns shown alongside.
The moving window can be any size but here it is set to 100 time-steps to
allow the IFH to stabilise.  The input-frequency is the
fraction of each iso-group in this window, and is represented by bar
height. The order of bars follows the iso-rule index, from all 1's
(left) to all 0's (right), and a missing bar denotes iso-groups that
have not been visited.
The bar height can be  represented in two ways: either the actual frequency
but (possibly) amplified to show up small bars of rarely visited iso-groups
while the most visited are subject to a maximum cut-off, or
alternatively as log$_2$ frequency to clearly show all bars
but still distinguishes between rare and frequent.
There are also two types of space-time pattern presentation: cells colored according to
their actual values, or colored\footnote{Matching bar/cell colors cycle through 14
contrasting colors, which can be shuffled.} 
according to the IFH bar responsible for a given cell, as in these figures.
These alternatives are toggled on-the-fly.

Figures~\ref{histograms for Life, Eppstein, SapinR} and
\ref{histograms for Variant, Precursor, Sayab} show histograms for
the the binary $v2k9$ glider-gun iso-rules in figure~\ref{significant binary iso-rules}
on a 3$\times$3 Moore neighborhood with 102 iso-groups.
The Sayab histogram is shown both according to actual frequency, and
according to log$_2$ frequency which will be employed in further examples.


DDLab applies the IFH for a number of supplementary functions.
Filtering permits cross referencing a particular iso-group index with
its occurrence in space-time pattern, as well as revealing structures
within a repetitive background. The consequences of mutations can be
monitored by flipping (and restoring) the output of random or selected
iso-groups to a different value.  The Shannon entropy of the histogram
and its variability are applied to automatically categorise rule-space
between order, complexity and chaos.  These functions are discussed
below.

\section{filtering}
\label{filtering}

\enlargethispage{7ex}
\vspace{-2ex}
\begin{figure}[htb]
\textsf{\small
\begin{minipage}[b]{1\linewidth}
\raisebox{1ex}{\fbox{\includegraphics[height=.17\linewidth, bb=6 6 315 273, clip=]{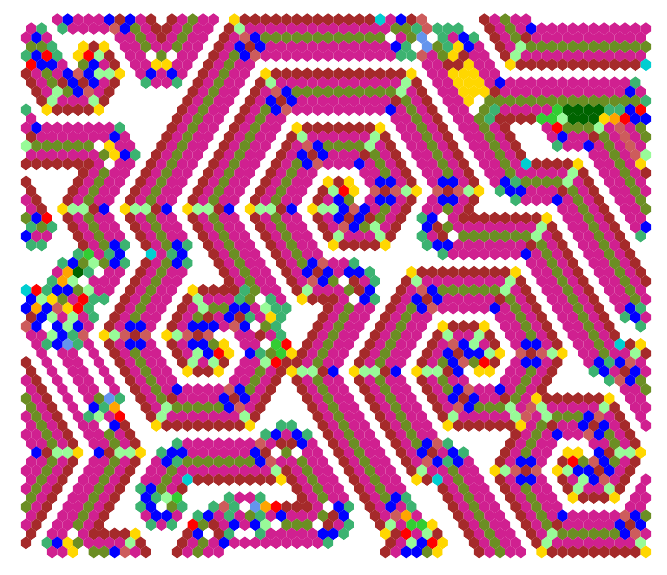}}}
\raisebox{1ex}{\fbox{\includegraphics[height=.17\linewidth, bb=6 6 315 273, clip=]{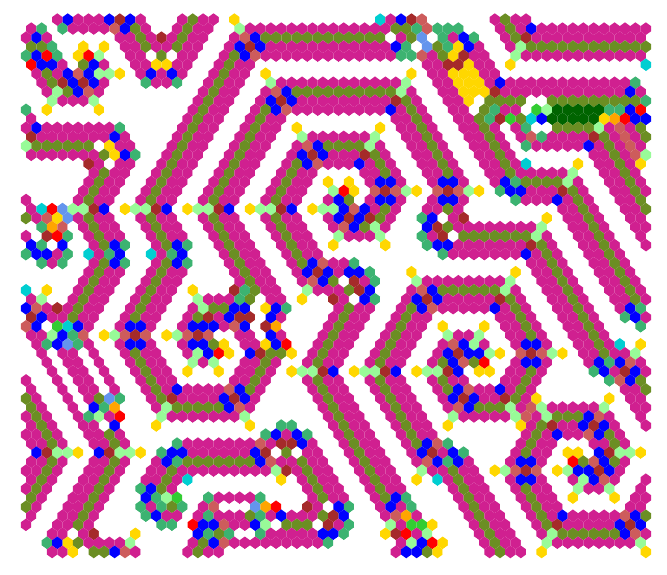}}}
\raisebox{1ex}{\fbox{\includegraphics[height=.17\linewidth, bb=6 6 315 273,bb=3 0 310 264, clip=]{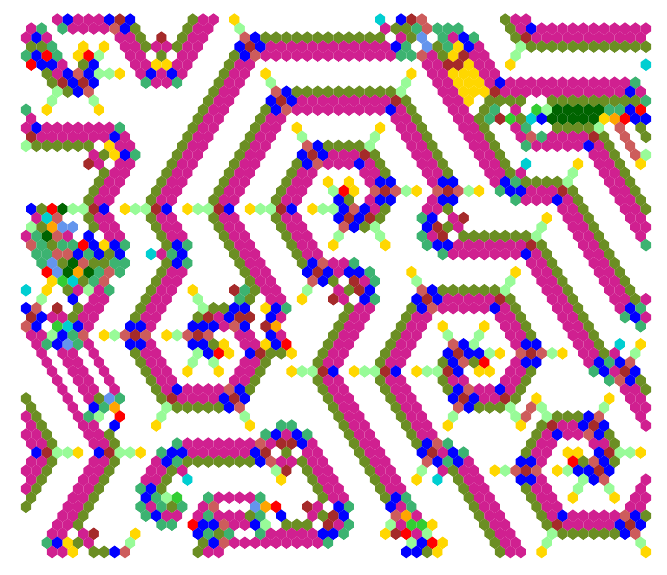}}}
\phantom{x}\raisebox{7ex}{f1-3}\phantom{xxx}\\
\raisebox{1ex}{\fbox{\includegraphics[height=.17\linewidth, bb=6 6 315 273, clip=]{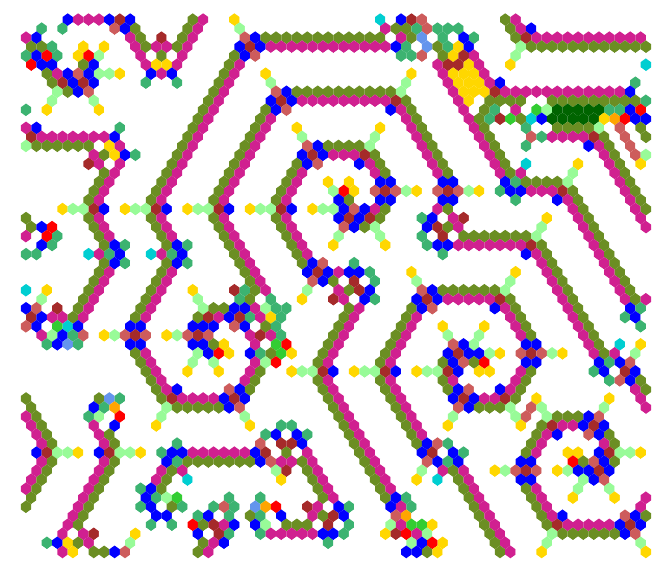}}}
\raisebox{1ex}{\fbox{\includegraphics[height=.17\linewidth, bb=6 6 315 273, clip=]{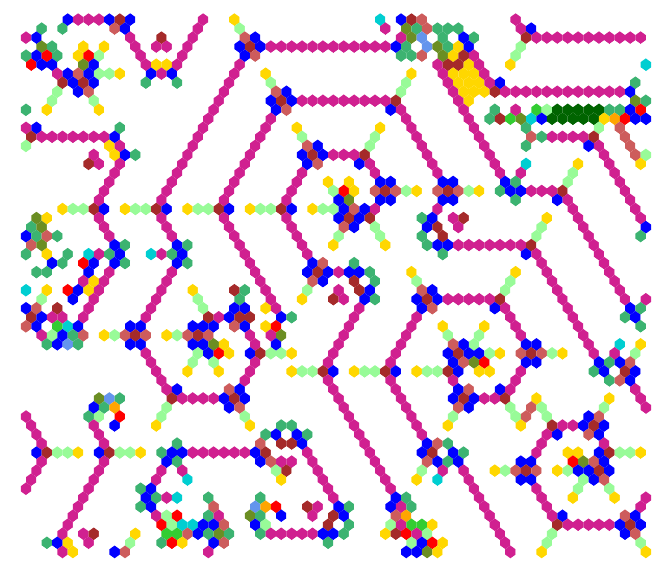}}}
\raisebox{1ex}{\fbox{\includegraphics[height=.17\linewidth, bb=6 6 315 273, clip=]{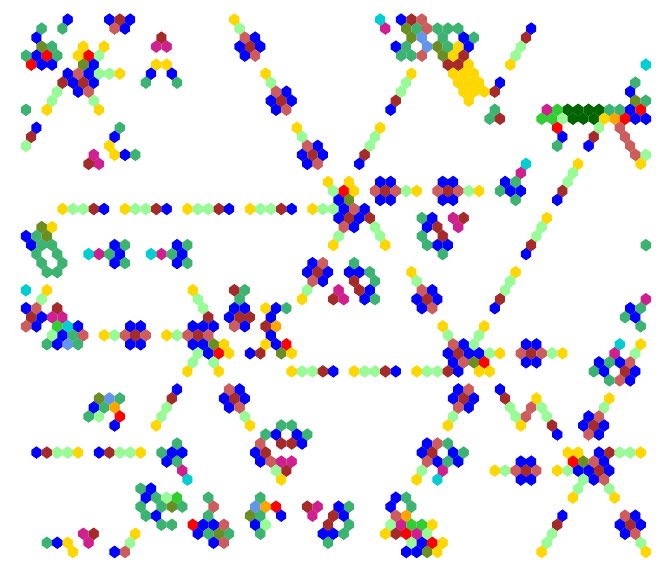}}}
\phantom{x}\raisebox{7ex}{f3-6}\phantom{xxx}
\end{minipage}
\begin{minipage}[b]{1\linewidth}
\includegraphics[width=1\linewidth,height=.25\linewidth]{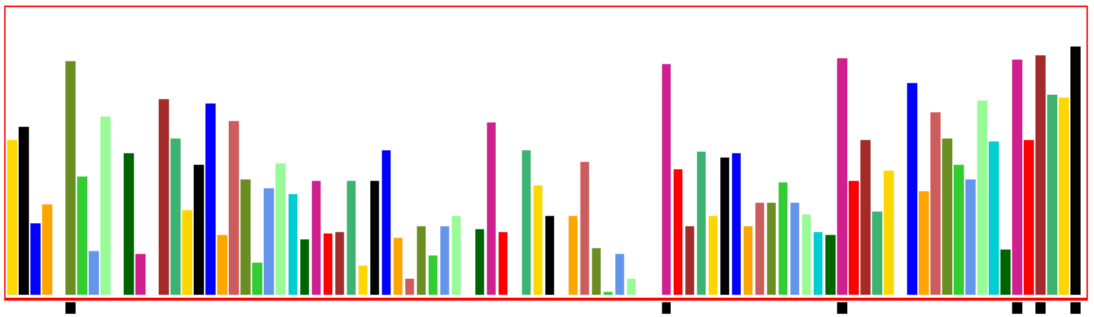}
\end{minipage}
}
\vspace{-4ex}
\caption[emergent spirals]
{\textsf{The $v3k6$ emergent spirals rule with a log$_2$ iso-rule IFH
    size 92 in a moving window of 100 time-steps. Space-time pattern
    cells are colored according to the histogram bar responsible for
    that cell. For the same time-step, 6 stages of filtering f1 to f6
    (high to low frequency) are marked by a black block below the relevant
    bar, and shown as a filtered snapshots. Filtering f1 makes no
    difference to its appearance because white cells are the most
    frequent, represented by the rightmost black bar. 
    In the dynamics of this rule gliders first emerge
    and gradually self-organise into stable spirals.}}
\label{emergent spirals}
\end{figure}
\clearpage

\noindent The IFH allows the progressive
filtering/unfiltering\cite[EDD:32.11.5]{EDD} of space-time patterns
on-the-fly by keyhits as space-time patterns iterate.  This applies
for any rule type according to frequency given by the height of
histogram bars.  Progressive filtering proceeds from high to low
frequency, unfiltering from low to high.  For each frequency
filtered/unfiltered, a black block appears/disappears at the base of
the relevant bar, and the corresponding cell disappears/reappears in
the space-time pattern, whether colored by histogram colors as in
figure~\ref{emergent spirals}, or by value as in
figure~\ref{reaction-diffusion CA}.  Keyhits can remove the entire
filter scheme or reverse (antifilter) the scheme for added
flexibility.  Bars can also be targeted to
filter/unfilter\cite[EDD:32.16.7]{EDD}, and mutated/restored described
in section~\ref{the IFH mutation/filter game} below.

When pattern colors correspond to IFH colors, filtering allows cross
referencing a particular iso-group index with its occurrence in the
space-time pattern, so helps to reveal how complex structures,
gliders, eaters and glider-guns are built and their sensitivity to
mutation. The IFH in figure~\ref{emergent spirals} is set for a moving
window of 100 time-steps to allow the bars to stabilise. However, if
the pattern itself has largely stabilised and the moving window size
is reduced (minimum one time-step), then the few structures that
remain dynamic can be picked out in the IFH by the bars that
continue to oscillate.

\begin{figure}[htb]
\textsf{\small
\includegraphics[width=.26\linewidth, bb=87 94 249 256, clip=]{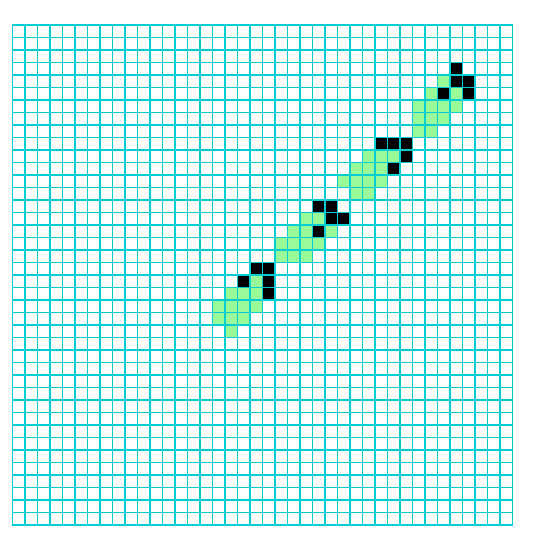}
\includegraphics[width=.26\linewidth, bb=87 94 249 256, clip=]{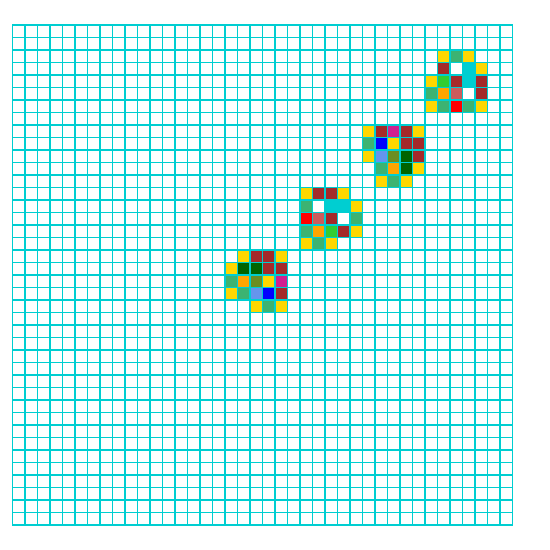}
\phantom{i}
\raisebox{18ex}{\begin{minipage}[t]{.35\linewidth}
The 4 phases of the game-of-Life glider, moving North East.\\
{\it far left}: colors by value with green dynamic trails.\\
{\it near left}: colors corresponding to histogram colors.
\end{minipage}}
\hfill\\
\includegraphics[width=1\linewidth,height=.25\linewidth]{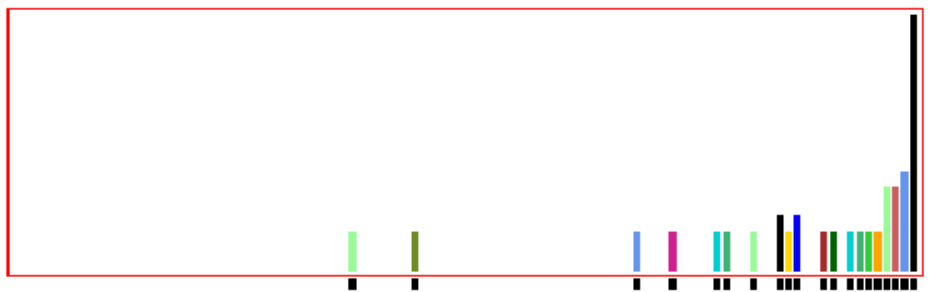}
}
\vspace{-4ex}
\caption[Life-glider iso-histogram]
{\textsf{The log$_2$ iso-rule IFH of the game-of-Life glider
in a moving window of 10 time-steps showing the 20 bars (from a maximum of 102) that are
responsible for the glider's existence. When these are filtered as marked by black blocks,
the iso-group indices, in descending frequency order
(63,56,31,27,22,21,18,15,14,13,10,9,7,6,5,4,3,2,1,0) can be viewed and amended
in a separate window.}}
\label{Life-glider iso-histogram}
\end{figure}

To determine the iso-groups responsible for any particular structure,
say the glider in the game-of-Life, an isolated glider is run to
generate its iso-rule IFH, which if fully filtered will provide the
complete list of the responsible iso-groups in descending frequency
order, as in figure~\ref{Life-glider iso-histogram}.

\vspace{1ex}

\section{the IFH mutation/filter game}
\label{the IFH mutation/filter game}

\begin{figure}[htb]
\textsf{\small
\fbox{\includegraphics[height=.2\linewidth, bb=55 39 251 177, clip=]{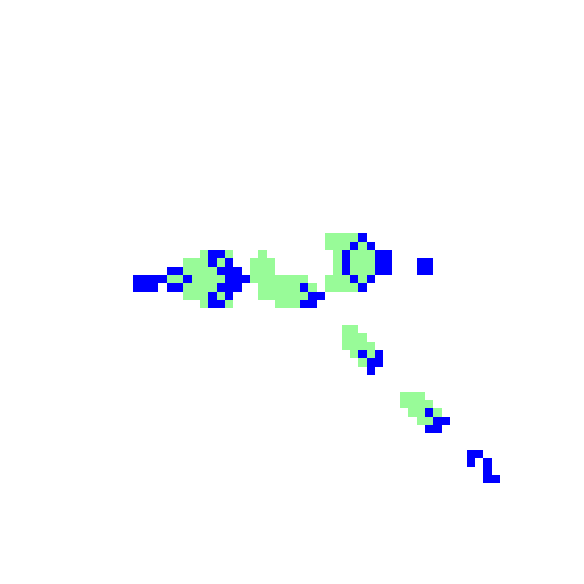}}
\fbox{\includegraphics[height=.2\linewidth, bb=55 39 251 177,  clip=]{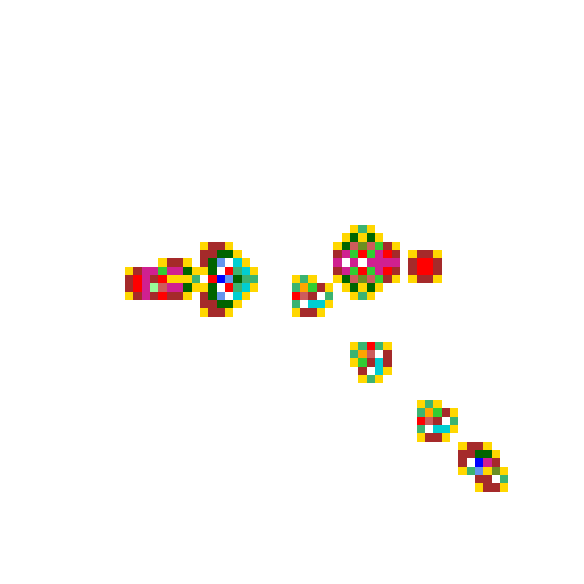}}
\phantom{i}
\raisebox{13ex}{\begin{minipage}[t]{.34\linewidth}
The game-of-Life glider-gun.\\
{\it far left}: colors by value with green dynamic trails.\\
{\it near left}: colors corresponding to IFH colors.
\end{minipage}}
\hfill\\
\includegraphics[width=1\linewidth,height=.25\linewidth]{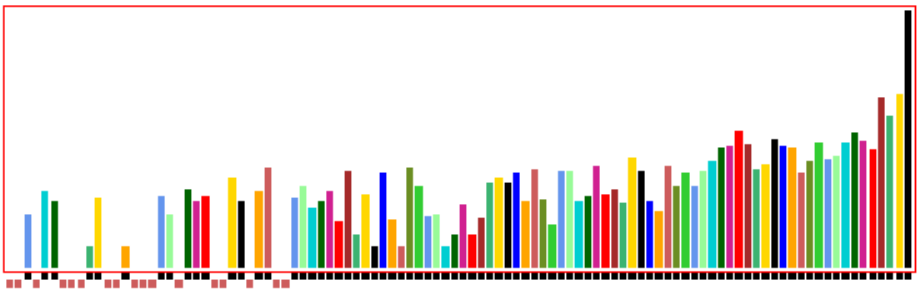}\\[2ex]
\includegraphics[width=.6\linewidth,bb= 1 1 165 29,  clip=]{pdf-figs/LifeM17-hist}
\phantom{i}
\raisebox{8ex}{\begin{minipage}[t]{.35\linewidth}
Blow up of the left lower corner showing indicator blocks: filter (black) and mutation (red).
\end{minipage}}
\hfill\\
The iso-rules (hex) are compared below:\\
\texttt{
00 00 00 00 00 60 03 1c 61 c6 7f 86 a0 ---game-of-Life.\\
34 e6 e4 64 c0 60 03 1c 61 c6 7f 86 a0 ---after all 17 neutral mutations.}\\
}
\vspace{-3ex}
\caption[Life-glider-gun iso-histogram]
{\textsf{The log$_2$ iso-rule IFH of the game-of-Life glider-gun/eater.
The all active bars were firstly filtered, then all 17 neutral bars were mutated. 
The glider-gun/eater is preserved, though any other dynamics would be drastically altered.
}}
\label{Life-glider-gun iso-histogram}
\end{figure}

\noindent The IFH allows interactive (or targeted) rule
mutations\cite[EDD:32.5.4]{EDD} while watching their effects on
space-time patterns, to make/restore single mutations on-the-fly with
keyhits, without the need to pause, in a sort of mutation/filter
game\cite[EDD:16.10.8]{EDD}.  Any number of mutations can be made in
sequence, and restored in reverse order to finally return to the start
rule.  When applied to an iso-rule, mutations conserve isotropy, of
course.

The mutation algorithm can operate in conjunction with on-the-fly
filtering described in section~\ref{filtering}. The keyhit to mutate
will preferentially select an unfiltered iso-group bar at random and
assign a random value different from the current output. For a binary
rule the output is simply flipped.  A red block is shown at the base
of the bar --- beside the black filter block if this is also active.
When the latest mutation is restored its red block is removed.

To observe the effects, the appearance of pattern filtering can be
toggled off/on with a keyhit while the IFH filtering scheme
remains visible.  The mutation game can be the most effective if all active
bars have been filtered because a mutation to an inactive iso-group
will be neutral for self-contained dynamics such as a glider-gun/eater
system.

\begin{figure}[t]
\textsf{\small
\fbox{\includegraphics[height=.23\linewidth, bb=8 14 226 231,  clip=]{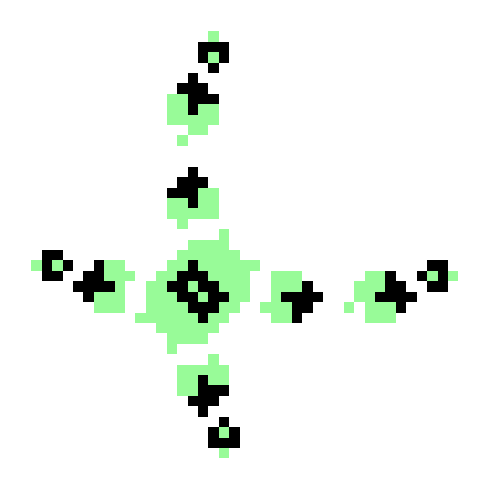}}
\fbox{\includegraphics[height=.23\linewidth, bb=7 10 182 186, clip=]{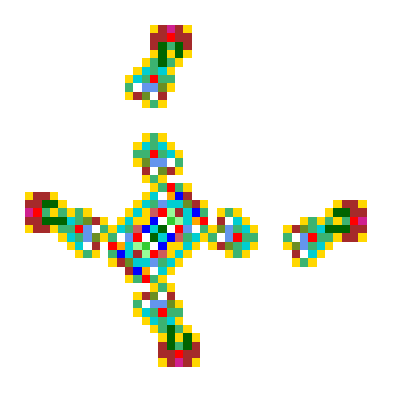}}
\phantom{i}
\raisebox{14ex}{\begin{minipage}[t]{.45\linewidth}
The SapinR glider-gun.\\
{\it far left}: colors by value with green \mbox{dynamic trails.}\\
{\it near left}: colors corresponding to IFH colors.
\end{minipage}}\\
\includegraphics[width=1\linewidth,height=.25\linewidth]{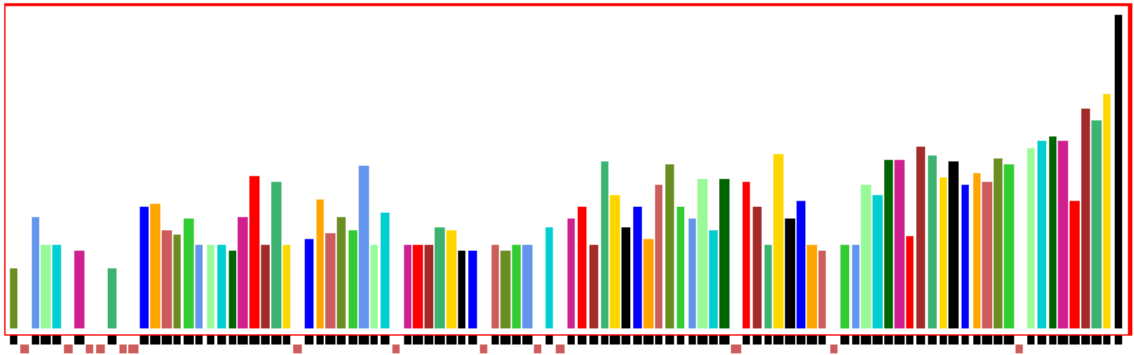}\\
The iso-rules (hex) are compared below:\\
\texttt{
24 01 13 1a 14 20 50 2c 45 05 48 e0 50 ---SapinR.\\
1a fe e9 c4 79 87 23 c3 4a 0d 48 e0 50 ---after all 14 neutral mutations.}
}
\vspace{-3ex}
\caption[SapinR-glider--gun iso-histogram]
{\textsf{The log$_2$ iso-rule IFH of the SapinR rule glider-gun/eaters.
All active bars were firstly filtered,  then all 14 neutral bars were mutated. 
The glider-gun/eaters are preserved, though other dynamics would be drastically altered.
}}
\label{SapinR-glider--gun iso-histogram}
\end{figure}

\begin{figure}[H]
\textsf{\small
\fbox{\includegraphics[height=.23\linewidth, bb=53 70 194 212, clip=]{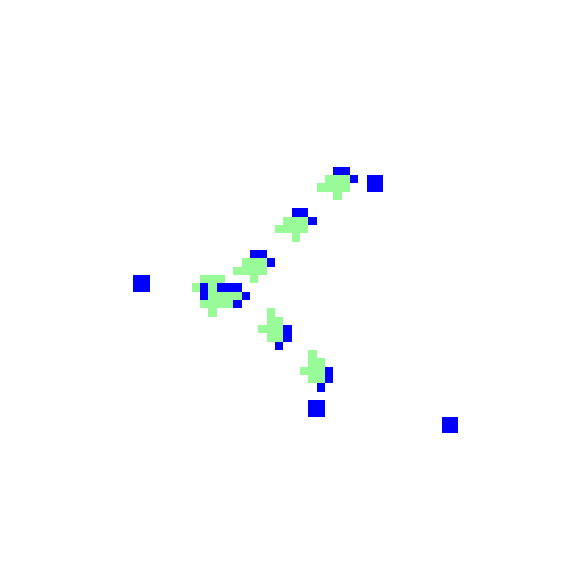}}
\fbox{\includegraphics[height=.23\linewidth, bb=53 70 194 212, clip=]{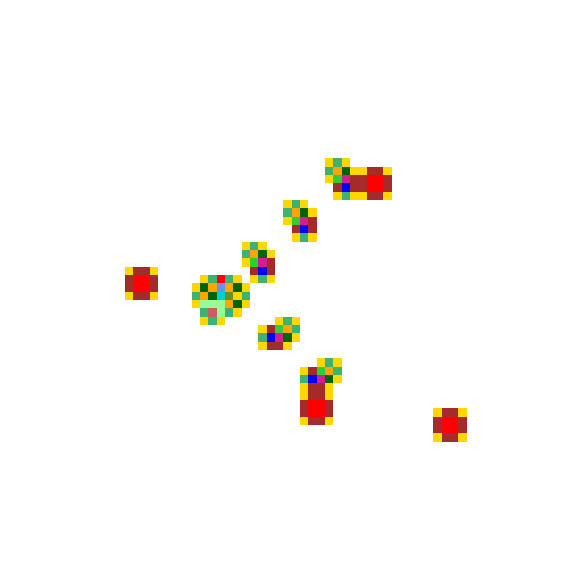}}
\phantom{i}
\raisebox{14ex}{\begin{minipage}[t]{.45\linewidth}
The Sayab glider-gun.\\
{\it far left}: colors by value with green \mbox{dynamic trails.}\\
{\it near left}: colors corresponding to IFH colors.
\end{minipage}}\\
\includegraphics[width=1\linewidth,height=.25\linewidth]{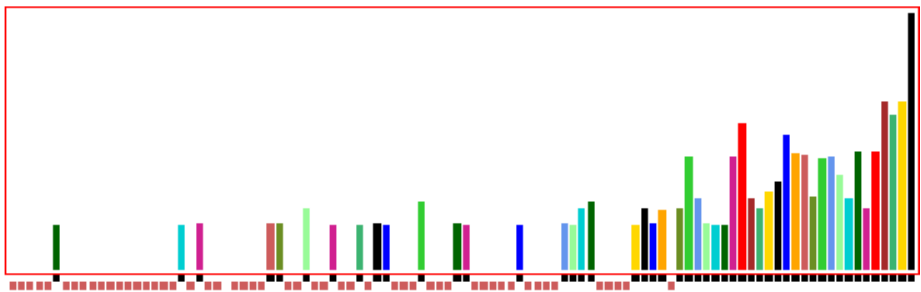}\\
The iso-rules (hex) are compared below:\\
\texttt{
24 01 13 1a 14 20 50 2c 45 05 48 e0 50 ---Sayab.\\
1a fe e9 c4 79 87 23 c3 4a 0d 48 e0 50 ---after all 52 neutral mutations.}
}
\vspace{-3ex}
\caption[Sayab-glider-gun iso-histogram]
{\textsf{The log$_2$ iso-rule IFH of the Sayab rule glider-gun.
All active bars were firstly filtered,  then all 52 neutral outputs mutated. 
The glider-guns are preserved, though other dynamics would be drastically  altered.
}}
\label{Sayab-glider-gun iso-histogram}
\end{figure}
\clearpage

\begin{figure}[htb]
\textsf{\small
\fbox{\includegraphics[width=.22\linewidth, bb=20 3 205 185, clip=]{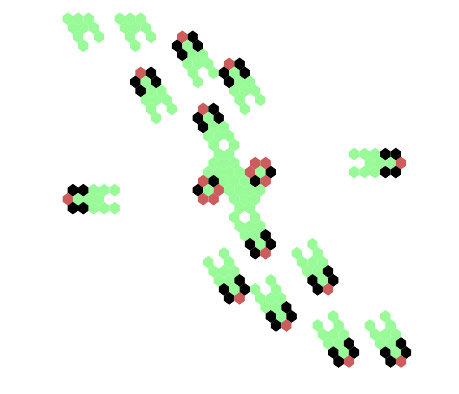}}
\fbox{\includegraphics[width=.22\linewidth, bb=20 3 205 185, clip=]{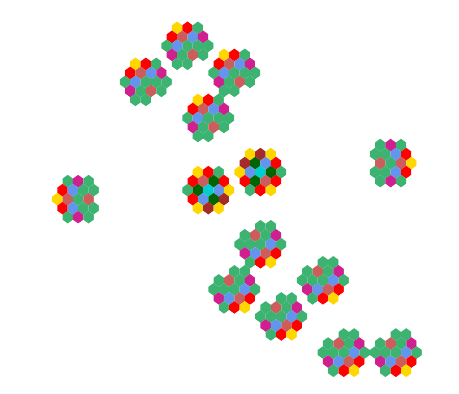}}
\phantom{i}
\raisebox{15ex}{\begin{minipage}[t]{.45\linewidth}
The Beehive glider-gun with NULL boundary conditions.\\
{\it far left}: colors by value with green \mbox{dynamic trails.}\\
{\it near left}: colors corresponding to IFH colors.
\end{minipage}}
\hfill\\
\includegraphics[width=1\linewidth,height=.25\linewidth]{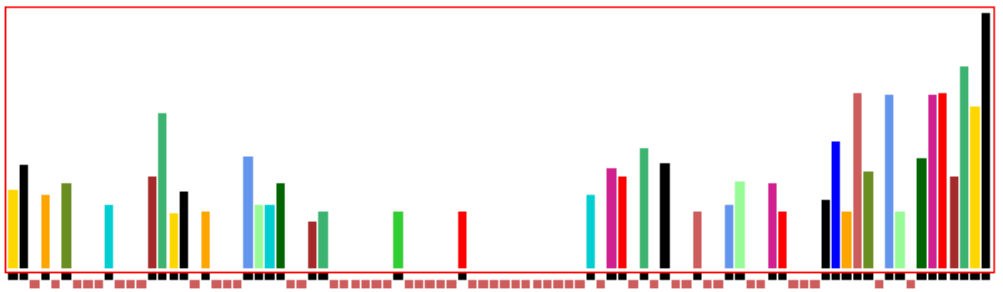}\\
{\it above}: the iso-rule IFH representing 92 iso-groups, 40  are active.\\
The iso-rules (hex) are compared below:\\
\texttt{\scriptsize
8a 20 01 60 2a 20 aa aa a9 aa 2a 99 2a a9 28 0a 2a 2a 69 80 2a 62 84 --Beehive\\[-1ex] 
86 20 85 a0 1a 10 a4 a4 08 69 98 05 62 a8 20 06 02 26 66 80 0a a2 84 --52 neutral mutations}\\[1ex]
\includegraphics[width=.4\linewidth,height=.25\linewidth]{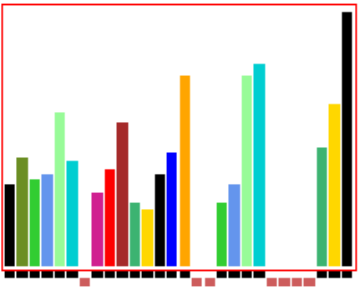}
\phantom{i}
\raisebox{18ex}{\begin{minipage}[t]{.55\linewidth}
{\it left}: The same Beehive rule glider-gun as a k-totalistic rule IFH
representing 28 combinations of totals,
The kcodes (hex) are compared below:\\
\texttt{\footnotesize
0a0682822a1424 ---Beehive\\
0a0282816a2264 ---after 7 neutral mutations.} All one-value mutation were explored
in \cite{beehive-webpage}.
\end{minipage}}\\
}
\vspace{-4ex}
\caption[Beehive glider-gun iso-histogram]
{\textsf{The $v3k6$ Beehive rule\cite{Wuensche05} glider-gun showing both
its log$_2$ iso-rule and k-totalistic IFH.
Because the Beehive rule does not have
eaters to stop gliders, boundary conditions were set to NULL.
The IFH were firstly filtered then neutral outputs randomly mutated, 
with each mutation making one of the two possible
changes because $v$=3.
The glider-gun is preserved, though other dynamics would  drastically be altered. 
}}
\label{Beehive-glider-gun iso-histogram}
\end{figure}

\begin{figure}[htb]
\textsf{\small
\fbox{\includegraphics[height=.23\linewidth, bb=13 8 203 179, clip=]{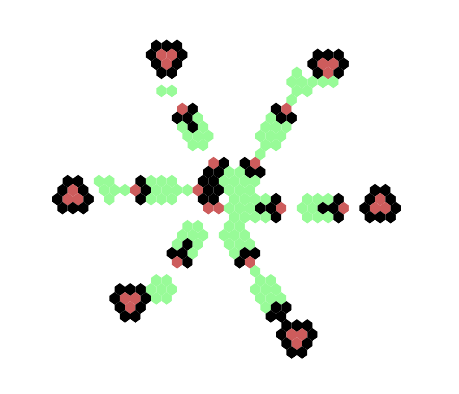}}
\fbox{\includegraphics[height=.23\linewidth, bb=13 8 203 179, clip=]{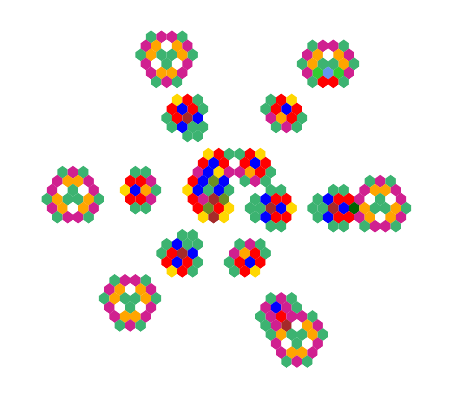}}
\phantom{i}
\raisebox{14ex}{\begin{minipage}[t]{.4\linewidth}
The Spiral glider-gun.\\
{\it far left}: colors by value with green \mbox{dynamic trails.}\\
{\it near left}: colors corresponding to \mbox{histogram colors.}
\end{minipage}}
\hfill\\
\includegraphics[width=1\linewidth,height=.25\linewidth]{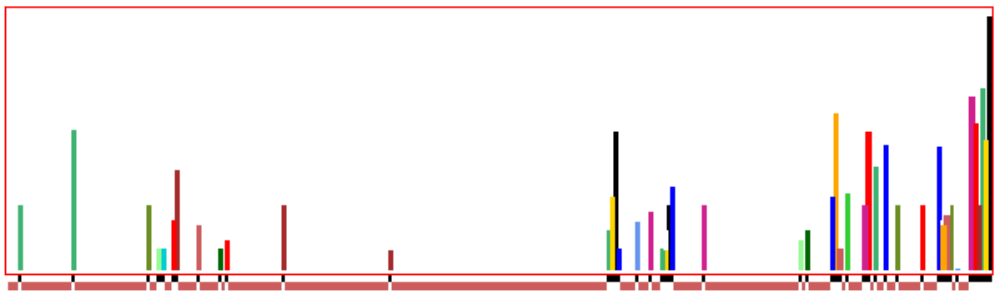}\\
{\it above}: the iso-histogram representing 276 iso-groups, 46 are active.\\
The iso-rules (hex) are compared below:\\
\texttt{\scriptsize
00 a6 58 a6 66 a6 6a aa a6 a8 02 90 08 96 28 02 92 08 98 a6 69 2a 66 65 9a 69 8a 9a a6 69 a2\\[-1.5ex] 
66 6a 96 96 6a a9 a9 a6 66 a9 69 08 82 68 28 02 a6 a2 69 a6 66 66 99 a2 62 a6 98 8a 26 08 8a\\[-1.5ex]  
26 64 9a 40 2a 62 84 --Spiral\\
98 08 06 21 15 50 94 45 59 50 52 60 51 58 94 49 44 95 25 60 96 84 89 10 44 94 16 00 50 85 19\\[-1.5ex] 
99 81 00 29 80 54 56 51 08 44 00 08 18 46 10 00 41 46 04 59 09 10 46 04 81 81 00 82 41 04 69\\[-1.5ex] 
a9 9a a5 00 18 02 84 ---after all 233 neutral mutations.}\\[1ex]
\includegraphics[width=.4\linewidth,height=.25\linewidth]{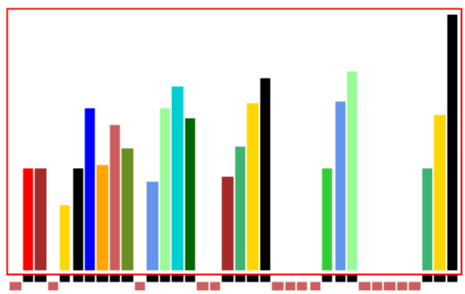}
\phantom{i}
\raisebox{15ex}{\begin{minipage}[t]{.55\linewidth}
{\it left}: the k-totalistic histogram representing 36 combinations of totals,
The kcodes (hex) are compared below:\\
\texttt{\scriptsize
020609a2982a68aa64 ---Spiral.\\
410605a01805a845a4 --after 14 neutral mutations.
}
\end{minipage}}\\
}
\vspace{-4ex}
\caption[Spiral glider-gun iso-histogram]
{\textsf{The Spiral glider-gun showing both
its log$_2$ iso- and k-totalistic histograms.
The histograms were firstly filtered then neutral outputs randomly mutated, 
with each mutation making one of the two possible
changes because $v$=3. 
The glider-gun is preserved, though other dynamics would be altered. 
}}
\label{Spiral-glider-gun iso-histogram}
\vspace{-1ex}
\end{figure}

\begin{figure}[htb]
\textsf{\small
\includegraphics[height=.26\linewidth, bb=15 4 492 482, clip=]{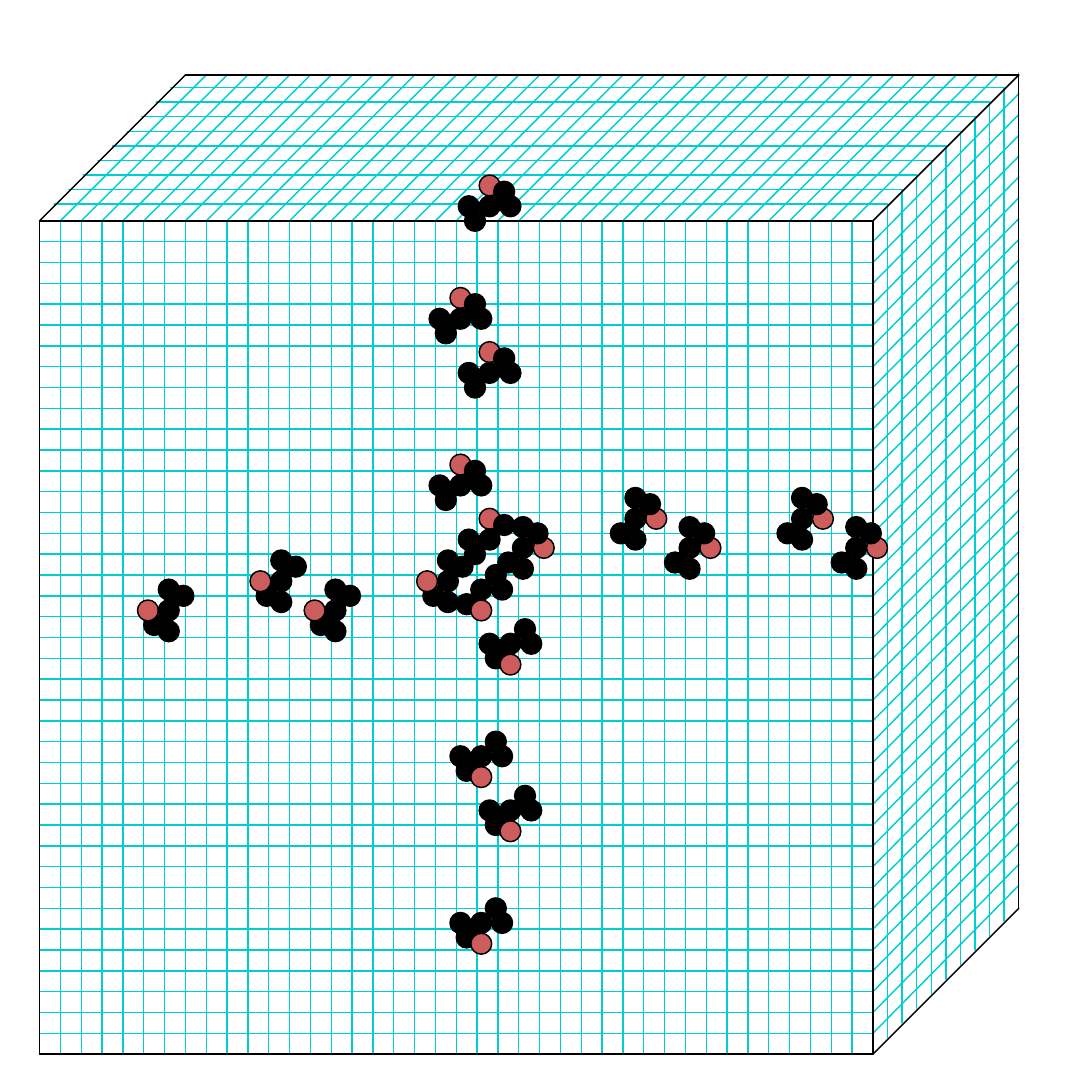}
\includegraphics[height=.26\linewidth]{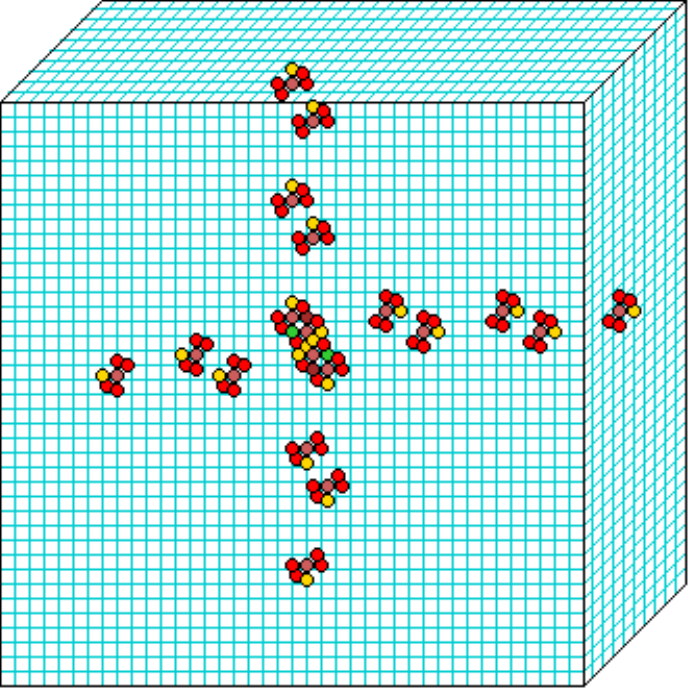}
\phantom{i}
\raisebox{14ex}{\begin{minipage}[t]{.45\linewidth}
The Spiral 3d glider-gun.\\
{\it far left}: colors by value.\\
{\it near left}: colors corresponding to IFH colors.
\end{minipage}}
\includegraphics[width=1\linewidth,height=.25\linewidth]{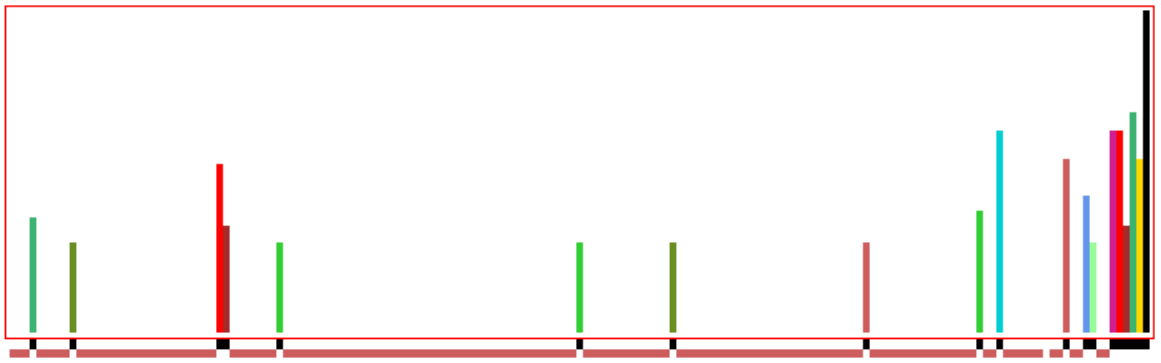}\\
{\it above}: the iso-histogram representing 171 iso-groups, 19 are active.\\
The iso-rules (hex) are compared below:\\
\texttt{\scriptsize
2a 04 8a 01 22 04 06 56 04 95 a4 96 44 84 82 22 52 25 01 06 25 51 8a 61 64 15 59 56 19 21 06\\[-1.5ex] 59 20 a1 60 18 02 45 01 12 00 02 84 --Spiral3d.\\
04 29 59 98 45 59 59 29 02 6a 69 49 22 59 58 58 04 90 58 50 90 82 61 06 91 0a 22 28 62 4a 69\\[-1.5ex] 
 84 65 26 89 86 51 82 58 8a 90 a2 84 ---after all 152 neutral mutations.}
}
\vspace{-1ex}
\caption[Spiral3d-glider-gun iso-rule IFH]
{\textsf{The log$_2$ iso-rule IFH of the Spiral3d glider-gun
with iso-rule size=171 --- shorter than Spiral2d size 276.
The 3d glider-gun emerges spontaneously, but stable structures (such as eaters)
are absent. The IFH were firstly filtered then neutral outputs mutated. 
The glider-guns are preserved, though other dynamics would be altered.
}}
\label{Spiral3d-glider-gun iso-histogram}
\end{figure}

In the examples in 
figures~\ref{Life-glider-gun iso-histogram} to 
\ref{Spiral3d-glider-gun iso-histogram}
the initial state is an isolated glider-gun, in most cases contained by
eaters.  All active bars are first filtered marking them with black
blocks, then mutations are made which automatically and randomly seek
out unfiltered inactive iso-groups represented by missing bars,
marking them with red blocks.  Because these are neutral mutations the
relevant glider-gun system must be preserved.

The mutation game is applied to the $v2k9$ rules: game-of-Life,
SapinR, Sayab, and also the hex lattice $v3k6$ Beehive rule and $v3k7$
Spiral-rule --- both of which feature a 3d glider-gun
(figure~\ref{beehive rule glider-gun}). Sayab, SapinR, the 3d
Beehive-rule and the Spiral-rule (both 2d and 3d) are special in the
sense that their glider-guns emerge spontaneously from random initial
states, whereas glider-guns for other rules covered in this paper need
careful construction.

\vspace{2ex}
The take-home message from these experiments is the existence of a vast web of 
connected but diverse iso-rules that support the same glider-gun systems.

Once all gaps have been mutated, further mutations will hit active
bars disrupting glider-guns, which can be retrieved with keyhits to
unmutate and eventually (or immediately) reinstate the original rule.
A keyhit can also reinstate the original glider-gun pattern.

Mutations to active iso-groups can be done progressively by
unfiltering the least active bars, or a specific mutation index can be
selected.  Any mutation to an active iso-group will change the current
space-time dynamics to a greater or lesser extent. If the change is
interesting, a new glider or eater, the mutation can be retained. An
undesirable change such as excessive disorder can be repaired.  Among
functions in DDLab that can assist in these experiments are on-the-fly
keyhits for a random pattern, and for a random central
block\cite[EDD:32.8.1]{EDD}, respecting densities previously
specified.  Space-time patterns can be paused at any time to edit or
save the current state or iso-rule and access other
functions\cite[EDD:32.16]{EDD}.

Exploring state-space genetically close to significant k-totalistic
rules was done for the Beehive-rule\cite{Wuensche99} and the
Spiral-rule\cite{Wuensche&Adamatzky2006}, looking at all possible
single k-totalistic mutants\cite{beehive-webpage,spiral-webpage} and
some significant alternative behaviours were discovered. A finer
grained search based on iso-rules is now possible.

\section{input-entropy and min-max variability}
\label{input-entropy and min-max variability}
\vspace{2ex}

\begin{figure}[htb]
\textsf{
\includegraphics[width=1\linewidth]{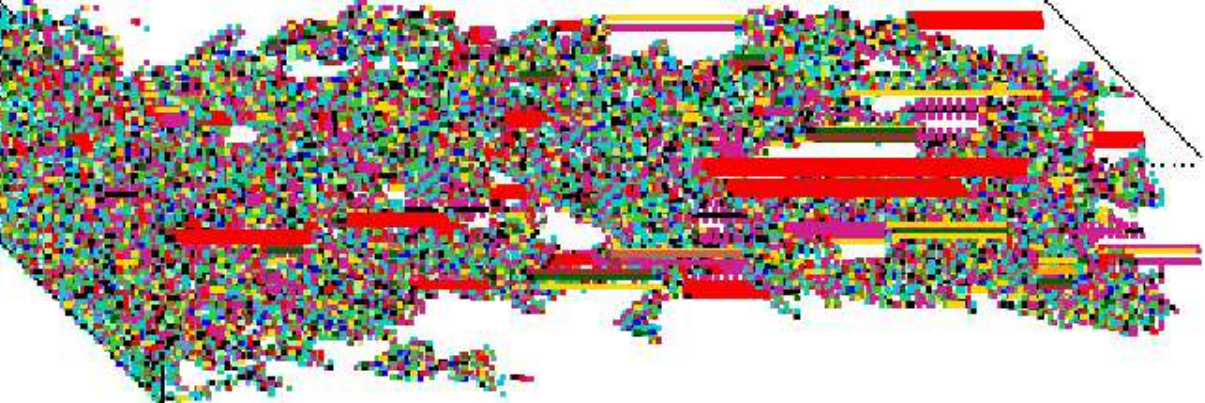}\\
\textcolor{white}{--------------}1$\longrightarrow$\textcolor{white}{------------------------------}%
time-steps $\longrightarrow$\textcolor{white}{------------------------------}250\\[.5ex]
\begin{minipage}[b]{.12\linewidth}
\textcolor{white}{entropy}\vspace{.8ex}1\\[11.5ex]
\phantom{x}entropy\\[13ex]
\phantom{entropy}\vspace{-1.2ex}0
\end{minipage}
\hfill
\fbox{
\begin{overpic}[height=.4\linewidth, width=.83\linewidth]{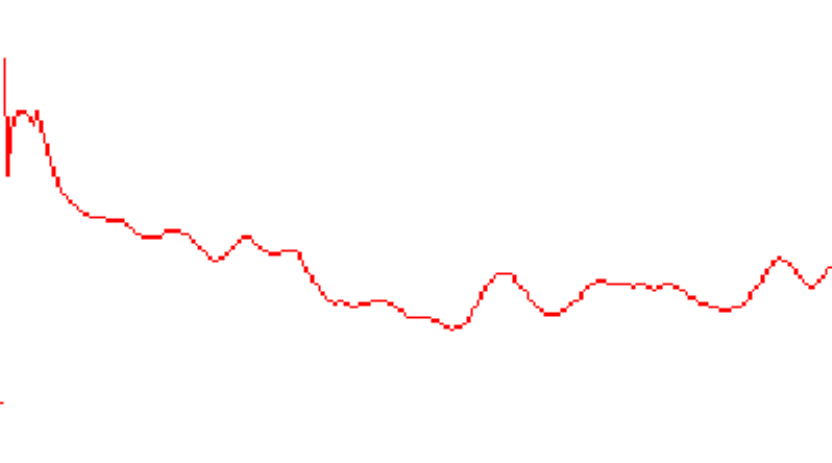}
\put(55,19.5){\large----------}
\put(50,12){\large----------}
\put(57,17){\large$\uparrow$}
\put(57,14.5){\large$\downarrow$}
\put(55,21.5){min-max}
\put(7,30){\large$\leftarrow$}
\put(12,30){ignore initial time-steps}
\end{overpic}
}}
\vspace{-2ex}
\caption[Life entropy time-plot]
{\textsf{{\it Top}: 250 time-steps of the game-of-Life from a random 40$\times$40 initial state
with a density of 30\% shown as a type of axonometric projection where
time-steps progress from left to right. Colors follow the iso-rule IFH.
{\it Above}: The entropy plot generated simultaneously: x-axis time-steps 1 to 250, 
y-axis normalised Shannon entropy $H_N$ from 0 to 1. The min-max entropy
is the greatest upslope after a short initial run,
22 time-steps for the scatter-plots in figure~\ref{entropy scatter-plots}.
}}
\label{Life-entropy}
\end{figure}

\noindent The entropy of the iso-rule IFH, the input-entropy, can be measured
and plotted over time.  The average entropy and its variability over a
window of time-steps indicates the quality --- ordered/complex/chaotic
--- of the dynamics, very roughly as follows,

\begin{center}
\begin{tabular}[b]{r|c|c|c} 
                    &  order & complexity & chaos \\\hline
      mean-entropy  & low    &  medium    & high  \\\hline           
entropy-variability & low    &  high      & low  
\end{tabular}
\end{center}

Both the mean-entropy and entropy-variability are measured from a run
of time-steps starting from a random (but possible biased) initial
state, discounting a short initial run to allow the dynamics
to settle into its typical behaviour. 

\enlargethispage{2ex}
The Shannon entropy of the IFH measures its heterogeneity\cite{Wuensche99}.
The input-entropy\cite[EDD:33.1]{EDD}
$H$, at time-step $t$, for one time-step ($w$=1), is given by 
$H^t = -\sum_{i=0}^{S-1} \left( {Q_{i}^{t}}/{n} \times
log_{2}\left({Q_{i}^{t}}/{n}\right)\right)$, where $Q_{i}^{t}$ is the
actual height of bar $i$ (input-frequency) at time-step~$t$. $S$ is the number of bars
(rule-table size) and $n$ is the CA lattice size.  The normalised
input-entropy $H_N$ is a value between 0 and 1, $H_N=H^t/log_{2}n$
used in the graphic display as in figure~\ref{Life-entropy} and is usually averaged
over a small moving window of time-steps (say $w$=10) to smooth an
otherwise jagged plot.

The mean entropy is the average $H_N$  over a longer run of time-steps.
The entropy-variability known as min-max\footnote{Variability by min-max is
  preferable to the previously adopted\cite{Wuensche97,Wuensche99} standard deviation
  which gives a high value for monotonic entropy decrease, characteristic of 
  a foreground pattern gradually dying out, which would be misleading to identify complex
  dynamics. Min-max is low for dying out dynamics so this problem is
  avoided.} is the maximum up-slope found in a run of time-steps --- the rise 
  in entropy following a lower value.

High entropy variability can be produced by glider dynamics,
because collisions create local chaos raising the entropy, from which
gliders re-emerge lowering entropy. The basic argument is that if
the entropy continues to vary sufficiently in typical dynamics, moving
both up and down, then some kind of large scale structural
interactions are unfolding.  As well as glider dynamics, this might
include competing zones of ordered domains, of order and chaos, of
domains of competing chaos, or some combination of the above.

Low entropy variability is a consequence of both steady chaos or
steady order especially when patterns stabilise or freeze.  However,
the mean entropy for chaos is high, for order low. In this way the
quality of dynamics can be distinguished.

\section{automatically classifying rule-space}
\label{automatically classifying rule-space}

The two measures, mean entropy and min-max entropy variability, are
applied for an automatic classification of rule-space by creating
scatter plots in DDLab for large samples of rules. The plots
distribute rule behaviour according to a merging continuum of
order/complexity/chaos on a 2d surface, and allow a targeted
examination of individual rules or rule sub-groups at characteristic
locations on the plot. Details and examples for creating, sorting,
probing and interpreting the scatter plots are provided in
\cite[EDD:33]{EDD}.

\begin{figure}[H]
\textsf{\small
\begin{center}
\begin{minipage}[b]{1\linewidth}
\includegraphics[height=.4\linewidth]{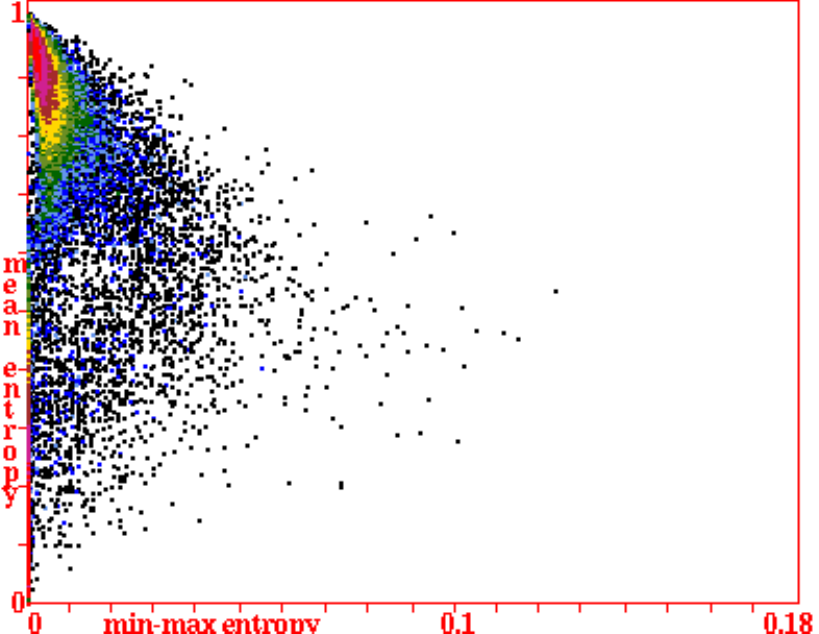}
\hfill
\includegraphics[height=.4\linewidth,bb=3 1 305 284, clip=]{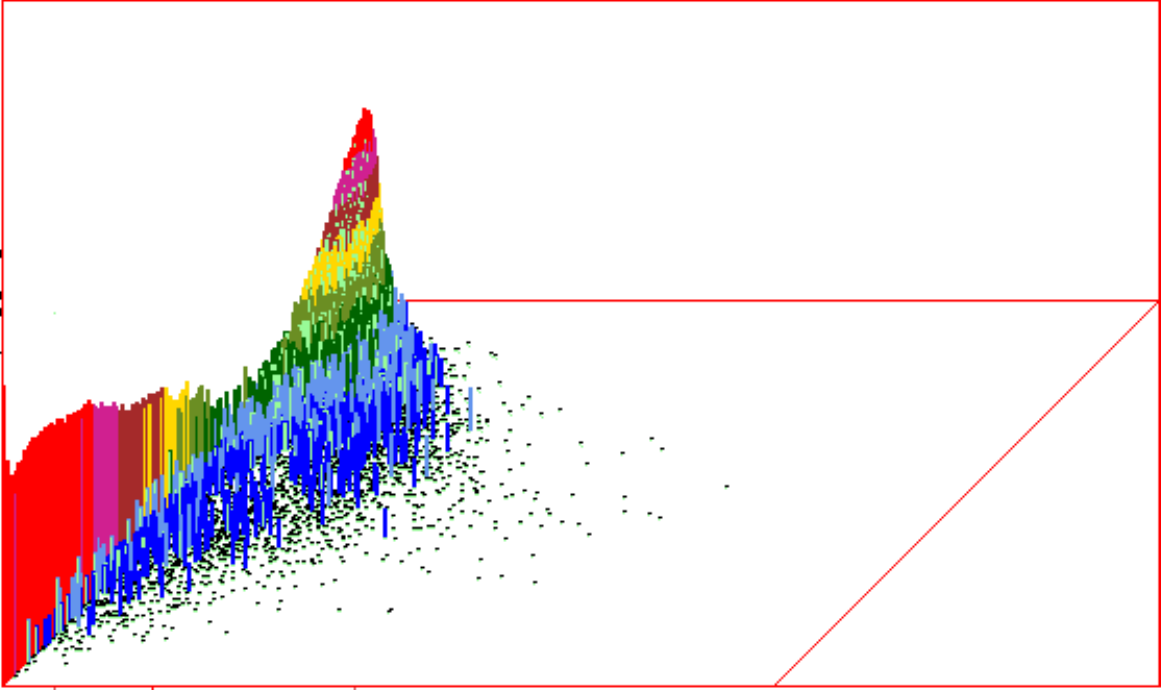}\\
(a) orthogonal lattice $v2k9$, approx density 0/1: seed=70/30, iso-rule=70/30.
\end{minipage}\\[2ex]
\begin{minipage}[b]{1\linewidth}
\includegraphics[height=.4\linewidth]{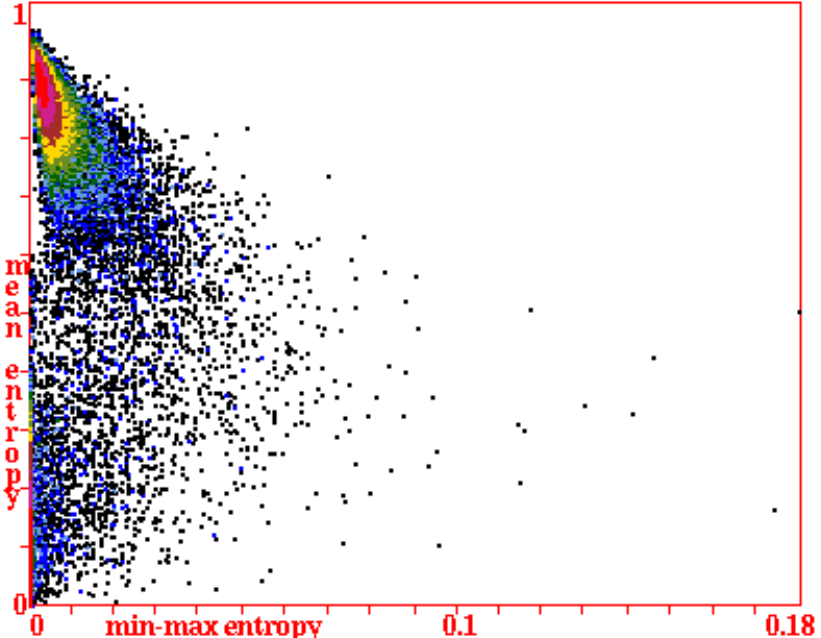}
\hfill
\includegraphics[height=.4\linewidth,bb=3 1 305 284,  clip=]{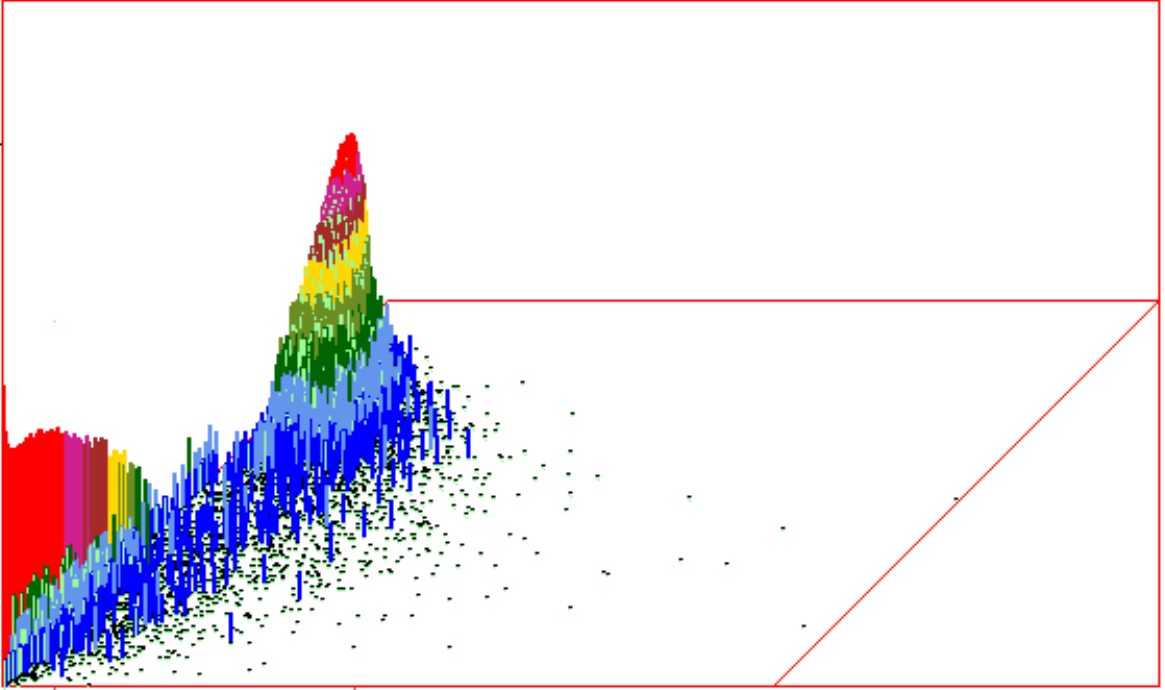}\\
(b) hexagonal lattice $v3k7$, approx density 0/1/2: seed=70/15/15, iso-rule=60,20,20.
\end{minipage}
\end{center}
}
\vspace{-3ex}
\caption[entropy scatter-plots]
       {\textsf{Mean-entropy/entropy-variability (min-max)
    scatter plots based on 2d 60$\times$60 2d the iso-rule IFH.
    The sample size is 50000, collected overnight on a laptop.
    (a) $v2k9$ square 2d as for the game-of-Life, and (b) $v3k7$ 
    hex 2d as for the Spiral rule. {\it Left}: Each dot represents one or more rules
    falling within the squares of a 256x256 grid where the x-axis is
    min-max entropy, y-axis is mean entropy, with dot colors
    indicating the pile-up frequency.  This presentation allows
    iso-rule-space to be probed to examine the rules and dynamics.
    {\it Right}: with a z-axis showing log$_2$ rule frequency.
    Measures were started after 22 initial time-steps and taken over
    the subsequent 200 time-steps. The approximate densities of the
    random initial state and the iso-rule are indicated.}}
       \label{entropy scatter-plots}
\end{figure}

Hitherto the scatter plots were based on full
rule-tables\cite{Wuensche97,Wuensche99} even when made
isotropic\cite{Gomez2015,Gomez2017}, or were based on totalistic
rules\cite{Wuensche05}.  Now the scatter plots can be based on the
iso-rule paradigm as in figures~\ref{entropy scatter-plots}.  Further
investigation of these recent plots will be held over for a subsequent
paper.

To construct the scatter plots in DDLab, random iso-rules and initial
states are generated but with biases in favour complex dynamics; in
spite of this the dynamics captured is mostly chaotic.  The bias
criteria can follow known logically universal rules or some other
conjecture.  For each successive rule, the space-time pattern is run
from a set of random initial states. For each initial state, after a
delay to allow the CA to settle into its typical behavior, the
variability of the input-entropy and the mean entropy are recorded.
Then the average results from the set of initial states are plotted
--- the entropy variability \mbox{($x$-axis)} against the mean entropy
($y$-axis) --- and the data for each rule is appended to a file.

\begin{figure}[htb]
\includegraphics[width=1\linewidth, bb=17 84 686 526, clip=]{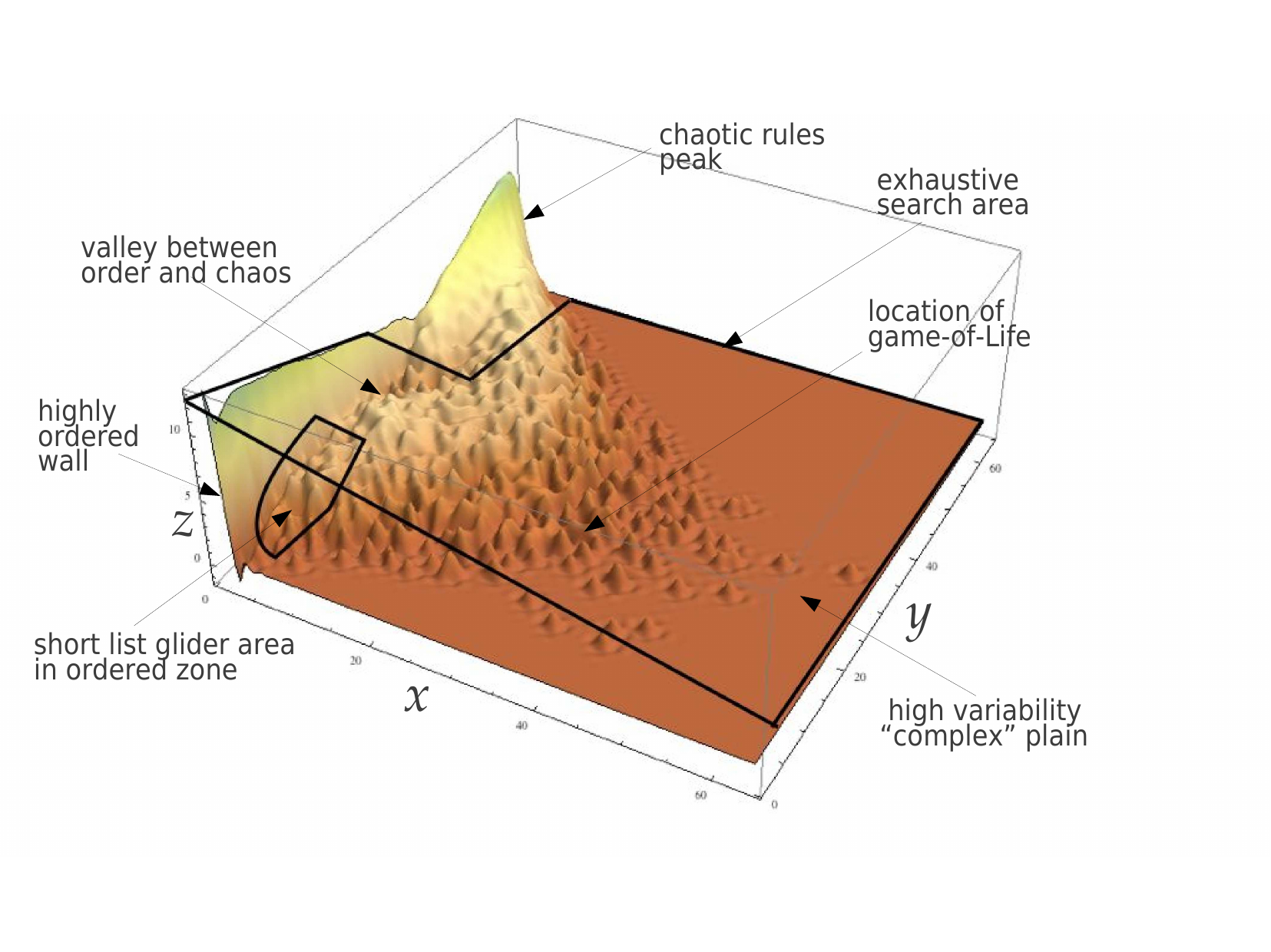}
\vspace{-6ex}
\caption[min-max-xyz.ps]
{\textsf{The typical shape of the x-y-z entropy-min-max scatter plot
    with characteristic dynamical behaviours found in different parts
    of the landscape and the basis for a probing search avoiding the
    chaotic peak. The high variability complex plain contains gliders
    and interacting mobile structures but is unlikely to support
    glider-guns because of the over-active dynamics and rare stable
    structure. This figure is taken from our 2015
    paper\cite{Gomez2015} for 93000 isotropic $v2k9$ 2d rules but
    where the entropy was based on the full 512 rule-table instead of
    the shorter 102 iso-rule-table in figure~\ref{entropy scatter-plots}.
}}  \label{min-max-xyz typical plot}
\end{figure}

Probing various locations of a sorted plot\cite[EDD:33.6]{EDD} with the pointer
selects rules or rule patches which can then be listed and run in sequence
to see the dynamics, or scanned automatically in
blocks of time-steps. Figure~\ref{min-max-xyz typical plot} shows
the locations of characteristic dynamical behaviours.

\section{summary}

Isotropy is arguably the proper canvas for CA logical universality
to play out based on glider-gun/eater dynamics, and we have investigated the
few such rules we are aware of --- the game-of-Life, Sayab, the Spiral rule,
and others, some of which were discovered with earlier analogous 
methods\cite{Adamatzky-beehive,Adamatzky&Wuensche2006,Gomez2015,Gomez2017,Gomez2018,Gomez2020}. 

Isotropic notations for binary CA exist such as the Hensel for
Golly\cite{Golly} and Sapin's\cite{Sapin2004,Sapin2010}, but a general
systematic approach to encompass multi-value in one, two and three
dimensions was missing, so we have proposed the iso-rule paradigm in
this paper.  Iso-rules are based on a lookup-table of \mbox{iso-groups},
assemblies of all rotated/reflected neigborhood configurations which
can be examined graphically in DDLab.  Iso-rules-tables are ordered in
the spirit of Wolfram's classical
convention\cite{wolfram83,wolfram2002}.

Iso-rules provide an intermediate granularity between isotropic rules
based on full lookup-tables, and isotropic subsets --- totalistic,
reaction-diffusion and survival/birth rules. DDLab is able to convert
these rule types to \mbox{iso-rules}, which then become subject to all
DDlab's other functions\cite{EDD}.

The input-frequency histogram and its mutation/filter game has the
potential for investigating the low-level drivers of
glider/glider-gun/eater dynamics.  Input-entropy and its variability
distinguish iso-rules according to ordered/complex/chaotic dynamics,
and allow the automatic collection of large samples of classified
iso-rule-space to search for new and interesting \mbox{iso-rules}.
These method are now available for future research.

\section{acknowledgements}  
\label{acknowledgements}
Figures and experiments were made with DDLab\cite{Wuensche-DDLab}.
J.M. G\'omez Soto acknowledges his residency at Discrete Dynamics Lab, 
and financial support from the Research Council of Mexico (CONACyT).

\end{document}